\documentclass[10pt, a4paper]{article}
\usepackage[left=1in, right=1in, top=1in, bottom=1in]{geometry}
\usepackage{setspace}
\usepackage{float}
\usepackage{setspace}
\usepackage{color,soul}

\usepackage[dvipsnames]{xcolor}
\usepackage{tcolorbox}

\newcommand{\gammadist}{\Gamma}

\usepackage{amsfonts,amsmath,amssymb,amsthm,enumerate,array}
\usepackage{bm,dsfont,multicol}
\usepackage{graphicx,color}
\usepackage{graphicx,graphics,psfrag,booktabs}
\usepackage[authoryear]{natbib}
\usepackage{caption}
\usepackage{rotating,threeparttable,booktabs}
\usepackage{mathrsfs}
\usepackage{tcolorbox}

\usepackage{titlesec}
\titleformat{\section}{\large}{\textbf\thesection.}{.5em}{\textbf}
\titlespacing{\section}{0pt}{*3}{*2}
\titleformat{\subsection}{\normalfont}{\textbf\thesubsection.}{.5em}{\textbf}
\titlespacing{\subsection} {0pt}{*3}{*2}
\titleformat{\subsubsection}{\normalfont}{\textbf\thesubsubsection.}{.5em}{\textbf}

\newtheorem{thm}{Theorem}

\newtheorem{cor}{Corollary}
\newtheorem{rmk}{Remark}
\newtheorem{proposition}{Proposition}

\renewcommand{\thesection}{$\S$\arabic{section}}

\renewcommand{\thesection}{\textsf{\arabic{section}}}
\renewcommand{\thesubsection}%
{\thesection.{\textsf{\arabic{subsection}}}}
\renewcommand{\thesubsubsection}%
{\thesubsection.{\textsf{\arabic{subsubsection}}}}
\def\d{\textup{d}}
\allowdisplaybreaks[4]

\begin{document}

\title{Efficient importance sampling for copula models\footnote{Supported by Zhejiang University K. P. Chao's High Technology Development Foundation (No. 2022RC002). Email addresses: chengsiang@zju.edu.cn, cdffuh@gmail.com, txpang@zju.edu.cn.
}}
\author{Siang Cheng$^{1}$, Cheng-Der Fuh$^{2}$, Tianxiao Pang$^{1}$}
\date{}	
\maketitle

\begin{center}
{\small \centerline{$^{1}$School of Mathematical Sciences, Zhejiang University, Hangzhou 310058, P. R. China}}
{\small \centerline{$^{2}$Graduate Institute of Statistics, National Central University, Taoyuan County 32049, Taiwan}}
\end{center}

\begin{abstract}
In this paper, we propose an efficient importance sampling algorithm for rare event simulation under copula models. In the algorithm, the derived optimal probability measure is based on the criterion of minimizing the variance of the importance sampling estimator within a parametric exponential tilting family. Since the copula model is defined by its marginals and a copula function, and its moment-generating function is difficult to derive, we apply the transform likelihood ratio method to first identify an alternative exponential tilting family, after which we obtain simple and explicit expressions of equations. Then, the optimal alternative probability measure can be calculated under this transformed exponential tilting family. The proposed importance sampling framework is quite general and can be implemented for many classes of copula models, including some traditional parametric copula families and a class of semiparametric copulas called regular vine copulas, from which sampling is feasible. The theoretical results of the logarithmic efficiency and bounded relative error are proved for some commonly-used copula models under the case of simple rare events. Monte Carlo experiments are conducted, in which we study the relative efficiency of the crude Monte Carlo estimator with respect to the proposed importance-sampling-based estimators, such that substantial variance reductions are obtained in comparison to the standard Monte Carlo estimators.
\end{abstract}
\medskip


\textbf{MSC2020 subject classifications}: 62D05, 62H05. \\


\section{Introduction}

Many statistical applications involve calculating a functional of the form $E[\Upsilon({\bm X})]$, where ${\bm X}=(X_{1},\ldots, X_{d})^{'}: \Omega \mapsto \mathbb{R}^{d}$ is a random vector on a probability space $(\Omega, \mathscr{F}, P)$ and $\Upsilon: \mathbb{R}^d \mapsto \mathbb{R}$ is a measurable function. Here the superscript $'$ denotes the transpose operation. If the components of $\bm X$ cannot be assumed to be independent, it is popular to model the distribution of $\bm X$ using a copula. That is, we have
\begin{align*}
P(X_1 \leq x_1, \ldots, X_d \leq x_d) = C(F_{1}(x_{1}), \ldots, F_{d}(x_{d}); \delta),
\end{align*}
where $F_{j} (x) := P( X_j \leq x), j = 1, \ldots, d$, are the marginal cumulative distribution functions (CDFs), $C: [0,1]^d \mapsto [0,1]$ is a copula function, and $\delta$ is a parameter. The copula model has become a popular, well-developed modeling tool in a variety of fields, as it couples multivariate distributions to their one-dimensional marginal CDFs. The reader is referred to Joe \cite{J1997}, Nelsen \cite{N2006}, and Embrechts \cite{EM2009} for details.

A standard approach to estimating $E[\Upsilon({\bm X})]$ is the Monte Carlo simulation. In statistical applications, often a set of outcomes of $\bm X$ with a low probability make a large contribution to the simulation of $E[\Upsilon({\bm X})]$. In this case, importance sampling can increase the number of samples in this set. Via a weighting approach, an unbiased estimator can be obtained with a reduced variance, cf. Owen et al. \cite{O2019} and Chen and Choe \cite{CC2019}.

Importance sampling for copulas has been investigated in simulation problems arising from quantitative finance in risk management or for pricing purposes. For instance, it is used to estimate portfolio risk or large portfolio loss probabilities, in which the joint CDF of the assets is modeled by a copula, cf. McNeil et al. \cite{M2015} and Dellaportas and Tsionas \cite{D2019}. In particular, to estimate value at risk (VaR), a popular risk measure which is a quantile of the return distribution or the loss distribution of the portfolio of interest corresponding to a small tail probability, the usual means is to employ the Monte Carlo method, cf. Glasserman \cite{G2004}, Nyquist \cite{N2017}, and Xie et al. \cite{X2019}, due to the implicit formula for VaR. However, VaR is ordinarily estimated with respect to a tail probability of $1\%$ or even $0.1\%$ such that the Monte Carlo method becomes inefficient, thus necessitating the use of importance sampling, cf. Glasserman and Li \cite{GL2005}, Huang et al. \cite{H2010}, Fuh et al. \cite{F2011}, and Bhamidi et al. \cite{Bh2015}. Note that Bee \cite{B2011} proposes an adaptive importance sampling for copulas that requires specification of the updated importance function and that has been applied to pricing first-to-default credit swaps.

In this paper we investigate efficient Monte Carlo simulation for copula models. There are four aspects to this study. First, since the copula model is defined by its marginals and a copula function, and its moment-generating function is difficult to derive, we firstly use the transform likelihood ratio (TLR) to obtain an alternative exponential tilting family, then obtain simple and explicit expressions of equations from which the optimal alternative probability measure can be calculated under this transformed exponential tilting family. Second, the proposed algorithm is general enough to cover some interesting copula models such as elliptical copula, Archimedean copula, and regular vine (R-vine) copula. Third, the logarithmic efficiency and bounded relative error for the importance sampling based on the moderate deviation tilting are proved for some commonly-used copula models under the case of simple rare events. Moreover, as a by-product, the bounded relative error for the importance sampling based on the large deviation tilting for $t$ distribution and $t$ copula is proved by applying the techniques used in this paper. Fourth, the derived tilting formula can be used to approximate the rare event probability which can provide guidance on risk management.

The rest of this paper is organized as follows. In Section~\ref{ISwithTLR}, we propose importance sampling via the TLR method for copula models. Then, in Section~\ref{ISforCM}, we derive the optimal tilting probability measures under corresponding simulation algorithms for Gaussian, $t$, and Clayton copulas, respectively. Simulation studies are presented in Section~\ref{Simulation}, in which we first consider the estimation of the probability of moderate deviation events under Gaussian, $t$, and Clayton copulas, and then illustrate how to apply our method to R-vine copula. We conclude in Section~\ref{Conclusion}. All proofs are deferred to the appendices.

\section{Importance Sampling with Transform Likelihood Ratio}\label{ISwithTLR}

Let ${\bm X}=(X_{1},\ldots, X_{d})^{'}$ be a $d$-dimensional random vector with continuous marginal CDFs $F_{1},\ldots,F_{d}$. Then by Sklar's theorem the joint CDF of ${\bm X}$ can be uniquely determined as
\begin{align*}
F( {x}) = C(F_{1}(x_{1}),\ldots ,F_{d}(x_{d});\delta), & \;~~~  {x}=(x_1,\ldots,x_d) \in \mathbb{R}^{d},
\end{align*}
where $C:[0,1]^{d}\mapsto [0,1]$ is a copula function, a $d$-dimensional CDF, with standard uniform margins, and $\delta$ is the copula parameter.

In this section, we seek to efficiently estimate
\begin{align*}
u = E[1\{{\bm X}\in A\}] = P({\bm X}\in A),
\end{align*}
where the expectation is calculated under the $P$-measure (we frequently write it as $E_P$ in the rest of this paper), $1\{\cdot\}$ is an indicator function, and $\{{\bm X}\in A\}$ is a rare event.

A useful tool in importance sampling for rare event simulation is exponential tilting, cf. Siegmund \cite{S1976}, Bucklew \cite{B2003}, Asmussen and Glynn \cite{A2007}, and the references therein. To apply this method, the moment-generating function must be computed, provided it exists. However, it is known that the copula model comprises a copula function and marginals, the complexity of which makes it difficult to compute the moment-generating function. It is thus also difficult to use exponential tilting directly to obtain an alternative probability measure. Furthermore, as the copula model is characterized only by a parameter that depicts dependency, changing the measure by shifting the location parameter or shrinking the scale parameter, cf. Bucklew \cite{B2003}, is also unsuccessful. To remedy this difficulty, we adopt the TLR method, which is first introduced by Kroese and Rubinstein \cite{KR2004} and also appearing in Rubinstein and Kroese \cite{RK2008}. The essential idea is to find a substitutional and simplified distributional family that has a functional relationship with the target distribution before the importance sampling process is executed. Then exponential twisting is employed on this substitutive family. In particular, write ${\bm X}=g({\bm V})$, so that we obtain an alternative distribution of $\bm V=(V_{1},\ldots,V_{d})^{'}$. Then the estimation  becomes
\begin{align*}
u &= E_{P}[1\{{\bm X}\in A\}] = E_P[1\{g({\bm V})\in A\}].
\end{align*}
Using the TLR method, ${\bm X}=g({\bm V})$ is either a linear or a nonlinear function of a random vector ${\bm V}$. Overall, importance sampling with the TLR method comprises two steps: we first consider a change of variables, and then employ importance sampling on the transformed random variables. Note that the transformed event $\{g({\bm V})\in A\}$ consisting of a function $g(\cdot)$ becomes a linear/nonlinear, complex event, with the fortunate consequence that the transformed underlying distribution function is easily computable.

Instead of sampling from the $P$-measure, an alternative $Q$-measure is incorporated such that the event of interest is simple to simulate, in which the density from the $Q$-measure is called an importance sampling density. Here we assume that $P$ is absolutely continuous with respect to $Q$. In addition, assume $P$ and $Q$ are absolutely continuous with respect to the Lebesgue measure, and denote their densities as $P$ and $Q$ as well. Then the target of estimation can be written as
\begin{align*}
u &= \int 1\{g({\bm V})\in A\}\frac{\d P}{\d Q}({\bm V})\d Q({\bm V}) = E_{Q}\left[1\{g({\bm V})\in A\}\frac{\d P}{\d Q}({\bm V})\right],
\end{align*}
where the subscript~$Q$ indicates that the expectation is taken with respect to the $Q$-measure. Accordingly, if ${\bm V}_{1}, \ldots, {\bm V}_{n}$ are
independently sampled from the $Q$-measure, then
\begin{align*}
\hat{u} = \frac{1}{n}\sum_{k=1}^{n}1\{g({\bm V}_{k})\in A\}\frac{\d P}{\d Q}({\bm V}_{k})
\end{align*}
is termed an \emph{importance sampling estimator}, in which the ratio of densities is referred to as the likelihood ratio or the Radon--Nikodym derivative evaluated at ${\bm V}_{k}$. If $\d P = \d Q$, then the likelihood ratio equals $1$, and it reduces to the ordinary crude Monte Carlo estimator.

What is crucial is to select an optimal alternative measure with the smallest variance in the coincident family. Here we employ exponential twisting, and then directly minimize the variance of the importance sampling estimator within this exponential tilting family. That is,
\begin{align*}
\frac{\d P}{\d Q}({\bm V}) := \frac{\d P}{\d Q_\theta}({\bm V}) = e^{-\theta^{'}  {\bm V} + \psi(\theta)},
\end{align*}
where $\theta=(\theta_{1},\ldots,\theta_{d})^{'}$ and $\psi(\theta)$ is the cumulant-generating function of $\bm V$, i.e., $\psi(\theta)=\ln \Psi(\theta)$
with $\Psi(\theta)=E_{P}[e^{\theta^{'} {\bm V}}]$ being the moment-generating function of $\bm V$ under the $P$-measure. The resulting variance of the importance sampling estimator is
\begin{align*}
\mathrm{Var}_{Q}(\hat{u})&= \frac{1}{n} \left\{ E_{Q} \left[1\{g({\bm V})\in A\}\left(\frac{\d P}{\d Q}({\bm V})\right)^{2}\right] - u^{2} \right\}.
\end{align*}
It then suffices to minimize the second-order moment of $\hat{u}$ under the $Q$-measure. Simple calculation suffices to minimize
\begin{align}\label{2ndmoment}
G(\theta):= E_{Q} \left[1\{g({\bm V})\in A\}\left(\frac{\d P}{\d Q}({\bm V})\right)^{2}\right] = E_{P}\left[1\{g({\bm V})\in A\}\frac{\d P}{\d Q}({\bm V})\right],
\end{align}
which is an expectation of $1\{g({\bm V})\in A\}$ multiplied by a likelihood ratio under the $P$-measure. Therefore, the optimal $\theta$ is obtained by solving
\begin{align}\label{eq10}
\frac{\partial G(\theta)}{\partial \theta_{i}}=\frac{\partial}{\partial \theta_{i}}E_{P}\left[1\{g({\bm V})\in A\}e^{-\theta^{'} {\bm V} +\psi(\theta)}\right]&=0, \quad \hbox{for}~ i=1, \ldots, d.
\end{align}
Define
\begin{align*}
\Theta=\{\theta: \Psi(\theta)<\infty\}.
\end{align*}
Note that, if
\begin{align}\label{eq11}
E_{P}\left[1\{g({\bm V})\in A\}\sup_{\theta\in \Theta}\left|\frac{\partial e^{-\theta^{'} {\bm V} +\psi(\theta)}}{\partial \theta_{i}}\right|\right]<\infty,
\end{align}
then (\ref{eq10}) can be written as
\begin{align*}
E_{P}\left[1\{g({\bm V})\in A\}\frac{\partial e^{-\theta^{'} {\bm V} +\psi(\theta)}}{\partial \theta_{i}}\right]=0, \quad \hbox{for}~ i=1, \ldots, d.
\end{align*}
The above equations can be further rewritten as
\begin{align}\label{eq_theta}
\frac{ E_{P}\left[1\{g({\bm V})\in A\}e^{-\theta^{'} {\bm V}}V_{i}\right] }{ E_{P}\left[1\{g( {\bm V})\in  A\}e^{-\theta^{'} {\bm V}}\right] } = \frac{\partial \psi(\theta)}{\partial \theta_{i}}, \quad \hbox{for}~ i=1, \ldots, d.
\end{align}

Using the idea of {\it conjugate measure} proposed in Fuh et al. \cite{F2013}, the above equations can be written in a concise version if $-\theta\in \Theta$. Define the
conjugate measure $\bar{Q}:=\bar{Q}_\theta$ of the measure~$Q$ as
\begin{align*}
\frac{\d \bar{Q}}{\d P}({\bm V}) = \frac{e^{-\theta^{'} {\bm V}} }{E_P[e^{-\theta^{'} {\bm V}}]} = e^{-\theta^{'} {\bm V}-\bar{\psi}(\theta)},
\end{align*}
where $\bar{\psi}(\theta)=\ln E_P[e^{-\theta^{'} {\bm V}}]=\psi(-\theta)$. Note that
\begin{align*}
\d\bar{Q}_\theta = e^{-\theta^{'} {\bm V}-\bar{\psi}(\theta)}\d P = e^{-\theta^{'} {\bm V}-\psi(-\theta)}\d P = \d Q_{-\theta}.
\end{align*}
Then, based on the $\bar{Q}$-measure, we provide an explicit expression for the left side of equation~(\ref{eq_theta}) and, in the following theorem, show the uniqueness for the optimal $\theta$ provided its existence. To this purpose, we make the following assumption, which is more restrictive than (\ref{eq11}):
\begin{align}\label{eq12}
& E_{P}\left[ 1\{g({\bm V})\in  A\} \sup_{\theta\in \Theta}\left|\left(-V_{i}+\frac{\partial \psi(\theta)}{\partial \theta_{i}}\right)
\left(-V_{j}+\frac{\partial \psi(\theta)}{\partial \theta_{j}}\right)\right|e^{-\theta^{'} {\bm V} +\psi(\theta)}\right] \notag \\
 &+ E_{P}\left[ 1\{g({\bm V})\in  A\}\sup_{\theta\in \Theta}\left|\frac{\partial^{2} \psi(\theta)}{\partial \theta_{i} \partial \theta_{j}}\right|e^{-\theta^{'} {\bm V} +\psi(\theta)}\right]<\infty.
\end{align}

\begin{thm}\label{thm_theta}
Suppose $g(\cdot)$ is a real-valued function and (\ref{eq12}) is fulfilled. Moreover, assume every component of $\bm V$ is non-degenerate.
If a solution $\theta$ for the optimization problem {\rm (\ref{2ndmoment})} exists, which satisfies a system of equations (\ref{eq_theta}), then the solution is unique. Hereafter, denote the uniquely optimal tilting point as $\theta_{o}=(\theta_{o1},\ldots,\theta_{od})^{'}$ provided its existence. If $-\theta_{o}\in\Theta$, then $\theta_{o}$ also satisfies
\begin{align}\label{eq_theta1}
E_{\bar{Q}_{\theta_o}}[V_{i}|g({\bm V})\in A] = \frac{\partial \psi(\theta_o)}{\partial \theta_{oi}}, \quad \hbox{for}~ i=1, \ldots, d.
\end{align}
\end{thm}

The proof of Theorem~\ref{thm_theta} is given in Appendix~1.

Since the left side of equation~(\ref{eq_theta1}) in Theorem~\ref{thm_theta} is an integral with a nonlinear and complicated integral domain, it is difficult to find an explicit solution. Here we apply a standard numerical method such as Newton's method or global methods for systems of nonlinear equations described in Dennis and Schnabel \cite{DS1996}. Furthermore, a stochastic algorithm such as generalized simulated annealing can be used to approximate the tilting parameter~$\theta$ recursively, cf. Tsallis and Stariolo \cite{TS1996} and Xiang et al. \cite{XGSH2012}.

The proposed importance sampling algorithm is described as follows.
\begin{center}
\textbf{Importance Sampling with TLR Method}
\end{center}
\begin{enumerate}
\item Change of variable according to a functional relationship, written as ${\bm X} = g({\bm V})$. Here ${\bm V}$ is a more tractable random vector in
comparison to ${\bm X}$. The event of interest becomes
\begin{align*}
\{{\bm X}\in A \} = \{g({\bm V})\in A\}.
\end{align*}
\item Change of measure based on exponential tilting for the transformed random vector~${\bm V}$. The optimal tilting measure $Q_{\theta}$ for
${\bm V}$ is obtained via (\ref{eq_theta}) or (\ref{eq_theta1}) in Theorem~\ref{thm_theta}.
\item Generate independent samples ${\bm V}_{1}, \ldots, {\bm V}_{n}$ from $Q_{\theta_o}$, and the importance sampling estimator of $u$ is
\begin{align*}
\hat{u} = \frac{1}{n}\sum_{k=1}^{n}1\{g({\bm V}_{k})\in A\}\frac{\d P}{\d Q_{\theta_o}}({\bm V}_{k}).
\end{align*}
\end{enumerate}

\section{Importance Sampling for Copula Models}\label{ISforCM}

To apply the proposed importance sampling with the TLR method, in this section, we consider the general copulas, including Gaussian, $t$, and Clayton copulas. To simulate copula models, a prototypical method for all absolutely continuous copulas is the conditional inverse method. However, this method is more suitable for low-dimensional copulas and is rather complicated in the high-dimensional case. For specific elliptical copulas such as Gaussian and $t$ copulas, one can simulate random variables from the corresponding multivariate Gaussian or $t$ distribution, and then use a standard transformation. To simulate a specific Archimedean copula, the Clayton copula, we will use the celebrated Marshall--Olkin method, cf. Marshall and Olkin \cite{MO1988} and Hofert \cite{H2008}. Here, we adopt the proposed importance sampling algorithm in Theorem~\ref{thm_theta} only with existing simulation methods for copula models. For corresponding simulation methods, the reader is referred to McNeil et al. \cite{M2015} for details. Note that for the above copula models, we shall prove that the solution to optimization problem (\ref{2ndmoment}) indeed exists under additionally mild conditions, thus the solution is unique by Theorem~\ref{thm_theta}.

To simplify the presentation, we present importance sampling for copula models in three subsections. Subsection~\ref{ISgeneral} concerns importance sampling for general copulas with the conditional inverse method, importance sampling for elliptical copulas is presented in Subsection~\ref{ISElliptical}, while importance sampling for the Clayton copula is presented in Subsection~\ref{ISArchimedean}.

\subsection{Importance Sampling for General Copulas}\label{ISgeneral}

An essential step in the conditional inverse method is to transform the underlying random vector to a uniform random vector on $(0,1)^d$. Under this situation, $\Theta=\mathbb{R}^{d}$. By applying the exponential embedding of $U(0,1)$ to conjugate truncated exponential distributions (see Proposition \ref{Pro1} below; the conjugate truncated exponential random variate $V_i$ means that $-V_i$ is the truncated exponential random variate) and using Theorem~\ref{thm_theta}, we have the following proposition.

\begin{proposition}\label{Pro1}
Under the assumptions of Theorem~\ref{thm_theta}, if $E[V_i|g( {\bm V}) \in A ] > E(V_i)$ is fulfilled for all $i=1, \ldots, d$, then the optimal $Q_{\theta}$-measure exists and is unique, and the optimal tilting point $\theta_{o}$ satisfies
\begin{align*}
E_{\bar{Q}_{\theta_{o}}}[V_{i}|g({\bm V})\in A] = \frac{\partial \psi(\theta_{o})}{\partial \theta_{oi}}, \quad \hbox{for}~ i=1, \ldots, d.
\end{align*}
Here $\bm V=(V_{1}, \ldots , V_{d})^{'}$, under the $\bar{Q}_{\theta_{o}}$-measure, are independently truncated exponentials on $(0,1)$ with rate parameters
$(\theta_{o1}, \ldots, \theta_{od})$, and $\psi(\theta_{o}) = \sum_{i=1}^{d} \ln  \left(\frac{e^{\theta_{oi}}-1}{\theta_{oi}}\right)$.
\end{proposition}

The proof of Proposition~\ref{Pro1} is given in Appendix~2.

Assume that $A$ is an upper corner event. Since such transformations (CDF transformations; see (\ref{CDF}) below for details) are continuous and non-decreasing functions, larger values of $\bm X$ correspond to larger values of $\bm V$, which imply $E[V_i|g({\bm V}) \in A ] > E(V_i)$. Therefore, the conditions $E[V_i|g( {\bm V}) \in A ] > E(V_i)$ for all $i=1, \ldots, d$ hold for upper corner events.

Note that in importance sampling for uniform distributions, one may consider a family of Beta distributions as an alternative exponential tilting embedding family. According to our simulation results, these two methods exhibit nearly the same simulation performance; here we consider only the family of truncated exponential distributions. Another issue for simulating copulas by the conditional inverse method is its inapplicability to high-dimensional copulas (cf. McNeil et al. \cite{M2015}) since all conditional distribution functions must be derived step by step.

The importance sampling algorithm via the conditional inverse method is summarized as follows.

\noindent
\begin{center}
\textbf{Importance Sampling via Conditional Inverse Method}
\end{center}
\begin{enumerate}
\item Change of variable: let
\begin{align}\label{CDF}
 {\bm X} = g({\bm V}) = (F_{1}^{-1}(V_{1}),F_{2}^{-1}[\Lambda^{-1}(V_{2}|V_{1})], \ldots, F_{d}^{-1}[\Lambda^{-1}(V_{d}|V_{1}, \ldots, V_{d-1})])^{'},
\end{align}
where ${\bm V}=(V_{1}, \ldots, V_{d})^{'}$ with $V_1, \ldots, V_d$ being independent uniform random variables on $(0,1)$, and $\Lambda^{-1}(V_{j}|V_{1}, \ldots, V_{j-1})$ for $j=2, \ldots, d$ is the inverse function of the conditional distribution function $\Lambda (V_{j}|V_{1}, \ldots, V_{j-1})$ with
\begin{align*}
\Lambda (v_{j}|v_{1}, \ldots, v_{j-1})&=P(V_{j}\leq v_{j}|V_{1}=v_{1}, \ldots, V_{j-1}=v_{j-1}) \nonumber \\
&= \lim_{\varepsilon\searrow 0} P(V_{j}\leq v_{j}|V_{1}\in (v_{1}, v_1+\varepsilon], \ldots, V_{j-1}\in (v_{j-1}, v_{j-1}+\varepsilon])\nonumber \\
&= \lim_{\varepsilon\searrow 0}\frac{P(V_{j}\leq v_{j}, V_{1}\in (v_{1}, v_1+\varepsilon], \ldots, V_{j-1}\in (v_{j-1}, v_{j-1}+\varepsilon])}{P(V_{1}\in (v_{1}, v_1+\varepsilon], \ldots, V_{j-1}\in (v_{j-1}, v_{j-1}+\varepsilon])}  \nonumber \\
&= \frac{\partial^{j-1}C_{j}(v_{1}, \ldots, v_{j})}{\partial v_{1}\cdots \partial v_{j-1}} \bigg{/} \frac{\partial^{j-1}C_{j-1}(v_{1},\ldots,v_{j-1})}{\partial v_{1}\cdots \partial v_{j-1}}, \quad \hbox{for}~j=2, \ldots, d.
\end{align*} Note that
\[C_{j}(v_{1}, \ldots, v_{j})=C(v_{1}, \ldots, v_{j}, 1, \ldots, 1)\]
is a $j$-dimensional marginal CDF.

\item Change of measure: employ an exponential tilting technique on $\bm V$. The optimal alternative measure $Q_{\theta}$ for ${\bm V}$ is an independently conjugate truncated exponential on $(0,1)^d$ with rate parameters $\theta_{o}=(\theta_{o1},\ldots,\theta_{od})^{'}$, which can be obtained by Proposition~\ref{Pro1}.

\item Using the optimal tilting parameters $\theta_{o}$ obtained in Step~2, we carry out the following steps for each of $n$ replications:

\begin{enumerate}
\item Generate ${\bm V}$ from an independently conjugate truncated exponential on $(0,1)^d$ with rate parameters $\theta_{o}=(\theta_{o1},\ldots,\theta_{od})^{'}$. That is, the probability density function of $V_i$ is
\begin{eqnarray*}
q_i(v_i)=\left\{
\begin{array}{ll}
\frac{\theta_{oi}e^{\theta_{oi}v_i}}{e^{\theta_{oi}}-1}, & 0<v_i<1, \\
0, & \hbox{otherwise}.
\end{array}\right.
\end{eqnarray*}

\item Multiply the indicator function and the likelihood ratio to get
\begin{align}\label{CIM1}
1\{g({\bm V})\in A\}e^{-\theta_{o}^{'} {\bm V}+\psi(\theta_{o})}.
\end{align}
\end{enumerate}

\item Average~(\ref{CIM1}) over $n$ independent replications to obtain the estimator.
\end{enumerate}

\subsection{Importance Sampling for Elliptical Copulas}\label{ISElliptical}

A simulation of an elliptical copula can be extracted from the related multivariate distribution. In this subsection, we take advantage of this property to propose an importance sampling for Gaussian and $t$ copulas. First, assume $X_{1}, \ldots, X_{d}$ are generated as Gaussian copula with a covariance matrix $\Sigma$ whose diagonal components are all $1$, then
\begin{align*}
F(x_{1},\ldots, x_{d}) = C(F_{1}(x_{1}),\ldots, F_{d}(x_{d});\Sigma) = {\Phi}_{\Sigma}[\Phi^{-1}(F_{1}(x_{1})),\ldots, \Phi^{-1}(F_{d}(x_{d}))],
\end{align*}
where $\Phi(\cdot)$ is the CDF of $N(0,1)$, ${\Phi}_{\Sigma}(\cdot)$ is the CDF of a multivariate normal with a covariance matrix $\Sigma$ (abbreviated
$\mathrm{MN}( {0},\Sigma)$), and $F_{1}(\cdot),\ldots, F_{d}(\cdot)$ are marginal CDFs of $(X_{1},\ldots, X_{d})^{'}$, respectively. Note that under this situation, $\Theta=\mathbb{R}^{d}$. Then, we have the following proposition.

\begin{proposition}\label{Pro2}
Under the assumptions of Theorem~\ref{thm_theta}, if $E[V_i|g( {\bm V}) \in A ] > E(V_i)$ is fulfilled for all $i=1, \ldots, d$, then the optimal
$Q_{\theta}$-measure exists and is unique, and the optimal tilting point $\theta_{o}$ satisfies
\begin{align}\label{eq_theta_2}
E_{\bar{Q}_{\theta_{o}}}[V_{i}|g({\bm V})\in A] = \frac{\partial \psi(\theta_{o})}{\partial \theta_{oi}}, \quad \hbox{for}~ i=1, \ldots, d,
\end{align}
where ${V}$, under the $\bar{Q}_{\theta_{o}}$-measure, is $\mathrm{MN}(-\Sigma\theta_{o},\Sigma)$, and $\psi(\theta_{o})=\frac{1}{2}\theta_{o}^{'}\Sigma\theta_{o}$.
\end{proposition}

The proof of Proposition~\ref{Pro2} is given in Appendix~2.

Consider an upper corner event $\{{\bm X}> {a} \}:=\{X_1 >a_1,\ldots,X_d > a_d\}$, where ${a}=(a_{1},\ldots,a_{d})^{'}$. Then the transformed event becomes
$\{ V_{1}>\Phi^{-1}(F_{1}(a_{1})),\ldots, V_{d}>\Phi^{-1}(F_{d}(a_{d})) \}$. To simplify the notations, let
\[
 {a}^{*}=(a_{1}^{*},\ldots ,a_{d}^{*})^{'} = \left(\Phi^{-1}(F_{1}(a_{1})),\ldots ,\Phi^{-1}(F_{d}(a_{d}))\right)^{'}.
\]
Then equation~(\ref{eq_theta_2}) can be considered as the first-order moment function of the truncated multivariate normal distribution, referring to Tallis \cite{T1961}. Therefore we have the following corollary.

\begin{cor}\label{Cor1}
Let $A$ be an upper corner event, and assume the covariance matrix is positive definite with all diagonal components~$1$ in a Gaussian copula. Then, under the assumptions of Proposition~\ref{Pro2}, equation~(\ref{eq_theta_2}) can be simplified as
\begin{align*}
\frac{\sum_{q=1}^{d}\rho_{iq}\phi(a_{q}^*+\rho_{q\cdot}\theta_{o})\bar{\Phi}_{d-1}(A_{qs};\Sigma_{q})}{\bar{\Phi}_{d}( {a}^*+\Sigma\theta_{o};\Sigma)} = 2\rho_{i\cdot}\theta_{o}, \quad i=1, \ldots, d,
\end{align*}
where $\rho_{iq}$ is the $(i,q)$-th component of $\Sigma$, $\phi(\cdot)$ is the probability density function of $N(0,1)$, $\rho_{ q \cdot}$ is the $q$-th row vector of $\Sigma$, $\bar{\Phi}_{d}$ is a $d$-variate survival function of $\Phi_{d}$, $A_{qs}=\frac{a_{s}^*-\rho_{sq}a_{q}^*}{\sqrt{1-\rho_{sq}^{2}}}$ for $s\neq q$ in $\bar{\Phi}_{d-1}$, and $\Sigma_{q}$ is the matrix of the first-order partial correlation coefficients of $V_{s}$ for $s\neq q$.
\end{cor}

The proof of Corollary~\ref{Cor1} is given in Appendix~2.

The importance sampling algorithm for Gaussian copulas is as follows.

\noindent
\begin{center}
\textbf{Importance Sampling for Gaussian Copulas}
\end{center}
\begin{enumerate}
\item Change of variable: let
\begin{align}\label{Gau1}
 {\bm X} = g({\bm V}) = \left[ F^{-1}_{1}(\Phi(V_{1})), \ldots ,F^{-1}_{d}(\Phi(V_{d})) \right]^{'},
\end{align}
where $F_{1}, \ldots, F_{d}$ are the marginal CDFs of ${\bm X}=(X_{1},\ldots ,X_{d})^{'}$, and ${\bm V}=(V_{1},\ldots,V_{d})^{'}$ is $\mathrm{MN}({0},\Sigma)$.

\item Change of measure: employ the exponential tilting embedding on ${\bm V}$. The optimal tilting probability measure $Q_{\theta}$ for ${\bm V}$ is $\mathrm{MN}(\Sigma\theta_{o},\Sigma)$, with $\theta_{o}$ obtained via Proposition~\ref{Pro2}.

\item Using the obtained optimal tilting parameters $\theta_{o}$, we carry out the following steps for each of $n$ replications:
\begin{enumerate}
\item Generate ${\bm V}$ from $\mathrm{MN}(\Sigma\theta_{o},\Sigma)$.
\item Multiply the indicator function and the likelihood ratio to get
\begin{align}\label{Gau}
1\{ g({\bm V})\in A\}e^{-\theta_{o}^{'} {\bm V}+\psi(\theta_{o})}.
\end{align}
\end{enumerate}

\item Average (\ref{Gau}) over $n$ independent replications to obtain the estimator.
\end{enumerate}


Next, we illustrate the logarithmic efficiency for the above importance sampling for Gaussian copula under the case of simple rare events. Since this result can be generalized to a more general tilting technique which is called sufficient exponential tilting (that is, the exponential embedding is based on two parameters corresponding to the sufficient statistic of the normal distribution, cf. Fuh and Wang \cite{F2024}), we present the logarithmic efficiency for the importance sampling for Gaussian copulas using this sufficient exponential tilting method in the following thoerem. Note that the probabilities of simple rare events of a Gaussian copula can be transformed into corresponding probabilities of the multivariate normal distribution, cf. equation (\ref{Gau1}), thus we state the logarithmic efficiency result for the two-dimensional standard normal distribution for simplicity.

\begin{thm}\label{SET}
Let $\bm{X}=(X_1, X_2)^{'}$ be a random vector from the two-dimensional standard normal distribution $\mathrm{MN}(0, I_2)$ with probability measure $P$. Consider the sufficient exponential tilting
\begin{align*}
      \frac{\d P}{\d Q_{\theta, M}}(\bm{X}) =\exp\left\{-\theta^{'}\bm{X} - \bm{X}^{'} M \bm{X}\right\} E_P \left[\exp\left\{\theta^{'}\bm{X} + \bm{X}^{'} M \bm{X}\right\}\right],
\end{align*}
where $\theta \in \mathbb{R}^2$ and $M$ is a $2$-by-$2$ matrix satisfying $M \preceq \frac{1}{2}I_2$ (that is, $\frac{1}{2}I_2-M$ is a nonnegative definite matrix). Then, for the case of approximating the probability $u= u(p)=P(\bm{X} > (p,p)^{'})$ with $p>0$, the sufficient exponential tilting entails the logarithmic efficiency. That is, for all $\epsilon>0$, the importance sampling estimator
$$\hat{u}=\frac{1}{n}\sum_{i=1}^n 1\{\bm X_i > (p,p)^{'}\}\frac{\d P}{\d Q_{\theta_o, M_o}}({\bm X_i})$$
satisfies
$$
\lim_{p \to \infty}\frac{\mathrm{Var}_{Q_{\theta_o, M_o}}(\hat{u})}{u^{2-\epsilon}}<\infty,
$$
where $\theta_o$ and $M_o$ are the optimal parameters of $\theta$ and $M$, respectively.
\end{thm}

The proof of Theorem \ref{SET} is given in Appendix 2.

\begin{rmk}
According to Asmussen and Kroese \cite{A2006}, $\hat{u}$ entails the bounded relative error if $\frac{\mathrm{Var}_{Q_{\theta_o, M_o}}(\hat{u})}{u^2}$ is bounded in $p$, and entails the logarithmic efficiency if $\frac{\mathrm{Var}_{Q_{\theta_o, M_o}}(\hat{u})}{u^{2-\epsilon}}$ is bounded in $p$ for all $\epsilon>0$. Note that the bounded relative error is a more nice property than the logarithmic efficiency.
\end{rmk}

The traditional optimal one-parameter exponential tilting also entails the logarithmic efficiency, and its proof is similar to that of Theorem \ref{SET}. We believe that the excess variance reduction of the sufficient exponential tilting method is relatively limited, let alone much heavier computational burden of obtaining the optimal tilting parameters. Consequently, we still adopt the traditional one-parameter exponential tilting method in this paper (except for the importance sampling via conditional inverse method) because the number of parameters for this method is significantly smaller than the sufficient exponential tilting method especially when the dimension goes larger.


Next, we consider the $t$ copula. Assume $X_{1}, \ldots, X_{d}$ are constructed from a $t$ copula with marginal CDFs $F_{i}$ for $i=1,\ldots,d$. Then the joint CDF is
\begin{align*}
F(x_{1},\ldots, x_{d}) = C(F_{1}(x_{1}),\ldots, F_{d}(x_{d});\nu ,\Sigma) =  {t}_{\nu ,\Sigma}(t^{-1}_{\nu}(F_{1}(x_{1})),\ldots, t^{-1}_{\nu}(F_{d}(x_{d}))),
\end{align*}
where $t_{\nu}$ is the CDF of a standard univariate $t$ distribution with a degree of freedom $\nu$, and ${t}_{\nu,\Sigma}$ is the CDF of a multivariate $t$ distribution $t_{\nu}( {0},\Sigma)$ with the diagonal components of $\Sigma$ all being~$1$.

Using a simulation technique similar to that for Gaussian copula and the fact that $t$ copula is extracted from a multivariate $t$ distribution, we produce $t$ copula by
\begin{align}\label{event_t1}
X_{i}=F^{-1}_{i}[t_{\nu}(T_{i})], &\quad \hbox{for $i=1,\ldots,d$,}
\end{align}
where ${\bm T}=(T_{1},\ldots,T_{d})^{'}$ is $t_{\nu}( {0},\Sigma)$. Since a multivariate $t$ distribution has a polynomial tail, standard exponential tilting cannot be applied directly. However, it is known that
\begin{align}\label{event_t2}
T_{i}=\frac{Z_{i}}{\sqrt{Y/\nu}}, &\quad \hbox{for $i=1,\ldots,d$,}
\end{align}
where ${\bm Z}=(Z_{1}, \ldots, Z_{d})^{'}$ is $\mathrm{MN}({0},\Sigma)$ and $Y$, independent of ${\bm Z}$, is a $\chi^2$ random variable with a degree of freedom $\nu$, denoted as $\chi_{\nu}$. That is, under the $P$-measure, $Y$ is from a Gamma distribution with a shape parameter $\nu/2$ and an inverse scale parameter $1/2$, and $\bm Z$ is from $\mathrm{MN}(0, \Sigma)$. Using the transformation (\ref{event_t2}), we can apply an exponential tilting scheme on ${\bm Z}$ and $Y$ simultaneously.

For an upper corner (or lower corner) event $\{{\bm X}> {a}\}$ (or $\{{\bm X}< {a}\}$) where ${a}=(a_{1}, \ldots, a_{d})^{'}$, the transformed event is $\{T_{1}>t_{\nu}^{-1}(F_{1}(a_{1})), \ldots, T_{d}>t_{\nu}^{-1}(F_{d}(a_{d}))\}$ (or $\{T_{1}<t_{\nu}^{-1}(F_{1}(a_{1})), \ldots, T_{d}<t_{\nu}^{-1}(F_{d}(a_{d}))\}$). To simplify the notations, denote $a_{i}^{*}=t_{\nu}^{-1}(F_{i}(a_{i}))$ for $i=1, \ldots, d$.
Through a detailed calculation, letting $W_{i}=\sqrt{\frac{Y}{\nu}}Z_{i}-\frac{Y}{\nu}a_{i}^{*}$ for $i=1,\ldots, d$, the transformed event becomes $\{{\bm W}> {0}\}$ (or $\{{\bm W}< {0}\}$), where ${\bm W}=(W_{1}, \ldots, W_{d})^{'}$. Therefore, under the tilting probability measure $Q$, $Y$ is from a Gamma distribution with a shape parameter $\nu/2$ and an inverse scale parameter $(1+\frac{2}{\nu}\theta^{'} {a}^{*}-\frac{1}{\nu}\theta^{'}\Sigma\theta)/2$, and ${\bm Z}$ is from  $\mathrm{MN}(\sqrt{\frac{y}{\nu}}\Sigma\theta,\Sigma)$ conditional on $Y=y$, where $\theta=(\theta_{1}, \ldots, \theta_{d})^{'}$ are tilting parameters; see the following proposition for more details.

\begin{proposition}\label{Pro3}
Under the assumptions of Theorem~\ref{thm_theta} with $\bm V$ replaced by $\bm W=\sqrt{\frac{Y}{\nu}} {\bm Z}-\frac{Y}{\nu} {a}^{*}$, if $E[W_i|g( {\bm W}) \in A ] > E(W_i)$ is fulfilled for all $i=1, \ldots, d$, then the optimal $Q_{\theta}$-measure exists and is unique, and the optimal tilting point $\theta_{o}$ satisfies
\begin{align*}
\frac{E_{P}\left[1\{{\bm W}> {0}\}e^{-\theta_{o}^{'} {\bm W}}W_{i}\right]}{E_{P}\left[1\{{\bm W}> {0}\}e^{-\theta_{o}^{'} {\bm W}}\right]}=\frac{\partial \psi(\theta_{o})}{\partial \theta_{oi}},\quad \hbox{for}~ i=1, \ldots, d,
\end{align*}
where $\psi(\theta_{o}) = -\frac{\nu}{2} \ln \left[1-\frac{2}{\nu}(\frac{1}{2}\theta_{o}^{'}\Sigma\theta_{o}-\theta_{o}^{'}a^{*})\right]$, and under the ${Q}_{\theta_{o}}$-measure, ${\bm Z}\sim \mathrm{MN}(\sqrt{\frac{y}{\nu}}\Sigma\theta_{o},\Sigma)$ conditional on $Y=y$, and $Y\sim \gammadist(\nu/2, ( 1+\frac{2}{\nu}\theta_{o}^{'}a^{*}-\frac{1}{\nu}\theta_{o}^{'}\Sigma\theta_{o})/2 )$. If $-\theta_{o}\in \Theta$, where $\Theta$ is the domain of $\theta$ that satisfies $E_P [e^{\theta^{'}\bm{W}}] < \infty$, then $\theta_{o}$ also satisfies
\begin{align}\label{eq14}
E_{\bar{Q}_{\theta_{o}}}\left[W_{i}| {\bm W}> {0}\right]=\frac{\partial \psi(\theta_{o})}{\partial \theta_{oi}},\quad \hbox{for}~ i=1, \ldots, d,
\end{align}
where, under the $\bar{Q}_{\theta_{o}}$-measure, ${\bm Z}\sim \mathrm{MN}(-\sqrt{\frac{y}{\nu}}\Sigma\theta_{o},\Sigma)$ conditional on $Y=y$, and $Y\sim \gammadist(\nu/2,$ $( 1-\frac{2}{\nu}\theta_{o}^{'}a^{*}-\frac{1}{\nu}\theta_{o}^{'}\Sigma\theta_{o})/2)$.
\end{proposition}

The proof of Proposition~\ref{Pro3} is given in Appendix~2.

\begin{rmk}[]
Although copula models are constructed by a dependence structure and marginal CDFs, the proposed importance sampling algorithms do not involve the marginal effects. Using the transform likelihood ratio method, the marginal effects are transformed to event effects.
\end{rmk}

The algorithm of importance sampling for $t$ copula with corner events is as follows. Hereafter we demonstrate an algorithm only with upper corner events. In the meantime, for lower corner events, it suffices to replace the symbol ``$>$'' by ``$<$''.

\noindent
\begin{center}
\textbf{Importance Sampling for $t$ Copulas with Corner Events}
\end{center}
\begin{enumerate}
\item Change of variable: let
\begin{align}\label{t1}
 {\bm X} = g({\bm T})=\left(F_{1}^{-1}\left[t_{\nu} \left( \frac{Z_{1}}{\sqrt{Y/\nu}} \right)\right] ,\ldots , F_{d}^{-1}\left[t_{\nu} \left( \frac{Z_{d}}{\sqrt{Y/\nu}} \right) \right] \right)^{'},
\end{align}
where $F_{1},\ldots,F_{d}$ are marginal CDFs of ${\bm X}$, ${\bm Z}\sim \mathrm{MN}({0},\Sigma)$, and $Y\sim \chi_{\nu}$. Then the event of interest $\{{\bm X}>{a} \}$ becomes $\{{\bm W}> {0} \}$ by the above-mentioned transformation.

\item Change of measure: employ an exponential tilting technique on ${Y}$ and $\bm Z$ simultaneously. The optimal tilting probability measure $Q_{\theta}$ for $(Y, \bm Z)$ is $Y\sim \gammadist(\nu/2, (1+\frac{2}{\nu}\theta_{o}^{'}a^{*}-\frac{1}{\nu}\theta_{o}^{'}\Sigma\theta_{o})/2)$ and $\bm Z\sim \mathrm{MN}(\sqrt{\frac{y}{\nu}}\Sigma\theta_{o},\Sigma)$ conditional on $Y=y$, with $\theta_{o}$ obtained via Proposition~\ref{Pro3}.

\item Using the obtained optimal tilting parameter $\theta_{o}$, we carry out the following steps for each of $n$ replications:
\begin{enumerate}
\item Generate $Y$ from $\gammadist(\nu/2, (1+\frac{2}{\nu}\theta_{o}^{'}a^{*}-\frac{1}{\nu}\theta_{o}^{'}\Sigma\theta_{o})/2) $ and ${\bm Z}$ from $\mathrm{MN}(\sqrt{\frac{y}{\nu}}\Sigma\theta_{o},\Sigma)$ provided that $Y=y$.
\item Set ${\bm W}= \sqrt{\frac{Y}{\nu}} {\bm Z}-\frac{Y}{\nu} {a}^{*}$.
\item Multiply the indicator function and the likelihood ratio to get
\begin{align}\label{t}
1\{{\bm W}> {0}\}e^{-\theta_{o}^{'} {\bm W}+\psi(\theta_{o})}.
\end{align}
\end{enumerate}

\item Average~(\ref{t}) over $n$ independent replications to obtain the estimator.
\end{enumerate}


Next, we present the bounded relative error for the above importance sampling for $t$ copula with corner events. Note that the probabilities of simple rare events of $t$ copula can be transformed into corresponding probabilities of the $t$ distribution, cf. equation (\ref{t1}). Thus, similar to Gaussian copula, in what follows we state the result of bounded relative error for the two-dimensional standard $t$ distribution for simplicity.

\begin{thm}\label{LEt}
Let $\bm{X}=(X_1, X_2)^{'}$ be a random vector from the two-dimensional standard $t$ distribution, with a degree of freedom $\nu\in (2,\infty)$ and probability measure $P$. Write $\bm{X}$ as
$$
\bm{X} = \frac{\bm{Z}}{\sqrt{Y/\nu}},
$$
where ${\bm Z}=(Z_{1}, Z_{2})^{'}$ is from $\mathrm{MN}({0},I_2)$ and $Y$, independent of ${\bm Z}$, is from $\chi_{\nu}$. Define a random vector $\bm{W}$ as
$$
\bm{W} = \sqrt{\frac{Y}{\nu}} {\bm Z}-\frac{Y}{\nu} (p, p)^{'},\quad p \in \mathbb{R}.
$$
Consider the exponential tilting
$$
\frac{\d P}{\d Q_{\theta}}(\bm{W}) = e^{-\theta^{'}\bm{W}} E_P [e^{\theta^{'}\bm{W}}], \quad \theta \in \Theta.
$$
Then for the case of approximating the probability $u=u(p)=P(\bm{X} > (p, p)^{'})=P(\bm{W} > 0)$ with $p>0$, the optimal exponential tilting entails the bounded relative error. That is, the importance sampling estimator
$$\hat{u}=\frac{1}{n}\sum_{i=1}^n 1\{\bm W_i > 0\}\frac{\d P}{\d Q_{\theta_o}}(\bm W_i)$$
satisfies
$$
\lim_{p \to \infty}\frac{\mathrm{Var}_{Q_{\theta_o}}(\hat{u})}{u^2}<\infty,
$$
where $\theta_o$ is the optimal parameter of $\theta$.
\end{thm}

The proof of Theorem~\ref{LEt} is given in Appendix~2.

\begin{rmk}
Theorem \ref{LEt} shows that the importance sampling based on the moderate deviation tilting for multivariate $t$ distribution and $t$ copula with corner events has the bounded relative error. Moreover, it can be found that Theorem \ref{LEt} is applicable to the one-dimensional $t$ distribution by checking the proof.
\end{rmk}

As a by-product, we shall show that the importance sampling based on the large deviation tilting for the one-dimensional $t$ distribution, multivariate $t$ distribution as well as $t$ copula also has the bounded relative error by applying the similar arguments in the proof of Theorem \ref{LEt}. We shall present this result formally in Proposition \ref{Pro5} below, taking the case of two-dimensional $t$ distribution for instance. To this end, we define
$$\tilde{\Theta}=\{\theta=(\theta_1,\theta_2)^{'}: E_P [e^{\theta^{'}\bm{W}}] < \infty, \theta_1>0, \theta_2>0\},$$
and, in what follows, find an upper bound for the second-order moment of $1\{\bm W > 0\}\frac{\d P}{\d Q_{\theta}}(\bm W)$ under $Q_{\theta}$-measure over the space $\tilde{\Theta}$. Noting that
$$E_{Q_\theta}\left[1\{\bm W > 0\}\frac{\d P}{\d Q_{\theta}}(\bm W)\right]^2=E_P\left[1\{\bm W > 0\}\frac{\d P}{\d Q_{\theta}}(\bm W)\right],$$
and
\begin{align*}
E_P\left[1\{\bm W > 0\}\frac{\d P}{\d Q_{\theta}}(\bm W)\right]=E_P\left[1\{\bm W > 0\}e^{-\theta^{'}\bm{W}+\psi(\theta)}\right]\le e^{\psi(\theta)},\quad \theta\in \tilde{\Theta},
\end{align*}
it is clear that $e^{\psi(\theta)}$ is what we are finding. Let $\tilde{\theta}_o=\underset{\theta\in \tilde{\Theta}}{\arg\min} \psi(\theta)$.

\begin{proposition}\label{Pro5}
Under the conditions of Theorem \ref{LEt}, consider the exponential tilting
$$
\frac{\d P}{\d Q_{\theta}}(\bm{W}) = e^{-\theta^{'}\bm{W}} E_P [e^{\theta^{'}\bm{W}}], \quad \theta \in \tilde{\Theta}.
$$
Then for the case of approximating the probability $u=u(p)=P(\bm{X} > (p, p)^{'})=P(\bm{W} > 0)$ with $p>0$, the optimal large deviation tilting entails the bounded relative error. That is, the importance sampling estimator
$$\hat{u}=\frac{1}{n}\sum_{i=1}^n 1\{\bm W_i > 0\}\frac{\d P}{\d Q_{\tilde{\theta}_o}}(\bm W_i)$$
satisfies
$$
\lim_{p \to \infty}\frac{\mathrm{Var}_{Q_{\tilde{\theta}_o}}(\hat{u})}{u^2}<\infty.
$$
\end{proposition}

The proof of Proposition~\ref{Pro5} is given in Appendix~2.

\begin{rmk}
Glasserman et al. \cite{GHS2002} shows that the importance sampling based on the large deviation tilting for the quadratic form of multivariate $t$ distribution has the bounded relative error; however, this result cannot be applied to the one-dimensional or multivariate $t$ distribution directly. As a result, Proposition~\ref{Pro5} is an addition to the literature on the importance sampling based on the large deviation tilting.
\end{rmk}


\subsection{Importance Sampling for Archimedean Copulas}\label{ISArchimedean}

A standard tool for the simulation of the Archimedean copula is the Marshall--Olkin method, which uses the properties of its constructed form,
\begin{align*}
C(u_{1},\ldots, u_{d})=\zeta^{-1}(\zeta(u_{1})+\cdots +\zeta(u_{d}))=\int e^{-v(\zeta(u_{1})+\cdots +\zeta(u_{d}))}\d G(v),
\end{align*}
where $\zeta$ is a Laplace--Stieltjes transform of a real-valued function $G(\cdot)$, defined as
\begin{align*}
\zeta^{-1}(t)=\int e^{-tv}\d G(v).
\end{align*}

Note that $G(\cdot)$ is a given CDF for a specific Archimedean copula. For example, for a Clayton copula with a dependence parameter $\delta$, $G(\cdot)$ is the CDF of $\gammadist(1/\delta, 1)$. It is known that when $U\sim U(0,1)$, the random variable $e^{-\zeta(U)}$ is uniformly distributed on $(0,1)$, which implies that $e^{-w\zeta(U)}$ is also uniform on $(0,1)$ conditional on $W=w$ sampling from $G(\cdot)$. Therefore, we can generate a $d$-variate sample $(U_{1},\ldots, U_{d})^{'}$ by
\begin{align*}
U_{i} = \zeta^{-1}(-\frac{1}{W}\ln V_{i}), \quad \hbox{for}~ i=1, \ldots, d,
\end{align*}
where $V_{1}, \ldots, V_{d}$ are independent uniform random variables on $(0,1)$. In different marginal settings, obtain $X_{i}=F_{i}^{-1}(U_{i})$ for $i=1,\ldots, d$, where $F_{i}(\cdot)$ is the CDF of $X_{i}$. Denote
$$\tilde{\bm V}=(W, V_1, \ldots, V_d)^{'}\quad {\rm and}\quad \theta=(\theta_W, \theta_1, \ldots, \theta_d)^{'}.$$
In what follows, we take the Clayton copula as an example to illuminate the importance sampling for Archimedean copulas. Note that, under this situation,
$$\Psi(\theta)=E_P [e^{\theta^{'}\tilde{\bm V}}]=(1-\theta_{oW})^{-1/\delta}\prod_{i=1}^d \frac{e^{\theta_{oi}}-1}{\theta_{oi}}.$$

\begin{proposition}\label{Pro4}
Under the assumptions of Theorem~\ref{thm_theta}, if $E[W|g( \tilde{\bm V}) \in A ] > E(W)$ and $E[V_i|g( \tilde{\bm V}) \in A ] > E(V_i)$ for $i=1, \ldots, d$ are fulfilled, then the optimal $Q_{\theta}$-measure exists and is unique, and the optimal tilting point $\theta_{o}=(\theta_{oW}, \theta_{o1}, \ldots, \theta_{od})^{'}$ satisfies
\begin{eqnarray*}
\begin{cases}
\displaystyle \frac{ E_{P}\left[1\{g(\tilde{\bm V})\in A\}e^{-\theta_o^{'} \tilde{\bm V}}W\right] }{ E_{P}\left[1\{g( \tilde{\bm V})\in  A\}e^{-\theta_o^{'} \tilde{\bm V}}\right] } = \frac{\partial \psi(\theta_o)}{\partial \theta_{oW}},\\
\displaystyle \frac{ E_{P}\left[1\{g(\tilde{\bm V})\in A\}e^{-\theta_o^{'} \tilde{\bm V}}V_{i}\right] }{ E_{P}\left[1\{g( \tilde{\bm V})\in  A\}e^{-\theta_o^{'} \tilde{\bm V}}\right] } = \frac{\partial \psi(\theta_o)}{\partial \theta_{oi}}, \quad \hbox{for}~ i=1, \ldots, d,
\end{cases}
\end{eqnarray*}
where $W, V_{1}, \ldots , V_{d}$, under the ${Q}_{\theta_{o}}$-measure, are independent, $W$ is from $\gammadist(\frac{1}{\delta}, 1-\theta_{oW})$ and $V_i$ is from the conjugate truncated exponential on $(0,1)$ with rate parameters $\theta_{oi}$ for $i=1,\ldots,d$, and $\psi(\theta_{o}) = -\frac{1}{\delta}\ln (1-\theta_{oW})+\sum_{i=1}^{d} \ln \left(\frac{e^{\theta_{oi}}-1}{\theta_{oi}}\right)$. If $-\theta_o\in \Theta$, where $\Theta$ is the domain of $\theta$ that satisfies $E_P [e^{\theta^{'}\tilde{\bm V}}] < \infty$, then $\theta_{o}$ also satisfies
\begin{eqnarray*}
\begin{cases}
\displaystyle E_{\bar{Q}_{\theta_{o}}}[W|g(\tilde{\bm V})\in A] = \frac{\partial \psi(\theta_{o})}{\partial \theta_{oW}},\\
\displaystyle E_{\bar{Q}_{\theta_{o}}}[V_{i}|g(\tilde{\bm V})\in A] = \frac{\partial \psi(\theta_{o})}{\partial \theta_{oi}}, \quad \hbox{for}~ i=1, \ldots, d,
\end{cases}
\end{eqnarray*}
where, under the $\bar{Q}_{\theta_{o}}$ measure,  $W$ is $\gammadist(\frac{1}{\delta}, 1+\theta_{oW})$ and $V_i$ is the truncated exponential on $(0,1)$ with rate parameters $\theta_{oi}$ for $i=1, \ldots, d$.
\end{proposition}

The proof of Proposition~\ref{Pro4} is given in Appendix~2.

Using the above results, we summarize an importance sampling for Clayton copula with the Marshall--Olkin method as follows.

\noindent
\begin{center}
\textbf{Importance Sampling for Clayton Copulas}
\end{center}
\begin{enumerate}
\item Change of variable: let
\begin{align*}
 {\bm X} = g(\tilde{\bm V})=\left(F_{1}^{-1}[\zeta_{\delta}^{-1}(-\frac{1}{W}\ln V_{1})], \ldots, F_{d}^{-1}[\zeta_{\delta}^{-1}(-\frac{1}{W}\ln V_{d})]\right)^{'},
\end{align*}
where $\tilde{\bm V}=(W, V_{1}, \ldots, V_{d})^{'}$ is an independent $(d+1)$-dimensional random vector, $W\sim \gammadist(\frac{1}{\delta},1)$, $V_{1}, \ldots, V_{d}$ are all $U(0,1)$ random variables, and
$$\zeta^{-1}_{\delta}(t):=\zeta^{-1}(t)=(t+1)^{-\frac{1}{\delta}}, \quad \zeta_{\delta}(t):=\zeta(t)=t^{-\delta}-1.$$

\item Change of measure: employ an exponential tilting technique on $\tilde{\bm V}$ with tilting parameters
$\theta=(\theta_{W},\theta_{1},\ldots,\theta_{d})^{'}$. The optimal tilting measure for $W$ is $\gammadist(\frac{1}{\delta}, 1-\theta_{oW})$,
and that for $V_i$ is the conjugate truncated exponential on $(0,1)$ with rate parameters $\theta_{oi}$ for $i=1,\ldots,d$, with $\theta_{o}$ obtained via Proposition~\ref{Pro4}.

\item Using the obtained optimal tilting parameters $\theta_{o}$, we carry out the following steps for each of $n$ replications:
\begin{enumerate}
\item Generate $W$ from the optimal alternative distribution $\gammadist(\frac{1}{\delta}, 1-\theta_{oW})$ and $V_{1}, \ldots, V_{d}$ from the conjugate truncated exponential distributions on $(0,1)$ with rate parameters $\theta_{o1}, \ldots, \theta_{od}$, respectively.
\item Multiply the indicator function and the likelihood ratio to get
\begin{align}\label{Gaus}
1\{ g(\tilde{\bm V})\in A\}e^{-\theta_{o}^{'} \tilde{\bm V}+\psi(\theta_{o})}.
\end{align}
\end{enumerate}

\item Average~(\ref{Gaus}) over $n$ independent replications to obtain the estimator.
\end{enumerate}


Next, we present the bounded relative error for the above importance sampling for Clayton copula with upper corner events.

\begin{thm}\label{LEC}
Let $\bm X=(X_1, X_2)^{'}$ be a random vector from the two-dimensional Clayton copula with parameter $\delta > 0$ and the marginal distributions both being $F$. Write $\bm X$ as
$$
\bm X = g(\tilde{\bm V}) = \left(F^{-1}[\zeta_{\delta}^{-1}(-\frac{1}{W}\ln V_{1})], F^{-1}[\zeta_{\delta}^{-1}(-\frac{1}{W}\ln V_{2})]\right)^{'},
$$
where $\tilde{\bm V}=(W, V_{1}, V_{2})^{'}$ is an independent three-dimensional random vector, with $W\sim \gammadist(\frac{1}{\delta},1)$, and
$V_{1}, V_{2}\sim U(0,1)$. Consider the exponential tilting
\begin{align*}
      \frac{\d P}{\d Q_{\theta}}(\tilde{\bm V}) = e^{-\theta^{'} \tilde{\bm V}} E_P [e^{ \theta^{'} \tilde{\bm V} }], \quad \theta \in (-\infty, 1) \times \mathbb{R}^2.
\end{align*}
Then for the case of approximating the probability $u=u(p)=P(\bm X > (p, p)^{'})$ with $p>0$, the optimal exponential tilting entails the bounded relative error. That is, the importance sampling estimator
$$\hat{u}=\frac{1}{n}\sum_{i=1}^n 1\{g(\tilde{\bm V}_i)> (p, p)^{'}\}\frac{\d P}{\d Q_{\theta_o}}(\tilde{\bm V}_i)$$
satisfies
$$
\lim_{p \to \infty}\frac{\mathrm{Var}_{Q_{\theta_o}}(\hat{u})}{u^2}<\infty,
$$
where $\theta_o$ is the optimal parameter of $\theta$.
\end{thm}

The proof of Theorem~\ref{LEC} is given in Appendix~2.


\begin{rmk}[]
(1) Theorems \ref{SET}, \ref{LEt} and \ref{LEC} suggest that it would be hard to have the bounded relative error if $u=u(p)$ decays with respect to $p$ at an exponential rate, while it would be not the case when $u=u(p)$ decays with respect to $p$ at the speed of a polynomial. (2) It would be possible to generalize Theorems \ref{SET}, \ref{LEt} and \ref{LEC} under the case of more general rare events. For example, in Theorem \ref{LEt}, if $u=u(p_1, p_2)=P(\bm{X} > (p_1, p_2)^{'})$ with $p_1,p_2>0$ such that $p_1 \asymp p_2$ (that is, there exist two constants $0<c_1<c_2<\infty$ such that $c_1\le p_1/p_2\le c_2$ for all large $p_1$ and $p_2$), then the importance sampling estimator $\hat{u}$ satisfies
$$
\lim_{\min\{p_1, p_2\} \to \infty}\frac{\mathrm{Var}_{Q_{\theta_o}}(\hat{u})}{u^2}<\infty.
$$
\end{rmk}

\section{Simulation Studies}\label{Simulation}

In this section, we present numerical results on relative efficiency using the method outlined in Theorem~\ref{thm_theta} for estimating tail probabilities.

\subsection{Importance Sampling for Gaussian, $t$ and Clayton Copulas}\label{efficiency}

First, we measure the performance in terms of the relative efficiency of crude Monte Carlo with respect to the proposed importance sampling, which is defined as
\begin{align*}
\hbox{sd\_eff(Naive, IS)} =\frac{\sqrt{\hat{var}(\hbox{Naive})}}{\sqrt{\hat{var}(\hbox{IS})}}=\frac{\mbox{sd}(\hbox{Naive})}{\mbox{sd}(\hbox{IS})}.
\end{align*}
Our proposed importance sampling estimator based on the conditional inverse method is denoted as IS$_{t1}$, while those based on the methods proposed in Subsections \ref{ISElliptical} and \ref{ISArchimedean} are all denoted as IS$_{t2}$. The relative efficiency of IS$_{t2}$ with respect to IS$_{t1}$ is defined as
\begin{align*}
\hbox{sd\_eff(IS$_{t2}$, IS$_{t1}$)} = \frac{\sqrt{\hat{var}(\hbox{IS$_{t2}$})}}{\sqrt{\hat{var}(\hbox{IS$_{t1}$})}}=\frac{\mbox{sd}(\hbox{IS$_{t2}$})}{\mbox{sd}(\hbox{IS$_{t1}$})}.
\end{align*}
Moreover, for a comparison purpose, we also apply the hazard rate twisting approach proposed by Juneja and Shahabuddin \cite{JS2002} and generalized by Ben Rached et al. \cite{Ben2018} in the simulations. Note that the hazard rate twisting approach in Juneja and Shahabuddin \cite{JS2002} and Ben Rached et al. \cite{Ben2018} is only applicable to independent random variables. Therefore, for the copula models considered in this paper, we first transform $\bm X$ to $\bm V$ by the conditional inverse method, then apply the hazard rate twisting approach to the components of $\bm V$. The importance sampling estimator based on the hazard rate twisting approach is denoted as IS$_{t3}$.

Second, in order to include the computing time in the comparisons, we adopt the work normalized relative variance (WNRV) metric which is proposed by Ben Rached et al. \cite{Ben2021}. The WNRV of an estimator $\hat{u}$ of $u$ is defined as follows:
$$\mbox{WNRV}(\hat{u})=\frac{\hat{var}(\hat{u})}{u^2}\cdot \frac{1}{M}\cdot \mbox{computing time in seconds},$$
where $M$ is the number of replications. Clearly, the smaller $\mbox{WNRV}(\hat{u})$ is, the better it performs in terms of considering variance reduction and computing time simultaneously. Then, similar to the measures $\hbox{sd\_eff(Naive, IS)}$ and $\hbox{sd\_eff(IS$_{t2}$, IS$_{t1}$)}$ defined previously, we define $\hbox{WNRV\_eff(Naive, IS)}$ and $\hbox{WNRV\_eff(IS$_{t2}$, IS$_{t1}$)}$ as follows:
\begin{align*}
\hbox{WNRV\_eff(Naive, IS)} =\frac{\mbox{WNRV(Naive)}}{\mbox{WNRV(IS)}},
\end{align*}
and
\begin{align*}
\hbox{WNRV\_eff(IS$_{t2}$, IS$_{t1}$)} = \frac{\mbox{WNRV(IS$_{t2}$)}}{\mbox{WNRV(IS$_{t1}$)}}.
\end{align*}

In the following examples, we first consider bivariate copulas. The simulation event is an equal upper corner and rare event $\{X_{1}>p, X_{2}>p\}$ for various $p$. Three examples---Gaussian, $t$, and Clayton copulas---are involved in this study. In each example, we perform importance sampling with the conditional inverse method, which is particularly useful for low-dimensional copulas. We also study specific importance sampling for Gaussian and $t$ copulas, and importance sampling with the Marshall--Olkin method for Clayton copulas. Note that in each example, the sample size~$n$ and the number of replications $M$ are set to $500$ and $5,000$, respectively.\\

\noindent
{\bf Example 1: Gaussian Copula}. Assume the joint CDF of $ {\bm X}=(X_{1},X_{2})^{'}$ is modeled by a bivariate Gaussian copula with a form of
\begin{align*}
F(x_{1},x_{2}) = C(F_{1}(x_{1}), F_{2}(x_{2}); \Sigma) =  {\Phi}_{\Sigma}[\Phi^{-1}(F_{1}(x_{1})),\Phi^{-1}(F_{2}(x_{2}))],
\end{align*}
where $\Phi(\cdot)$ is the CDF of $N(0,1)$, ${\Phi}_{\Sigma}(\cdot)$ is the CDF of a bivariate normal with mean zero and covariance matrix
$\Sigma=\left(
\begin{array}{cc}
1 & \rho \\
\rho & 1
\end{array}\right)$, and $F_{1}(\cdot)$ and $F_{2}(\cdot)$ are the marginal CDFs of $X_{1}$ and $X_{2}$, respectively.

Two importance sampling methods are used in this simulation study. One is the conditional inverse method, which describes
\begin{align*}
X_{1} = F^{-1}_{1}(V_{1}),  \quad X_{2} = F^{-1}_{2}[\Lambda^{-1}(V_{2}|V_{1})],
\end{align*}
where $V_{1}$ and $V_{2}$ are independent uniform random variables on $(0,1)$ and
\begin{align*}
\Lambda^{-1}(V_{2}|V_{1}) = \Phi[\sqrt{1-\rho^{2}}\Phi^{-1}(V_{2})+\rho\Phi^{-1}(V_{1})]
\end{align*}
is the inverse of conditional distribution function. Note that the conditional distribution function of the bivariate Gaussian copula is
\begin{align*}
\Lambda (V_{2}|V_{1}) = \Phi \left( \frac{\Phi^{-1}(V_{2}) -\rho\Phi^{-1}(V_{1})} {\sqrt{1-\rho^{2}}}\right).
\end{align*}
After the transformation, the event of interest becomes $\{V_{1}>F_{1}(p),V_{2}>\Lambda(F_{2}(p)|V_{1})\}$, which turns out to be an involved event, comprising non-linearity and intractability. The optimal tilting probability measure is obtained via Proposition~\ref{Pro1}. The importance sampling estimator is denoted as IS$_{t1}$, and the optimal tilting point is denoted as $\theta_{t1}$.

Next, we transform $ {X}$ as a function of the bivariate normal random variable ${\bm V}=(\tilde{V}_{1},\tilde{V}_{2})^{'}\sim \mathrm{MN}({0},\Sigma)$,
\begin{align*}
X_{1}=F^{-1}_{1}[\Phi(\tilde{V}_{1})], \quad X_{2}=F^{-1}_{2}[\Phi(\tilde{V}_{2})].
\end{align*}
The event of interest becomes $\{\tilde{V}_{1}>\Phi^{-1}(F_{1}(p)), \tilde{V}_{2}>\Phi^{-1}(F_{2}(p))\}$. Then we employ the importance sampling algorithm on the transformed random variable $\bm V$ as shown in Proposition~\ref{Pro2}. We denote the importance sampling estimator as IS$_{t2}$ and the optimal tilting point as $\theta_{t2}$.

\begin{table}[]\footnotesize
\caption{Numerical results for Gaussian copulas with $\rho=0$ whose margins are both $N(0,1)$.}
\label{ISg1}\centering
\begin{tabular}{ccccc}
      \toprule
      $p$ & 0.760 & 1.282 & 1.471 & 1.857  \\
      \toprule
      Naive estimator &4.99E-02	 &1.00E-02	 &5.00E-03	 &1.00E-03 \\
      sd(Naive)      &9.56E-03   &4.44E-03	 &3.15E-03	 &1.41E-03 \\
      WNRV(Naive)    &3.79E-05   &1.74E-04   &4.49E-04   &1.81E-03 \\
      \hline
       $\theta_{t1}$        &(7.09, 7.09)    &(15.95, 15.95)   &(22.56, 22.56) 	 &(50.34, 50.34)   \\
      IS$_{t1}$ estimator    &5.00E-02	     &1.00E-02	       &5.00E-03	     &1.00E-03 \\
      sd(IS$_{t1}$)         &2.63E-03  	     &5.29E-04	       &2.65E-04         &5.20E-05 \\
      WNRV(IS$_{t1}$)       &1.69E-05        &1.99E-05         &1.66E-05         &1.52E-05 \\
      sd$\_$eff(Naive, IS$_{t1}$)   &3.64            &8.43 	           &12.01            &27.13  \\
      WNRV$\_$eff(Naive, IS$_{t1}$) &2.24            &8.74 	           &27.03            &119.21  \\
      \hline
      $\theta_{t2}$         &(1.14, 1.14)    &(1.58, 1.58) 	  &(1.74, 1.74)      &(2.09, 2.09) \\
      IS$_{t2}$ estimator    &5.02E-02	    &1.00E-02	      &5.00E-03          &1.00E-03 \\
      sd(IS$_{t2}$)         &4.18E-03	    &1.05E-03	      &5.66E-04          &1.41E-04 \\
      WNRV(IS$_{t2}$)       &7.46E-06	    &1.39E-05	      &1.38E-05          &1.96E-05 \\
      sd$\_$eff(Naive, IS$_{t2}$)   &2.29 	        &4.21             &5.61 	         &10.68 \\
      WNRV$\_$eff(Naive, IS$_{t2}$) &5.08 	        &12.48            &32.54 	         &92.27 \\
      \hline
      $\theta_{t3}$         &0.33           &0.57       	  &0.62              &0.71 \\
      IS$_{t3}$ estimator    &5.01E-02	    &1.00E-02	      &5.00E-03          &9.98E-04 \\
      sd(IS$_{t3}$)         &6.41E-03	    &1.71E-03	      &9.49E-04          &2.43E-04 \\
      WNRV(IS$_{t3}$)       &1.30E-05	    &2.26E-05	      &3.07E-05          &4.45E-05 \\
      sd$\_$eff(Naive, IS$_{t3}$)   &1.49 	        &2.60             &3.32 	         &5.80 \\
      WNRV$\_$eff(Naive, IS$_{t3}$) &2.92 	        &7.70             &14.63 	         &40.67 \\
      \hline
      sd$\_$eff(IS$_{t2}$, IS$_{t1}$)   &1.59       &2.00 	          &2.14 	         &2.54  \\
      WNRV$\_$eff(IS$_{t2}$, IS$_{t1}$) &0.44       &0.70 	          &0.83 	         &1.29  \\
      \bottomrule
\end{tabular}
\end{table}

\begin{table}[]\footnotesize
\caption{Numerical results for Gaussian copulas with $\rho=0.5$ whose margins are both $N(0,1)$.}
\label{ISg2}\centering
\begin{tabular}{ccccc}
      \toprule
      $p$ & 1.100 & 1.712 & 1.936 & 2.395  \\
      \toprule
      Naive estimator &5.02E-02	 &1.00E-02	&5.10E-03	&1.00E-03 \\
      sd(Naive)      &9.93E-03	 &4.42E-03	&3.19E-03	&1.40E-03 \\
      WNRV(Naive)    &4.15E-05	 &2.17E-04	&4.53E-04	&2.24E-03 \\
      \hline
       $\theta_{t1}$       &(13.94, 4.02)   &(44.93, 6.48)    &(74.50, 7.78) 	 &(240.44, 11.66)   \\
      IS$_{t1}$ estimator   &5.00E-02	    &1.00E-02	      &5.00E-03	         &1.00E-03 \\
      sd(IS$_{t1}$)        &2.35E-03  	    &4.69E-04	      &2.45E-04          &4.99E-05 \\
      WNRV(IS$_{t1}$)      &1.53E-05  	    &1.68E-05	      &1.70E-05          &1.97E-05 \\
      sd$\_$eff(Naive, IS$_{t1}$)    &4.22	    &9.34        &13.38        &28.12 \\
      WNRV$\_$eff(Naive, IS$_{t1}$)  &2.71	    &12.91       &26.58        &113.55 \\
      \hline
      $\theta_{t2}$        &(1.01, 1.01)    &(1.36, 1.36) 	  &(1.49, 1.49)      &(1.77, 1.77) \\
      IS$_{t2}$ estimator   &5.00E-02	    &1.00E-02	      &5.00E-03	         &1.00E-03 \\
      sd(IS$_{t2}$)        &3.59E-03	    &9.00E-04	      &4.69E-04          &1.00E-04 \\
      WNRV(IS$_{t2}$)      &7.28E-06	    &1.14E-05	      &1.22E-05          &1.58E-05 \\
      sd$\_$eff(Naive, IS$_{t2}$)    &2.77 	    &4.92       &6.73	      &12.75 \\
      WNRV$\_$eff(Naive, IS$_{t2}$)  &5.70 	    &19.00      &37.23	      &141.94 \\
      \hline
      $\theta_{t3}$         &0.40           &0.60       	  &0.65              &0.73 \\
      IS$_{t3}$ estimator    &5.00E-02	    &1.00E-02	      &4.99E-03          &9.95E-04 \\
      sd(IS$_{t3}$)         &5.85E-03	    &1.56E-03	      &8.70E-04          &2.13E-04 \\
      WNRV(IS$_{t3}$)       &1.21E-05	    &2.14E-05	      &2.87E-05          &4.19E-05 \\
      sd$\_$eff(Naive, IS$_{t3}$)   &1.70 	        &2.83             &3.67 	         &6.57 \\
      WNRV$\_$eff(Naive, IS$_{t3}$) &3.43 	        &10.14            &15.78 	         &53.46 \\
      \hline
      sd$\_$eff(IS$_{t2}$, IS$_{t1}$)    &1.52       &1.90       &1.99 	    &2.21  \\
      WNRV$\_$eff(IS$_{t2}$, IS$_{t1}$)  &0.48       &0.68       &0.71 	    &0.80  \\
      \bottomrule
\end{tabular}
\end{table}

\begin{table}[]\footnotesize
\caption{Numerical results for Gaussian copulas with $\rho=-0.5$ whose margins are both $N(0,1)$.}
\label{ISg3}\centering
\begin{tabular}{ccccc}
      \toprule
      $p$  & 0.411  & 0.806  & 0.947  & 1.233  \\
      \toprule
      Naive estimator &5.03E-02	 &1.01E-02	&5.00E-03	&1.00E-03 \\
      sd(Naive)      &9.90E-03	 &4.45E-03	&3.10E-03	&1.37E-03 \\
      WNRV(Naive)    &4.78E-05	 &2.34E-04	&4.80E-04	&2.39E-03 \\
      \hline
       $\theta_{t1}$       &(3.58, 8.20)    &(6.25, 24.25)    &(7.68, 39.2) 	 &(12.09, 121.57)   \\
      IS$_{t1}$ estimator   &5.01E-02	    &1.00E-02	      &5.00E-03	         &1.00E-03 \\
      sd(IS$_{t1}$)        &3.51E-03  	    &7.35E-04	      &3.74E-04          &1.00E-4 \\
      WNRV(IS$_{t1}$)      &4.02E-05  	    &4.18E-05	      &4.28E-05          &4.65E-5 \\
      sd$\_$eff(Naive, IS$_{t1}$)    &2.82	       &6.06	      &8.30          &13.75 \\
      WNRV$\_$eff(Naive, IS$_{t1}$)  &1.19	       &5.61	      &11.21         &51.39 \\
      \hline
      $\theta_{t2}$        &(1.44, 1.44)    &(2.07, 2.07) 	  &(2.31, 2.31)      &(2.81, 2.81) \\
      IS$_{t2}$ estimator   &5.00E-02	    &1.00E-02	      &5.00E-03	         &1.00E-03 \\
      sd(IS$_{t2}$)        &4.96E-03	    &1.30E-03	      &6.86E-04          &1.73E-04 \\
      WNRV(IS$_{t2}$)      &1.40E-05	    &2.33E-05	      &2.60E-05          &3.35E-05 \\
      sd$\_$eff(Naive, IS$_{t2}$)      &1.99 	     &3.41         &4.51	     &7.94 \\
      WNRV$\_$eff(Naive, IS$_{t2}$)    &3.43 	     &10.04        &18.45	     &71.19 \\
      \hline
      $\theta_{t3}$         &0.20           &0.51       	  &0.58              &0.68 \\
      IS$_{t3}$ estimator    &4.99E-02	    &1.00E-02	      &5.00E-03          &9.94E-04 \\
      sd(IS$_{t3}$)         &7.69E-03	    &2.06E-03	      &1.15E-03          &3.05E-04 \\
      WNRV(IS$_{t3}$)       &2.19E-05	    &3.76E-05	      &4.68E-05          &8.51E-05 \\
      sd$\_$eff(Naive, IS$_{t3}$)   &1.29 	        &2.16             &2.70 	         &4.49 \\
      WNRV$\_$eff(Naive, IS$_{t3}$) &2.18 	        &6.22             &10.26 	         &28.08 \\
      \hline
      sd$\_$eff(IS$_{t2}$, IS$_{t1}$)      &1.41       &1.78	      &1.84	      &1.73  \\
      WNRV$\_$eff(IS$_{t2}$, IS$_{t1}$)    &0.35       &0.56	      &0.61	      &0.72  \\
      \bottomrule
\end{tabular}
\end{table}

\begin{table}[]\footnotesize
\caption{Numerical results for Gaussian copulas with $\rho=0$ whose margins are both exponential distributions with rate parameter $1$.}
\label{ISg4}\centering
\begin{tabular}{ccccc}
      \toprule
      $p$    &1.498     &2.303     &2.649	   &3.454  \\
      \toprule
      Naive estimator  &4.99E-02	 &0.99E-02	&5.07E-03	&1.04E-03 \\
      sd(Naive)       &9.69E-03	 &4.50E-03	&3.17E-03	&1.45E-03 \\
      WNRV(Naive)     &4.45E-05	 &2.32E-04	&4.55E-04	&2.20E-03 \\
      \hline
      $\theta_{t1}$        & (7.09, 7.09)	& (15.94, 15.94)	& (22.53, 22.53)	& (50.40, 50.40)   \\
      IS$_{t1}$ estimator   & 5.00E-02	    & 1.00E-02          & 5.00E-03          & 1.00E-03 \\
      sd(IS$_{t1}$)        & 2.61E-03       & 5.29E-04          & 2.65E-04	        & 5.31E-05 \\
      WNRV(IS$_{t1}$)      & 1.83E-05       & 1.85E-05          & 1.89E-05	        & 1.82E-05 \\
      sd$\_$eff(Naive, IS$_{t1}$)    & 3.71 	        & 8.59             & 12.01 	        & 27.31 \\
      WNRV$\_$eff(Naive, IS$_{t1}$)  & 2.43 	        & 12.52            & 24.14 	        & 120.90 \\
      \hline
      $\theta_{t2}$        & (1.14, 1.14)	& (1.58, 1.58)      & (1.74, 1.74)	    & (2.09, 2.09) \\
      IS$_{t2}$ estimator   & 5.00E-02	    & 1.00E-02	        & 5.00E-03	        & 0.99E-03 \\
      sd(IS$_{t2}$)        & 4.19E-03	    & 1.07E-03	        & 5.74E-04	        & 1.41E-04 \\
      WNRV(IS$_{t2}$)      & 9.77E-06	    & 1.51E-05	        & 1.69E-05	        & 2.17E-05 \\
      sd$\_$eff(Naive, IS$_{t2}$)   & 2.32 	        & 4.21          & 5.54        & 10.98  \\
      WNRV$\_$eff(Naive, IS$_{t2}$) & 4.55 	        & 15.32         & 26.98       & 101.41  \\
      \hline
      $\theta_{t3}$         &0.33           &0.57       	  &0.62              &0.71 \\
      IS$_{t3}$ estimator    &5.00E-02	    &1.00E-02	      &5.00E-03          &1.00E-03 \\
      sd(IS$_{t3}$)         &6.50E-03	    &1.77E-03	      &9.52E-04          &2.39E-04 \\
      WNRV(IS$_{t3}$)       &1.34E-05	    &2.31E-05	      &2.76E-05          &4.32E-05 \\
      sd$\_$eff(Naive, IS$_{t3}$)   &1.49 	        &2.54             &3.33 	         &6.07 \\
      WNRV$\_$eff(Naive, IS$_{t3}$) &3.32 	        &10.04            &16.49 	         &50.93 \\
      \hline
      sd$\_$eff(IS$_{t2}$, IS$_{t1}$)    & 1.60    & 2.02    & 2.17    & 2.49  \\
      WNRV$\_$eff(IS$_{t2}$, IS$_{t1}$)  & 0.53    & 0.82    & 0.89    & 1.19  \\
      \bottomrule
\end{tabular}
\end{table}

\begin{table}[]\footnotesize
\caption{Numerical results for Gaussian copulas with $\rho=0.5$ whose margins are both exponential distributions with rate parameter $1$.}
\label{ISg5}\centering
\begin{tabular}{ccccc}
      \toprule
      $p$ & 1.997    &	3.137    &	3.633	& 4.791  \\
      \toprule
      Naive estimator &5.02E-02	 &1.01E-02	&5.06E-03	&0.99E-03 \\
      sd(Naive)      &9.79E-03	 &4.49E-03	&3.19E-03	&1.41E-03 \\
      WNRV(Naive)    &4.79E-05	 &2.22E-04	&4.56E-04	&2.28E-03 \\
      \hline
      $\theta_{t1}$        & (13.93, 4.02)	& (44.97, 6.49)	    & (74.49, 7.78)	    & (240.62, 11.66)   \\
      IS$_{t1}$ estimator   & 5.00E-02	    & 1.00E-02          & 5.00E-03          & 1.00E-03  \\
      sd(IS$_{t1}$)        & 2.32E-03       & 4.80E-04          & 2.45E-04	        & 4.82E-05  \\
      WNRV(IS$_{t1}$)      & 1.61E-05       & 1.69E-05          & 1.70E-05	        & 1.77E-05  \\
      sd$\_$eff(Naive, IS$_{t1}$)   & 4.23 	        & 9.43             & 13.34 	        & 29.33      \\
      WNRV$\_$eff(Naive, IS$_{t1}$) & 2.98 	        & 13.15            & 26.75 	        & 128.84      \\
      \hline
      $\theta_{t2}$        & (1.01, 1.01)	& (1.36, 1.36)      & (1.49, 1.49)	    & (1.77, 1.77)\\
      IS$_{t2}$ estimator   & 5.01E-02	    & 1.00E-02          & 5.00E-03          & 1.00E-03  \\
      sd(IS$_{t2}$)        & 3.59E-03	    & 9.00E-04	        & 4.80E-04	        & 1.00E-04 \\
      WNRV(IS$_{t2}$)      & 7.09E-06	    & 1.09E-05	        & 1.23E-05	        & 1.46E-05 \\
      sd$\_$eff(Naive, IS$_{t2}$)   & 2.73 	        & 4.98             & 6.66	        & 13.01  \\
      WNRV$\_$eff(Naive, IS$_{t2}$) & 6.76 	        & 20.30            & 37.01	        & 156.36  \\
      \hline
      $\theta_{t3}$         &0.40           &0.60       	  &0.65              &0.73 \\
      IS$_{t3}$ estimator    &5.00E-02	    &9.99E-03	      &5.01E-03          &1.00E-03 \\
      sd(IS$_{t3}$)         &5.89E-03	    &1.53E-03	      &8.68E-04          &2.09E-04 \\
      WNRV(IS$_{t3}$)       &1.25E-05	    &2.12E-05	      &2.66E-05          &3.84E-05 \\
      sd$\_$eff(Naive, IS$_{t3}$)   &1.66 	        &2.93             &3.68 	         &6.75 \\
      WNRV$\_$eff(Naive, IS$_{t3}$) &3.83 	        &10.47            &17.14 	         &59.38 \\
      \hline
      sd$\_$eff(IS$_{t2}$, IS$_{t1}$)   & 1.55    & 1.88    & 1.96    & 2.07  \\
      WNRV$\_$eff(IS$_{t2}$, IS$_{t1}$) & 0.44    & 0.65    & 0.72    & 0.82  \\
      \bottomrule
\end{tabular}
\end{table}

\begin{table}[]\footnotesize
\caption{Numerical results for Gaussian copulas with $\rho=-0.5$ whose margins are both exponential distributions with rate parameter $1$.}
\label{ISg6}\centering
\begin{tabular}{ccccc}
      \toprule
      $p$ & 1.078    &	1.560    &	1.761	&2.218  \\
      \toprule
      Naive estimator & 5.01E-02	 &0.98E-02	&4.95E-03	&1.03E-03 \\
      sd(Naive)      & 9.73E-03	 &4.33E-03	&3.19E-03	&1.42E-03 \\
      WNRV(Naive)    & 4.49E-05	 &2.18E-04	&4.74E-04	&2.20E-03 \\
      \hline
      $\theta_{t1}$        & (3.58, 8.21)	& (6.25, 24.25)	    & (7.68, 39.17)	    & (12.08, 121.46)   \\
      IS$_{t1}$ estimator   & 5.00E-02	    & 1.00E-02          & 5.00E-03          & 1.00E-03  \\
      sd(IS$_{t1}$)        & 3.55E-03       & 7.48E-04          & 3.87E-04	        & 1.00E-04  \\
      WNRV(IS$_{t1}$)      & 3.68E-05       & 4.07E-05          & 4.22E-05	        & 4.20E-05  \\
      sd$\_$eff(Naive, IS$_{t1}$)    & 2.74 	      & 5.79          & 8.35 	       & 18.83      \\
      WNRV$\_$eff(Naive, IS$_{t1}$)  & 1.22 	      & 5.35          & 11.22 	       & 52.31      \\
      \hline
      $\theta_{t2}$        & (1.44, 1.44)	& (2.07, 2.07)      & (2.31, 2.31)	    & (2.81, 2.81) \\
      IS$_{t2}$ estimator   & 5.00E-02	    & 1.00E-02          & 5.02E-03          & 1.00E-03  \\
      sd(IS$_{t2}$)        & 4.96E-03	    & 1.26E-03	        & 7.00E-04	        & 1.73E-04 \\
      WNRV(IS$_{t2}$)      & 1.32E-05       & 2.04E-05          & 2.53E-05	        & 3.37E-05  \\
      sd$\_$eff(Naive, IS$_{t2}$)      & 1.96 	      & 3.43           & 4.55          & 8.66  \\
      WNRV$\_$eff(Naive, IS$_{t2}$)    & 3.41 	      & 10.67          & 18.72         & 65.21  \\
      \hline
      $\theta_{t3}$         &0.20           &0.51       	  &0.58              &0.68 \\
      IS$_{t3}$ estimator    &5.02E-02	    &1.00E-02	      &4.98E-03          &9.99E-04 \\
      sd(IS$_{t3}$)         &7.71E-03	    &2.10E-03	      &1.16E-03          &3.08E-04 \\
      WNRV(IS$_{t3}$)       &2.27E-05	    &4.03E-05	      &5.02E-05          &8.58E-05 \\
      sd$\_$eff(Naive, IS$_{t3}$)   &1.26 	        &2.06             &2.75 	         &4.61 \\
      WNRV$\_$eff(Naive, IS$_{t3}$) &1.98 	        &5.41             &9.44 	         &25.64 \\
      \hline
      sd$\_$eff(IS$_{t2}$, IS$_{t1}$)   & 1.40    & 1.69    & 1.81    & 1.73  \\
      WNRV$\_$eff(IS$_{t2}$, IS$_{t1}$) & 0.36    & 0.50    & 0.60    & 0.80  \\
      \bottomrule
\end{tabular}
\end{table}

The marginal CDFs are assumed to be either all $N(0,1)$ or all exponential distributions with rate parameter 1. The simulation results with various $p$ and $\rho$ are reported in Tables~\ref{ISg1}--\ref{ISg6}. It can be found from Tables~\ref{ISg1}--\ref{ISg6} that the improvement seems irrelevant to the marginal CDFs since the relative efficiency has no significant difference with respect to marginal CDFs. For general events a complicated marginal leads to a complicated event. For an upper corner event transformed using the conditional inverse method, $\{V_{1}>F_{1}(p), V_{2}>\Lambda(F_{2}(p)|V_{1})\}$ is not too difficult to calculate even if $F_{i}(\cdot )$ is quite complicated. The case of an upper corner event transformed using the simulation method for Gaussian copulas (see Subsection \ref{ISElliptical}) is similar. We also note that the variance of IS$_{t1}$ is significantly smaller than that of IS$_{t2}$, and the variance of IS$_{t2}$ is significantly smaller than that of IS$_{t3}$. The finding that the performance of IS$_{t3}$ is the worst in terms of variance reduction is expected since the hazard rate twisting approach is originally designed for heavy-tailed random variables, cf. Juneja and Shahabuddin \cite{JS2002} and Ben Rached et al. \cite{Ben2018}. Moreover, roughly speaking, the performance of IS$_{t3}$ is still the worst while that of IS$_{t2}$ is the best in terms of the WNRV metric which considers the variance reduction and computational time simultaneously. This finding becomes more clear as the simulated probability decreases.

\begin{figure}[]
\small
\centering
  \includegraphics[scale=0.22]{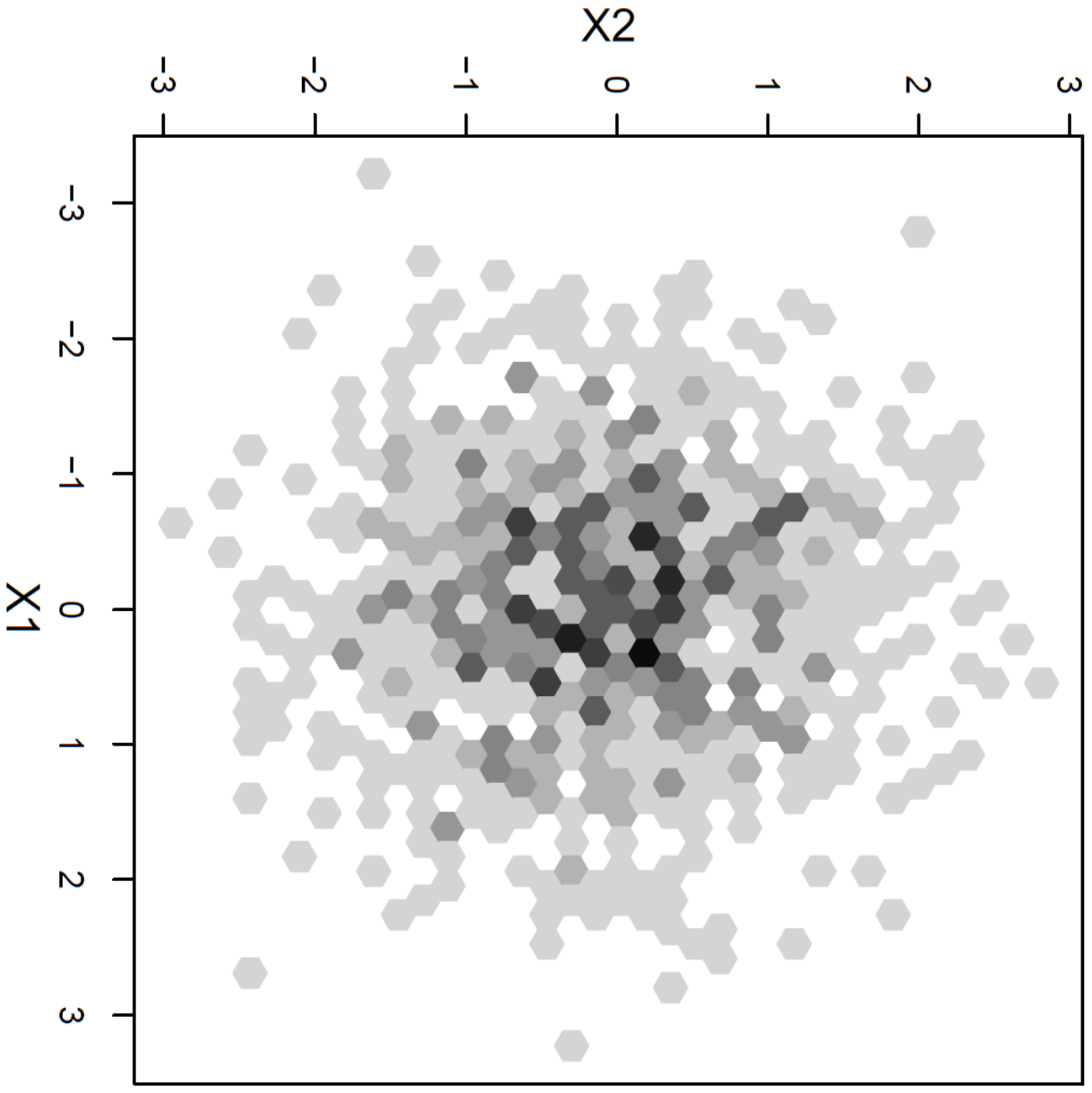}
  \includegraphics[scale=0.22]{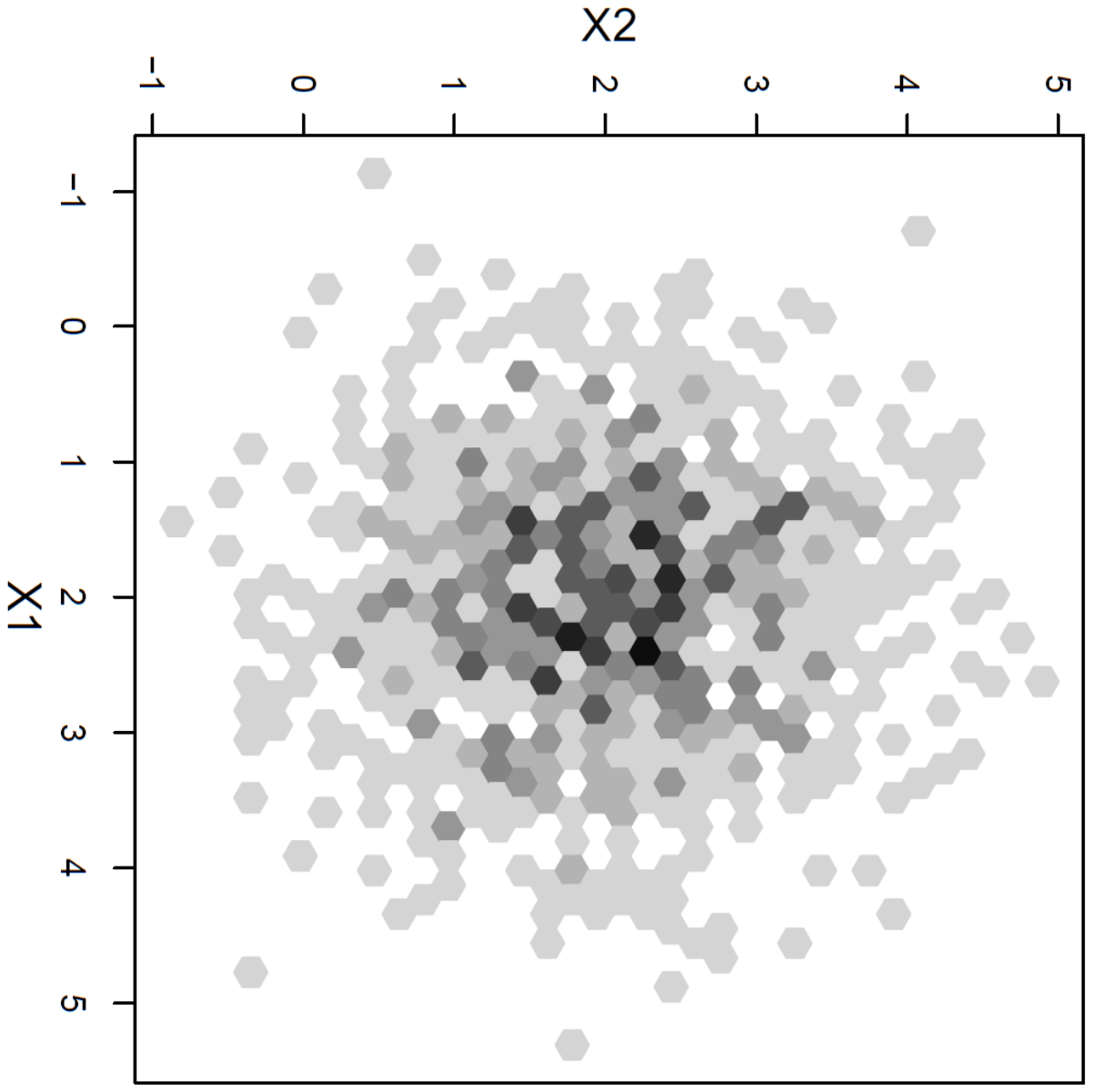}
  \includegraphics[scale=0.22]{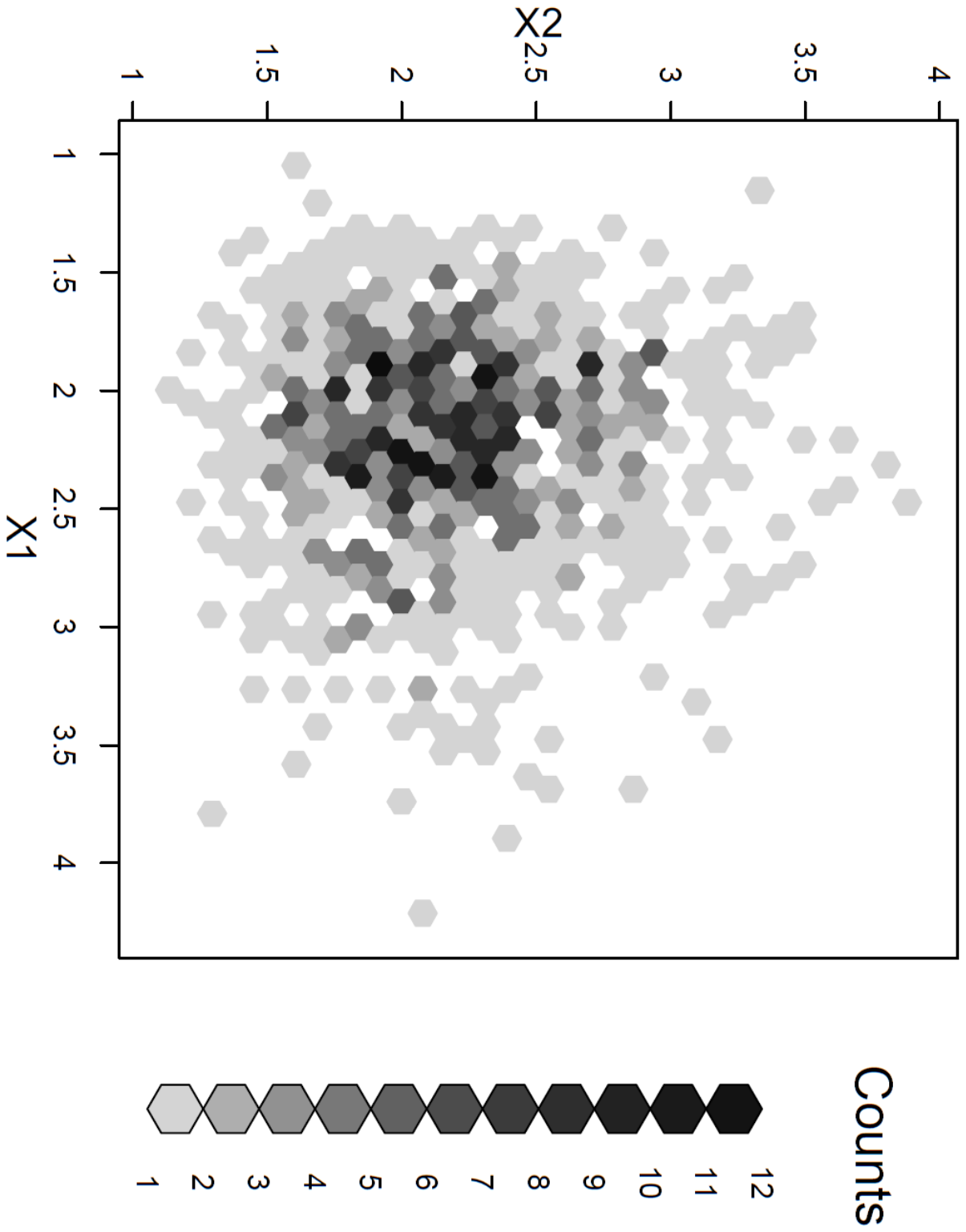}
  \caption{Left: a scatter plot of a Gaussian copula with $\rho=0$ and standard normal margins under the $P$-measure; middle: under the
  $Q_{\theta_{t2}}$-measure; right: under the $Q_{\theta_{t1}}$-measure.}
  \label{LTR}
\end{figure}

Next, we focus on the comparison of the performance of IS$_{t1}$ and IS$_{t2}$. From the tables, it can be found that the relative efficiency of IS$_{t2}$ with respect to IS$_{t1}$ is at least $1.40$ in terms of variance reduction, and IS$_{t1}$ outperforms IS$_{t2}$ as the simulated probability decreases. One reason for this outcome is that one has mean shift and variance shrinkage simultaneously in this sampling distribution for exponential tilting in the conditional inverse method; while there is only mean shift for exponential tilting in the other method. We also demonstrate the corresponding scatter plots of the sampling distributions of Gaussian copulas with standard normal margins before and after the change of measure. The demonstration uses the method of bivariate binning into hexagon cells in Figure~\ref{LTR}, cf. Carr et al. \cite{C1987}: darker colors mean that more sample points lie in the hexagon cell. In addition to mean shift, the right panel in Figure~\ref{LTR} is more concentrated than the other panels. Therefore, different alternative measures selected within their corresponding alternative families result in different efficiency improvements. However, if we compare the performance of IS$_{t1}$ and IS$_{t2}$ in term of the WNRV metric which considers the variance reduction and computational time simultaneously, it can be found that IS$_{t2}$ outperforms IS$_{t1}$ in most cases.

For higher-dimensional Gaussian copulas, we demonstrate a relative efficiency between the variance of the Monte Carlo estimator and that of the importance sampling estimator via the method described in Subsection \ref{ISElliptical} for Gaussian copulas (note that for high-dimensional copulas, the conditional inverse method becomes complex and unrealistic). As an illustration, we consider an equal and upper corner event again, $\{X_{1}>p, \ldots, X_{d}>p\}$ with $d=4$. Let
\begin{align}\label{Sigma}
\Sigma=\left(
\begin{array}{cccc}
1 & \frac{1}{2} & 0 & 0  \\
\frac{1}{2} & 1 & \frac{1}{2} & 0  \\
0 & \frac{1}{2} & 1 & \frac{1}{2}  \\
0 & 0 & \frac{1}{2} & 1
\end{array}
\right).
\end{align}
The marginal CDFs are assumed to be either all $N(0,1)$ or all exponential with rate parameter $1$. Denote this importance sampling estimator as IS$_{t2}$. The optimal $\theta_{t2}$ can be obtained by means of Proposition~\ref{Pro2}. The simulation results are reported in Tables~\ref{ISdim41} and~\ref{ISdim42}. These results indicate that the importance sampling estimator has a great deal of variance reduction as well as time saving in comparison to the crude Monte Carlo estimator, especially for the events with very small probabilities.

\begin{table}[]\footnotesize
\caption{Numerical results for Gaussian copulas with $\Sigma$ defined by (\ref{Sigma}) whose margins are all $N(0,1)$.}
\label{ISdim41}\centering
\begin{tabular}{ccccc}
      \toprule
      $p$   & 0.394   & 0.886   & 1.064   & 1.428\\
      \hline
      Naive estimator & 4.98E-02  & 1.01E-02   & 5.03E-03   & 1.00E-03 \\
      sd(Naive)      & 9.62E-03   & 4.44E-03    & 3.17E-03   & 1.40E-03 \\
      WNRV(Naive)    & 2.38E-05   & 1.20E-04    & 2.54E-04   & 1.21E-03 \\
      \midrule
      $\theta_{t2}$     & (0.70,0.47,0.47,0.70)     & (0.99,0.62,0.62,0.99)      & (1.11,0.68,0.68,1.11)     & (1.35,0.81,0.81,1.35) \\
      IS$_{t2}$ estimator & 5.01E-02   & 1.00E-02    & 5.01E-03   & 1.00E-03 \\
      sd(IS$_{t2}$)      & 4.54E-03   & 1.20E-03    & 6.57E-04   & 1.64E-04 \\
      WNRV(IS$_{t2}$)    & 6.95E-05   & 6.87E-05    & 7.46E-05   & 1.02E-04 \\
      sd$\_$eff(Naive, IS$_{t2}$)   & 2.12    & 3.71    & 4.83     & 8.52  \\
      WNRV$\_$eff(Naive, IS$_{t2}$) & 0.34    & 1.75    & 3.40     & 11.80 \\
      \bottomrule
\end{tabular}
\end{table}

\begin{table}[]\footnotesize
\caption{Numerical results for Gaussian copulas with $\Sigma$ defined by (\ref{Sigma}) whose margins are all exponential distributions with rate parameter $1$.}
\label{ISdim42}\centering
\begin{tabular}{ccccc}
      \toprule
      $p$   & 1.059   & 1.673   & 1.941   & 2.569\\
      \hline
      Naive estimator & 5.02E-02   & 1.01E-02   & 5.03E-03   & 1.00E-03 \\
      sd(Naive)      & 9.91E-03   & 4.52E-03    & 3.13E-03   & 1.40E-03 \\
      WNRV(Naive)    & 2.44E-05   & 1.26E-04    & 2.40E-04   & 1.25E-03 \\
      \midrule
      $\theta_{t2}$     & (0.70,0.47,0.47,0.70)     & (0.99,0.62,0.62,0.99)      & (1.11,0.68,0.68,1.11)     & (1.35,0.81,0.81,1.35) \\
      IS$_{t2}$ estimator & 5.00E-02   & 1.00E-02    & 5.00E-03   & 1.00E-03 \\
      sd(IS$_{t2}$)      & 4.44E-03   & 1.21E-03    & 6.69E-04   & 1.61E-04 \\
      WNRV(IS$_{t2}$)    & 6.52E-05   & 6.99E-05    & 7.80E-05   & 1.00E-04 \\
      sd$\_$eff(Naive, IS$_{t2}$)   & 2.23    & 3.74    & 4.68     & 8.68  \\
      WNRV$\_$eff(Naive, IS$_{t2}$) & 0.37    & 1.80    & 3.08     & 12.45 \\
      \bottomrule
\end{tabular}
\end{table}

\bigskip
\noindent
{\bf Example 2: $t$ Copula}. To describe symmetric tail dependence, a commonly-used elliptical copula is the $t$ copula, which is extracted from a multivariate $t$ distribution. The bivariate joint CDF of a $t$ copula for $X_{1}$ and $X_{2}$ with marginal CDFs $F_{1}$ and $F_{2}$ is
\begin{align*}
F(x_{1}, x_{2}) = C(F_{1}(x_{1}), F_{2}(x_{2}); \nu ,\Sigma) =  {t}_{\nu ,\Sigma}(t^{-1}_{\nu}(F_{1}(x_{1})),t^{-1}_{\nu}(F_{2}(x_{2}))),
\end{align*}
where $t_{\nu}$ is the CDF of a univariate $t$ distribution and ${t}_{\nu,\Sigma}$ is the joint CDF of a bivariate $t$ distribution $t_{\nu}({0},\Sigma)$ with $\Sigma=\left(
\begin{array}{cc}
1 & \rho \\
\rho & 1
\end{array}\right)$.

In this simulation study we employ three importance sampling methods. The first is the conditional inverse method with the exponential twisting, in which the algorithm is the same as Example~1 except for the inverse of the conditional distribution
\begin{align*}
\Lambda^{-1}(V_{2}|V_{1}) = t_{\nu}\left[ t^{-1}_{\nu +1}(V_{2})\sqrt{\frac{\nu + [t^{-1}_{\nu}(V_{1})]^{2}}{\nu +1}(1-\rho^{2})} + \rho t^{-1}_{\nu}(V_{1}) \right].
\end{align*}
Note that the conditional distribution function of a two-dimensional $t$ copula is
\begin{align*}
\Lambda(V_{2}|V_{1}) = t_{\nu+1}\left[ \sqrt{\frac{\nu+1}{\nu+[t^{-1}_{\nu}(V_1)]^2}}\frac{t^{-1}_{\nu}(V_2)-\rho t^{-1}_{\nu}(V_1)}{\sqrt{1-\rho^2}} \right].
\end{align*}
The importance sampling estimator is denoted as IS$_{t1}$ and the optimal tilting point is denoted as $\theta_{t1}$. The second algorithm is based on applying the exponential tilting scheme on ${\bm Z}$ and $Y$ simultaneously, described in Proposition~\ref{Pro3}. The importance sampling estimator is denoted as IS$_{t2}$, and the optimal tilting point is denoted as $\theta_{t2}$. The last algorithm is the conditional inverse method with the hazard rate twisting, and the importance sampling estimator is denoted as IS$_{t3}$ with the optimal tilting point being denoted as $\theta_{t3}$. Suppose the marginal CDFs are both $t_2$, then the simulation results with $\nu=5$ and $\rho\in \{0, 0.5, -0.5\}$ are reported in Tables~\ref{IStcoupla1}--\ref{IStcoupla3}.

\begin{table}[]\footnotesize
\caption{Numerical results for $t$ copulas with degree of freedom $\nu=5$ and $\rho=0$ whose margins are both $t_{2}$.}
\label{IStcoupla1}\centering
\begin{tabular}{ccccc}
      \toprule
      $p$            &1.000       &2.268	    &3.066	    &6.128  \\
      \toprule
      Naive estimator &5.02E-02	  &1.00E-02	    &5.00E-03   &1.00E-03\\
      sd(Naive)      &9.75E-03    &4.50E-03	    &3.11E-03	&1.43E-03\\
      WNRV(Naive)    &1.33E-04    &5.49E-04	    &1.51E-03	&7.42E-03\\
      \hline
      $\theta_{t1}$        &(8.19, 6.83)	&(25.68, 11.35)  &(44.15, 12.88)    &(169.88, 15.22) \\
      IS$_{t1}$ estimator   &4.99E-02	    &1.00E-02	     &5.00E-03          &1.00E-03 \\
      sd(IS$_{t1}$)        &2.45E-03	    &4.80E-04	     &2.45E-04          &5.11E-05 \\
      WNRV(IS$_{t1}$)      &3.96E-05	    &3.05E-05	     &4.04E-05          &4.65E-05 \\
      sd$\_$eff(Naive, IS$_{t1}$)    &3.98 	        &9.43 	         &12.94 	        &27.93   \\
      WNRV$\_$eff(Naive, IS$_{t1}$)  &3.35 	        &17.98 	         &37.49 	        &159.36   \\
      \hline
      $\theta_{t2}$        &(1.25, 1.25)	&(2.09, 2.09)	 &(2.51, 2.51)      &(3.68, 3.68) \\
      IS$_{t2}$ estimator   &4.99E-02	    &1.00E-02	     &5.00E-03          &1.00E-03 \\
      sd(IS$_{t2}$)        &4.31E-03	    &9.85E-04	     &5.10E-04	        &1.00E-04 \\
      WNRV(IS$_{t2}$)      &1.76E-03	    &2.00E-03	     &2.60E-03          &2.99E-03 \\
      sd$\_$eff(Naive, IS$_{t2}$)     &2.26            &4.57           &6.07             &13.16   \\
      WNRV$\_$eff(Naive, IS$_{t2}$)   &0.08            &0.27           &0.58             &2.48   \\
      \hline
      $\theta_{t3}$         &0.36           &0.59       	  &0.65              &0.73 \\
      IS$_{t3}$ estimator    &5.00E-02	    &9.99E-03	      &4.97E-03          &9.95E-04 \\
      sd(IS$_{t3}$)         &6.12E-03	    &1.53E-03	      &8.38E-04          &2.01E-04 \\
      WNRV(IS$_{t3}$)       &2.75E-05	    &4.15E-05	      &4.99E-05          &7.20E-05 \\
      sd$\_$eff(Naive, IS$_{t3}$)   &1.59 	        &2.94             &3.71 	         &7.11 \\
      WNRV$\_$eff(Naive, IS$_{t3}$) &4.84 	        &13.23            &30.26 	         &103.06 \\
      \hline
      sd$\_$eff(IS$_{t2}$, IS$_{t1}$)     &1.76 	&2.05 	&2.08 	&1.96   \\
      WNRV$\_$eff(IS$_{t2}$, IS$_{t1}$)   &44.49 	&65.56 	&64.43 	&64.17   \\
      \bottomrule
\end{tabular}
\end{table}

\begin{table}[]\footnotesize
\caption{Numerical results for $t$ copulas with degree of freedom $\nu=5$ and $\rho=0.5$ whose margins are both $t_{2}$.}
\label{IStcoupla2}\centering
\begin{tabular}{ccccc}
      \toprule
      $p$            &1.592       &3.677		    &5.111	    &10.938  \\
      \toprule
      Naive estimator &5.01E-02	  &1.00E-02	    &5.10E-03   &1.00E-03 \\
      sd(Naive)      &9.77E-03    &4.42E-03	    &3.20E-03	&1.44E-03 \\
      WNRV(Naive)    &1.42E-04    &7.01E-04	    &1.45E-03	&7.92E-03 \\
      \hline
      $\theta_{t1}$        &(15.61, 3.58)	&(63.35, 4.56)   &(118.94, 4.82)    &(537.92, 5.22) \\
      IS$_{t1}$ estimator   &5.00E-02	    &1.00E-02	     &5.00E-03          &1.00E-03 \\
      sd(IS$_{t1}$)        &2.23E-03	    &4.69E-04	     &2.45E-04          &4.73E-05 \\
      WNRV(IS$_{t1}$)      &3.45E-05	    &3.62E-05	     &3.71E-05          &3.94E-05 \\
      sd$\_$eff(Naive, IS$_{t1}$)    &4.38 	        &9.41	       &13.46 	      &30.47 \\
      WNRV$\_$eff(Naive, IS$_{t1}$)  &4.13 	        &19.37	       &39.08 	      &201.11 \\
      \hline
      $\theta_{t2}$        &(1.15, 1.15)	&(1.88, 1.88)	 &(2.25, 2.25)      &(3.27, 3.27) \\
      IS$_{t2}$ estimator   &5.00E-02	    &1.00E-02	     &5.00E-03          &1.00E-03 \\
      sd(IS$_{t2}$)        &3.67E-03	    &8.25E-04	     &4.36E-04	        &1.00E-04 \\
      WNRV(IS$_{t2}$)      &1.40E-03	    &1.73E-03	     &1.96E-03          &2.08E-03 \\
      sd$\_$eff(Naive, IS$_{t2}$)    &2.66         &5.37           &7.42           &15.98   \\
      WNRV$\_$eff(Naive, IS$_{t2}$)  &0.10         &0.40           &0.74           &3.81   \\
      \hline
      $\theta_{t3}$         &0.42           &0.61       	  &0.66              &0.73 \\
      IS$_{t3}$ estimator    &5.00E-02	    &1.00E-02	      &5.00E-03          &1.00E-03 \\
      sd(IS$_{t3}$)         &5.81E-03	    &1.49E-03	      &8.30E-04          &1.99E-04 \\
      WNRV(IS$_{t3}$)       &3.03E-05	    &5.03E-05	      &6.11E-05          &8.57E-05 \\
      sd$\_$eff(Naive, IS$_{t3}$)   &1.68 	        &2.97             &3.86 	         &7.24 \\
      WNRV$\_$eff(Naive, IS$_{t3}$) &4.69 	        &13.94            &23.73 	         &92.42 \\
      \hline
      sd$\_$eff(IS$_{t2}$, IS$_{t1}$)    &1.65 	    &1.76 	&1.78 	&2.11   \\
      WNRV$\_$eff(IS$_{t2}$, IS$_{t1}$)  &40.68 	&47.87 	&52.74 	&52.76   \\
      \bottomrule
\end{tabular}
\end{table}

\begin{table}[]\footnotesize
\caption{Numerical results for $t$ copulas with a degree of freedom $\nu=5$ and $\rho=-0.5$ whose margins are both $t_{2}$.}
\label{IStcoupla3}\centering
\begin{tabular}{ccccc}
      \toprule
      $p$            &0.502       &1.197	    &1.573	    &2.842  \\
      \toprule
      Naive estimator &5.01E-02	  &0.99E-02	    &5.00E-03   &1.00E-03 \\
      sd(Naive)      &9.68E-03    &4.39E-03	    &3.17E-03	&1.41E-03 \\
      WNRV(Naive)    &1.44E-04    &7.36E-04	    &1.52E-03	&7.62E-03 \\
      \hline
      $\theta_{t1}$        &(3.92, 9.03)	&(8.61, 27.21)   &(12.54, 40.87)    &(35.43, 78.93) \\
      IS$_{t1}$ estimator   &5.01E-02	    &1.00E-02	     &5.00E-03          &1.00E-03 \\
      sd(IS$_{t1}$)        &3.14E-03	    &5.48E-04	     &2.65E-04          &4.94E-05 \\
      WNRV(IS$_{t1}$)      &7.14E-05	    &5.49E-05	     &4.66E-05          &4.34E-05 \\
      sd$\_$eff(Naive, IS$_{t1}$)     &3.08 	     &7.97           &12.34 	        &28.65  \\
      WNRV$\_$eff(Naive, IS$_{t1}$)   &2.01 	     &13.41          &32.72 	        &175.64  \\
      \hline
      $\theta_{t2}$        &(1.48, 1.48)	&(2.60, 2.60)	 &(3.15, 3.15)      &(4.68, 4.68) \\
      IS$_{t2}$ estimator   &5.00E-02	    &1.00E-02	     &5.00E-03          &1.00E-03 \\
      sd(IS$_{t2}$)        &5.15E-03	    &1.22E-03	     &6.48E-04	        &1.41E-04 \\
      WNRV(IS$_{t2}$)      &3.57E-03	    &4.63E-03	     &4.00E-03          &4.53E-03 \\
      sd$\_$eff(Naive, IS$_{t2}$)     &1.88        &3.58        &4.88          &10.22   \\
      WNRV$\_$eff(Naive, IS$_{t2}$)   &0.04        &0.16        &0.38          &1.68   \\
      \hline
      $\theta_{t3}$         &0.24           &0.55       	  &0.62              &0.72 \\
      IS$_{t3}$ estimator    &4.99E-02	    &9.98E-03	      &5.00E-03          &1.00E-03 \\
      sd(IS$_{t3}$)         &7.36E-03	    &1.79E-03	      &9.58E-04          &2.25E-04 \\
      WNRV(IS$_{t3}$)       &5.00E-05	    &7.20E-05	      &8.20E-05          &1.08E-04 \\
      sd$\_$eff(Naive, IS$_{t3}$)   &1.32 	        &2.45             &3.31 	         &6.27 \\
      WNRV$\_$eff(Naive, IS$_{t3}$) &2.88 	        &10.22            &18.54 	         &70.56 \\
      \hline
      sd$\_$eff(IS$_{t2}$, IS$_{t1}$)     &1.64 	&2.24 	&2.45 	&2.87   \\
      WNRV$\_$eff(IS$_{t2}$, IS$_{t1}$)   &49.97 	&84.37 	&85.91 	&104.37   \\
      \bottomrule
\end{tabular}
\end{table}

It follows from Tables~\ref{IStcoupla1}--\ref{IStcoupla3} that: (1) The conditional inverse method with the exponential twisting, the conditional inverse method with the hazard rate twisting, and the importance sampling method for $t$ copulas introduced in Subsection \ref{ISElliptical} all outperform the crude Monte Carlo method in terms of variance reduction, where IS$_{t1}$ outperforms IS$_{t2}$ while IS$_{t2}$ outperforms IS$_{t3}$, and as the estimated probability decreases, the relative efficiency of IS$_{t1}$, IS$_{t2}$ as well as IS$_{t3}$ with respect to the Naive estimator increases. (2) In terms of the WNRV metric which considers the variance reduction and computational time simultaneously, IS$_{t1}$ and IS$_{t3}$ still outperform the Naive estimator, but IS$_{t2}$ outperforms the Naive estimator only for events with very small probabilities. In addition, overall speaking, IS$_{t1}$ outperforms IS$_{t3}$.

\begin{rmk}
Note that in the simulations for elliptical copulas, the performance of importance sampling based on independent uniform random variables is better than that of importance sampling based on multivariate normal or $t$ random variables in terms of variance reduction. Therefore, the estimator based on importance sampling with the conditional inverse method is more efficient, statistically speaking. However, given the easy simulation of the multivariate normal or $t$ distributions, and seeing that the nested conditional distributions must be calculated using the conditional inverse method, importance sampling based on multivariate normal or $t$ random variables is more computationally efficient.
\end{rmk}

\noindent
{\bf Example 3: Clayton Copula}. In this example, we consider a specific Archimedean copula---the bivariate Clayton copula---in which the CDF of $(X_{1}, X_{2})^{'}$ with marginal CDFs $F_{1}$ and $F_{2}$ is
\begin{align*}
F(x_{1},x_{2}) = C(F_{1}(x_{1}),F_{2}(x_{2}); \delta) = \left( F_{1}(x_{1})^{-\delta}+F_{2}(x_{2})^{-\delta}-1 \right)^{-\frac{1}{\delta}},
\end{align*}
where $\delta\in (0,\infty)$.

Here, the event of interest is also an equal upper corner event $\{X_{1}>p, X_{2}>p\}$. Based on the conditional inverse method, the conditional distribution function of this bivariate Clayton copula is
\begin{align*}
\Lambda(V_{2}|V_{1}) = V_1^{-(\delta+1)}(V_1^{-\delta}+V_2^{-\delta}-1)^{-(\frac{1}{\delta}+1)},
\end{align*}
and its inverse function is
\begin{align*}
\Lambda^{-1}(V_{2}|V_{1}) = [(V_{1}^{\delta +1}V_{2})^{-\frac{\delta}{\delta +1}}+1-V_{1}^{-\delta}]^{-\frac{1}{\delta}}.
\end{align*}
The importance sampling estimator with the exponential twisting is denoted as IS$_{t1}$ and the corresponding optimal tilting point is denoted as $\theta_{t1}$, while the importance sampling estimator with the hazard rate twisting is denoted as IS$_{t3}$ and the corresponding optimal tilting point is denoted as $\theta_{t3}$.

Another method is the Marshall--Olkin algorithm, in which $(X_{1},X_{2})^{'}$ can be written as
\begin{align*}
(X_{1},X_{2})' = \left(F_{1}^{-1}[\zeta_{\delta}^{-1}(-\frac{1}{W}\ln V_{1})],F_{2}^{-1}[\zeta_{\delta}^{-1}(-\frac{1}{W}\ln V_{2})]\right)^{'},
\end{align*}
where $V_{1}$ and $V_{2}$ are independent $U(0,1)$ random variables, $W$ is a $\gammadist(1/\delta, 1)$ random variable, and
$$\zeta^{-1}_{\delta}(t):=\zeta^{-1}(t)=(t+1)^{-\frac{1}{\delta}}, \quad \zeta_{\delta}(t):=\zeta(t)=t^{-\delta}-1.$$
Note that $V_{1}, V_{2}$ and $W$ are mutually independent, and the cumulant-generating function of $\tilde{\bm V}=(W, V_1, V_2)^{'}$ is
$\psi(\theta_{o})=-\frac{1}{\delta}\ln(1-\theta_{oW})+\sum_{i=1}^2 \ln\left(\frac{e^{\theta_{oi}}-1}{\theta_{oi}}\right)$. Then the upper corner event becomes
\begin{align*}
\{X_{1}>p, X_{2}>p\} & = \left\{F_{1}^{-1}[\zeta_{\delta}^{-1}(-\frac{1}{W}\ln V_{1})]>p, F_{2}^{-1}[\zeta_{\delta}^{-1}(-\frac{1}{W}\ln V_{2})]>p \right\} \nonumber \\
&= \left\{V_{1}>e^{-W([F_{1}(p)]^{-\delta}-1)}, V_{2}>e^{-W([F_{2}(p)]^{-\delta}-1)} \right\}.
\end{align*}
According to Proposition~\ref{Pro4}, $\theta_{o}=(\theta_{oW}, \theta_{o1}, \theta_{o2})^{'}$ satisfies
\begin{eqnarray*}
\left\{
\begin{array}{l}
\displaystyle \frac{ E_{P}\left[1\{g(\tilde{\bm V})\in A\}e^{-\theta_o^{'} \tilde{\bm V}}W\right] }{ E_{P}\left[1\{g( \tilde{\bm V})\in  A\}e^{-\theta_o^{'} \tilde{\bm V}}\right] } = \frac{\partial \psi(\theta_o)}{\partial \theta_{oW}}, \\
\displaystyle \frac{ E_{P}\left[1\{g(\tilde{\bm V})\in A\}e^{-\theta_o^{'} \tilde{\bm V}}V_{i}\right] }{ E_{P}\left[1\{g( \tilde{\bm V})\in  A\}e^{-\theta_o^{'} \tilde{\bm V}}\right] } = \frac{\partial \psi(\theta_o)}{\partial \theta_{oi}}, \quad \hbox{for}~ i=1, 2.\\
\end{array}
\right.
\end{eqnarray*}
In this example, the importance sampling algorithm employs an exponential tilting on $W$ and $(V_{1}, V_{2})^{'}$ simultaneously, in which the optimal tilting parameters are denoted as $\theta_{W}$ and $\theta_{t2}$, respectively.

Here we assume that the marginal CDFs both follow the standard normal distribution. The simulation results with $\delta=3$ are reported in Table~\ref{IScla}. Note from Table~\ref{IScla} that: (1) The conditional inverse method with the exponential twisting, the conditional inverse method with the hazard rate twisting, and the importance sampling method for Clayton copulas introduced in Subsection \ref{ISArchimedean} all outperform the crude Monte Carlo method in terms of variance reduction, where IS$_{t1}$ outperforms IS$_{t2}$ while IS$_{t2}$ outperforms IS$_{t3}$, and the relative efficiency of IS$_{t1}$, IS$_{t2}$ and IS$_{t3}$ with respect to the Naive estimator all increases as the estimated probability decreases. (2) In terms of the WNRV metric which considers the variance reduction and computational time simultaneously, overall speaking, IS$_{t1}$, IS$_{t2}$ and IS$_{t3}$ still outperform the Naive estimator, where the performance of IS$_{t1}$ is better than that of IS$_{t3}$, while the performance of IS$_{t3}$ is better than that of IS$_{t2}$. Note that, through the transformation in the Marshall--Olkin method, we use $(d+1)$ random variables to deal with the $d$-variate case, which constitutes an inefficiency in terms of statistics even though it is easy to simulate a high-dimensional Archimedean copula. Hence, its improvements are not better than those yielded by the conditional inverse method.

\begin{table}[]\footnotesize
\caption{Numerical results for Clayton copulas with $\delta=3$, the margins of which are all $N(0,1)$.}
\label{IScla}\centering
\begin{tabular}{ccccc}
      \toprule
      $p$            &1.115	    &1.600	    &1.780	    &2.130  \\
      \toprule
      Naive estimator &5.03E-02	&1.03E-02	&5.10E-03	&1.00E-03 \\
      sd(Naive)      &9.81E-03	&4.52E-03	&3.18E-03	&1.44E-03 \\
      WNRV(Naive)    &3.66E-05	&1.62E-04	&3.30E-04	&1.81E-03 \\
      \hline
      $\theta_{t1}$         &(12.74, 4.03)	&(29.93, 8.57)	 &(43.33, 11.94)	&(97.00, 25.36)\\
      IS$_{t1}$ estimator    &5.04E-02       &1.03E-02	     &5.10E-03          &1.00E-03 \\
      sd(IS$_{t1}$)         &2.45E-03	    &5.20E-04	     &2.65E-04          &5.46E-05 \\
      WNRV(IS$_{t1}$)       &1.20E-05	    &1.27E-05	     &1.27E-05          &1.62E-05 \\
      sd$\_$eff(Naive, IS$_{t1}$)    &4.00          &8.68 	    &12.27           &26.33   \\
      WNRV$\_$eff(Naive, IS$_{t1}$)  &3.07          &12.76 	    &26.11           &112.08   \\
      \hline
      $\theta_{t2}$         &(2.53, 2.53)	&(5.31, 5.31)	 &(7.22, 7.22)	    &(14.58, 14.58)\\
      $\theta_{W}$          &0.775          &0.828	         &0.837             &0.848 \\
      IS$_{t2}$ estimator    &5.04E-02	    &1.03E-02	     &5.10E-03          &1.00E-03 \\
      sd(IS$_{t2}$)         &4.78E-03	    &1.14E-03	     &5.57E-04	        &1.41E-04 \\
      WNRV(IS$_{t2}$)       &9.52E-05	    &1.24E-04	     &1.24E-04          &1.68E-04 \\
      sd$\_$eff(Naive, IS$_{t2}$)     &2.05       &3.96	     &5.71 	       &11.67  \\
      WNRV$\_$eff(Naive, IS$_{t2}$)   &0.38       &1.31	     &2.67 	       &10.80  \\
      \hline
      $\theta_{t3}$         &0.36           &0.57       	  &0.63              &0.71 \\
      IS$_{t3}$ estimator    &5.05E-02	    &1.03E-02	      &5.05E-03          &1.05E-03 \\
      sd(IS$_{t3}$)         &6.26E-03	    &1.75E-03	      &9.46E-04          &2.47E-04 \\
      WNRV(IS$_{t3}$)       &1.46E-05	    &2.75E-05	      &3.25E-05          &5.17E-04 \\
      sd$\_$eff(Naive, IS$_{t3}$)   &1.57 	        &2.58             &3.36 	         &5.83 \\
      WNRV$\_$eff(Naive, IS$_{t3}$) &2.51 	        &5.89             &10.15 	         &3.50 \\
      \hline
      sd$\_$eff(IS$_{t2}$, IS$_{t1}$)     &1.95 	    &2.19	     &2.10 	      &2.59  \\
      WNRV$\_$eff(IS$_{t2}$, IS$_{t1}$)   &7.97 	    &9.75	     &9.77 	      &10.38  \\
      \bottomrule
\end{tabular}
\end{table}

\subsection{Importance Sampling for More General Copulas---R-Vine Copulas}\label{RVcopula}

As mentioned earlier, for higher-dimensional (greater than 2 dimension) copula models, importance sampling based on the conditional inverse method may become infeasible because the calculation of the conditional inverse function could be very challenging. However, the R-vine copula models can avoid this issue. Note that Theorem \ref{thm_theta}, Proposition \ref{Pro1}, as well as the importance sampling algorithm via conditional inverse method introduced in Subsection \ref{ISgeneral} are general enough such that they are all applicable to the R-vine copula models.

The R-vine copula is first proposed by Bedford and Cooke \cite{B2002}, and the problems related to this topic such as matrix expression, tree construction, parametric estimation and simulation are intensively studied by Aas et al. \cite{A2009}, Di{\ss}mann et al. \cite{D2013}, etc. Some books have systematically introduced and studied the R-vine copula models, and the reader is referred to Kurowicka and Joe \cite{KJ2010}, Joe \cite{J2014}, and Czado \cite{C2019} for more details.

R-vine copulas take bivariate copulas as building blocks and use pair copula decompositions to construct multivariate copulas. Consider a random vector $\bm{X} = (X_1,\ldots,X_d)^{'}$ with joint density $f(x_1, \ldots, x_d)$. We can decompose the joint density into a series of conditional densities, such that the joint density can be written as
\begin{align}\label{cdd}
f(x_1, \ldots, x_d) = f_1(x_1) \prod_{i = 2}^{d} f_{i \vert 1, \ldots, i-1}(x_i \vert x_1, \ldots, x_{i-1}),
\end{align}
and this decomposition is unique up to a re-labeling of the variables. For every conditional density $f_{i \vert 1,\ldots,i-1}(x_i \vert x_1,\ldots, x_{i-1}), i = 2,\ldots,d$, denote $F_{i-1,i \vert 1,\ldots,i-2}(x_{i-1},x_i \vert x_1,\ldots, x_{i-2})$ as the conditional CDF of $(X_{i-1}, X_i)$ given $(X_1,\ldots,X_{i-2})$ and
$$
C_{i-1,i; 1,\ldots,i-2}(F_{i-1 \vert 1,\ldots, i-2}(x_{i-1}\vert x_1,\ldots,x_{i-2}), F_{i\vert 1,\ldots, i-2}(x_i\vert x_1,\ldots,x_{i-2}))
$$
as its copula expression. Denote $c_{i-1,i; 1,\ldots,i-2}(\cdot,\cdot)$ as the density of $C_{i-1,i; 1,\ldots,i-2}(\cdot,\cdot)$. In what follows, we omit the arguments $x_1,\ldots,x_d$ for brevity. It shows that for every $i = 2,\ldots,d$,
\begin{align}
      f_{i \vert 1,\ldots,i-1} & = \frac{f_{i-1,i \vert 1,\ldots,i-2}}{f_{i-1 \vert 1,\ldots,i-2}}  \nonumber \\
       & = \frac{c_{i-1,i; 1,\ldots,i-2}(F_{i-1 \vert 1,\ldots, i-2}, F_{i\vert 1,\ldots, i-2}) \cdot f_{i-1 \vert 1,\ldots, i-2} \cdot f_{i\vert 1,\ldots, i-2}}{f_{i-1 \vert 1,\ldots,i-2}}  \nonumber \\
       & = c_{i-1,i; 1,\ldots,i-2}(F_{i-1 \vert 1,\ldots, i-2}, F_{i\vert 1,\ldots, i-2}) \cdot f_{i\vert 1,\ldots, i-2} \nonumber \\
       & \quad \vdots \nonumber \\
       & = f_i\prod_{j = 1}^{i-1} c_{i-j,i;1,\ldots,i-j-1}(F_{i-1 \vert 1,\ldots,i-j-1}, F_{i \vert 1,\ldots,i-j-1}).
      \label{cd}
\end{align}
Combining (\ref{cdd}) and (\ref{cd}) leads to a pair copula decomposition
\begin{align*}\label{pcd}
      f(x_1,\ldots,x_d) = \left( \prod_{i = 2}^{d}\prod_{j = 1}^{i-1} c_{i-j,i;1,\ldots,i-j-1}(F_{i-1 \vert 1,\ldots,i-j-1}, F_{i \vert 1,\ldots,i-j-1}) \right) \cdot \left( \prod_{k = 1}^{d}f_k \right).
\end{align*}
The complexity and flexibility of R-vine copulas lie in the diverse decompositions of conditional densities (\ref{cd}). One type of pair copula decompositions similar to ours is a special case of R-vine copulas called canonical vine copulas.

\begin{figure}[]
      \small
      \centering
        \includegraphics[scale=0.45]{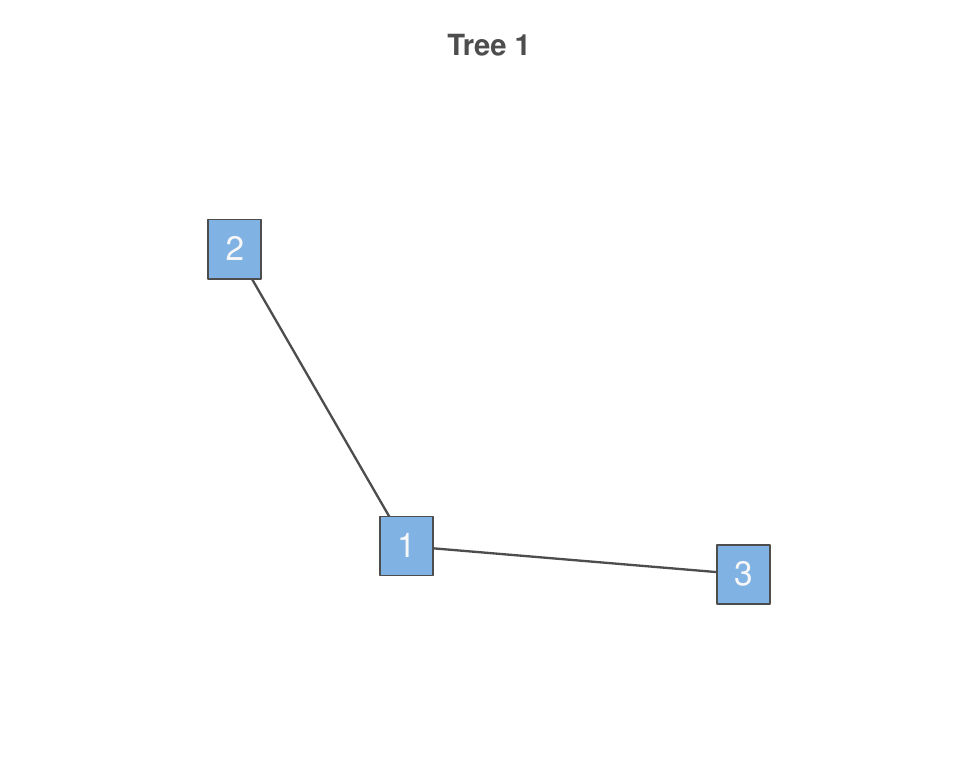}
        \includegraphics[scale=0.45]{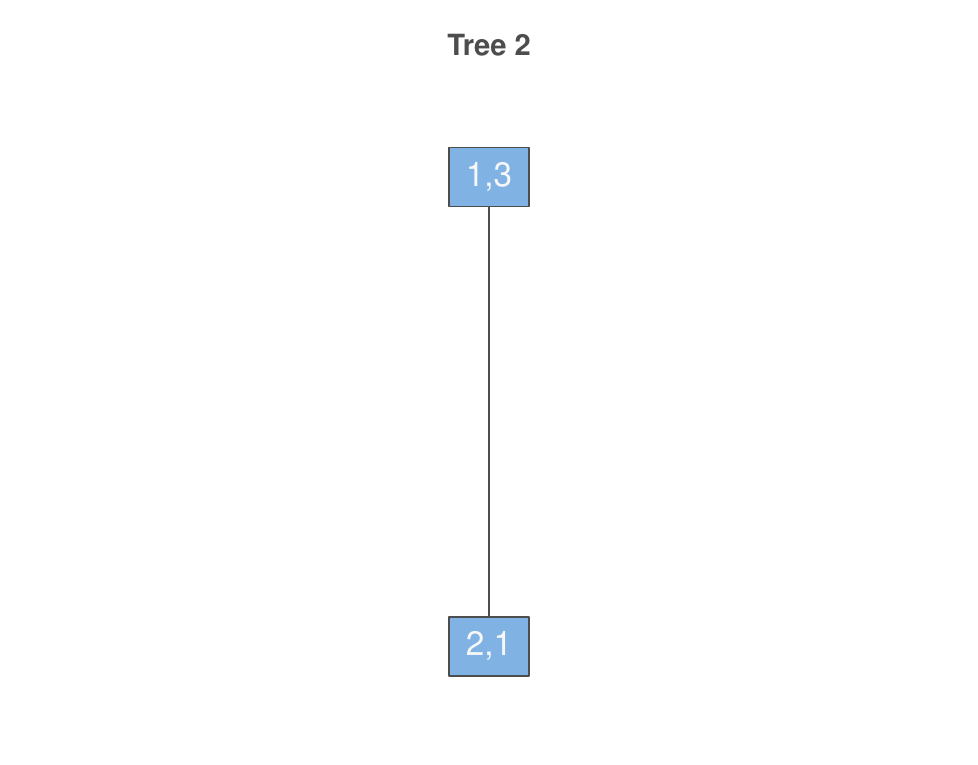}
        \caption{R-vine trees corresponding to the three-dimensional R-vine copula model.}
        \label{3drvt}
\end{figure}

Next, we illustrate how to execute important sampling for R-vine copulas by two elementary examples. Since the copula model is irrelevant to marginal distributions, we assume the marginal distributions of $\bm{X} = (X_1, \ldots, X_d)^{'}$ are all $U(0,1)$. The first example is a three-dimensional R-vine copula whose tree expression is displayed in Figure~\ref{3drvt}. Using the $h$ function defined in Czado \cite{C2019}, the relationship between $\bm{X}$ and $\bm{V}$ can be expressed as
\begin{align*}
      & V_1 = X_1, \\
      & V_2 = h_{2 \vert 1}(X_2 \vert X_1), \\
      & V_3 = h_{3 \vert 2 ;1}\left(h_{3\vert 1}(X_3 \vert X_1)\vert h_{2\vert 1}(X_2\vert X_1)\right).
\end{align*}
Then, the conditional inverse transformation is
\begin{equation*}
      \begin{aligned}
            & X_1 = V_1, \\
            & X_2 = h_{2 \vert 1}^{-1}(V_2 \vert X_1), \\
            & X_3 = h_{3\vert 1}^{-1}\left(h_{3\vert 2;1}^{-1}(V_3 \vert h_{2\vert 1}(X_2\vert X_1)) \vert X_1\right),
      \end{aligned}
      \label{3dcit}
\end{equation*}
which completes the Step 1 of the importance sampling algorithm via conditional inverse method. Continuing with the rest of the steps in Subsection \ref{ISgeneral} will finish the importance sampling.

\begin{figure}[]
      \small
      \centering
        \includegraphics[scale=0.25]{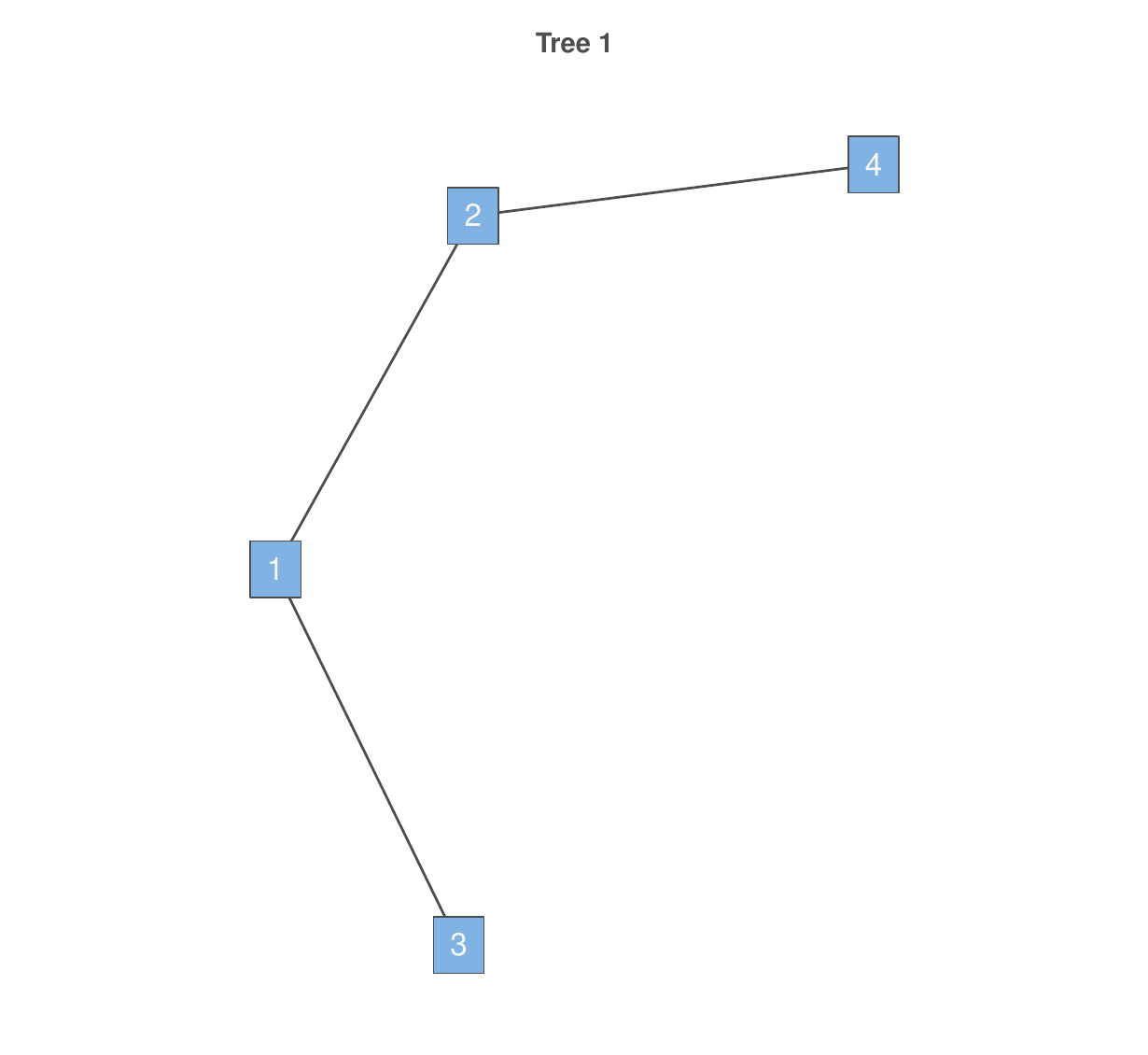}
        \includegraphics[scale=0.25]{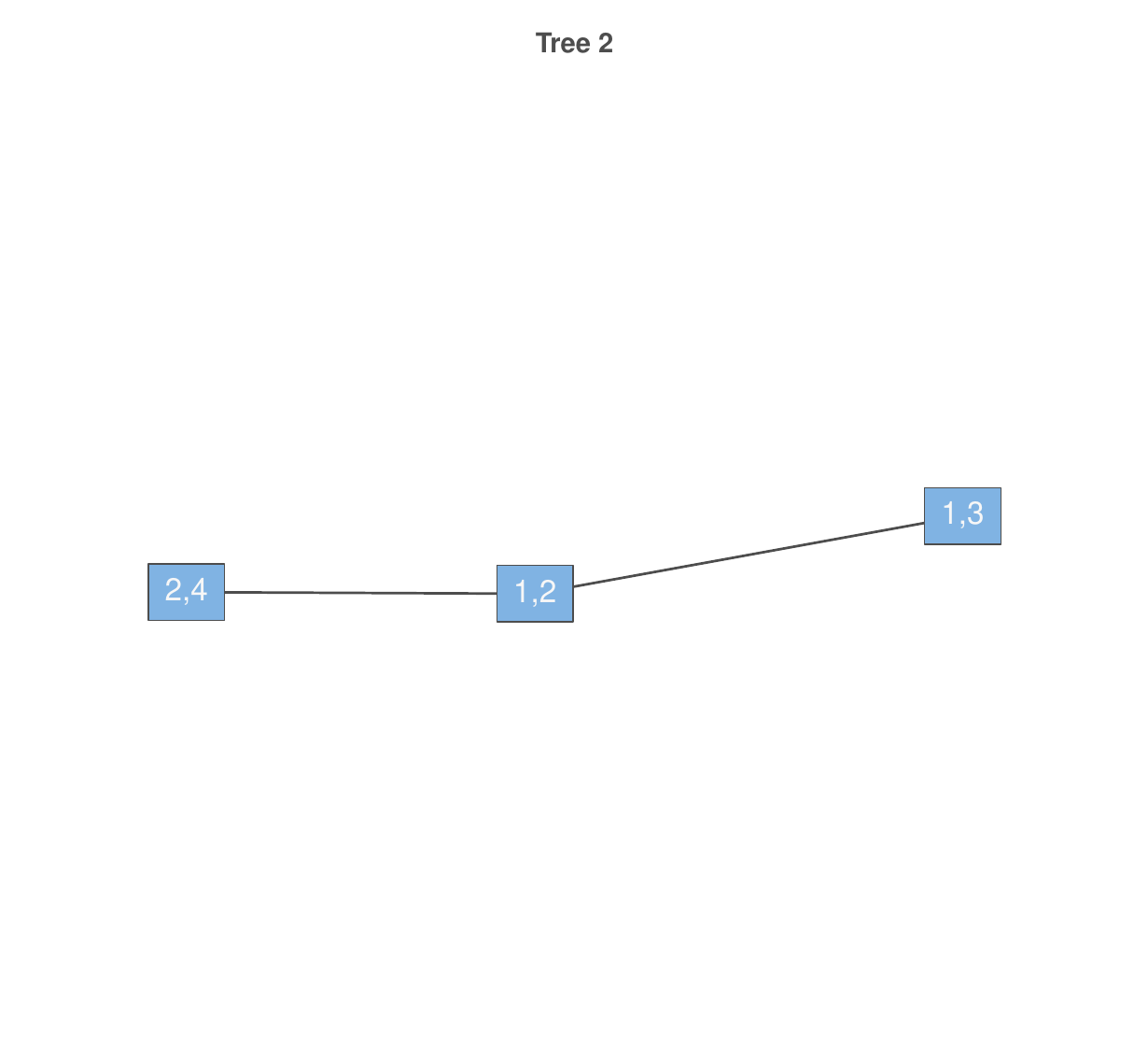}
        \includegraphics[scale=0.25]{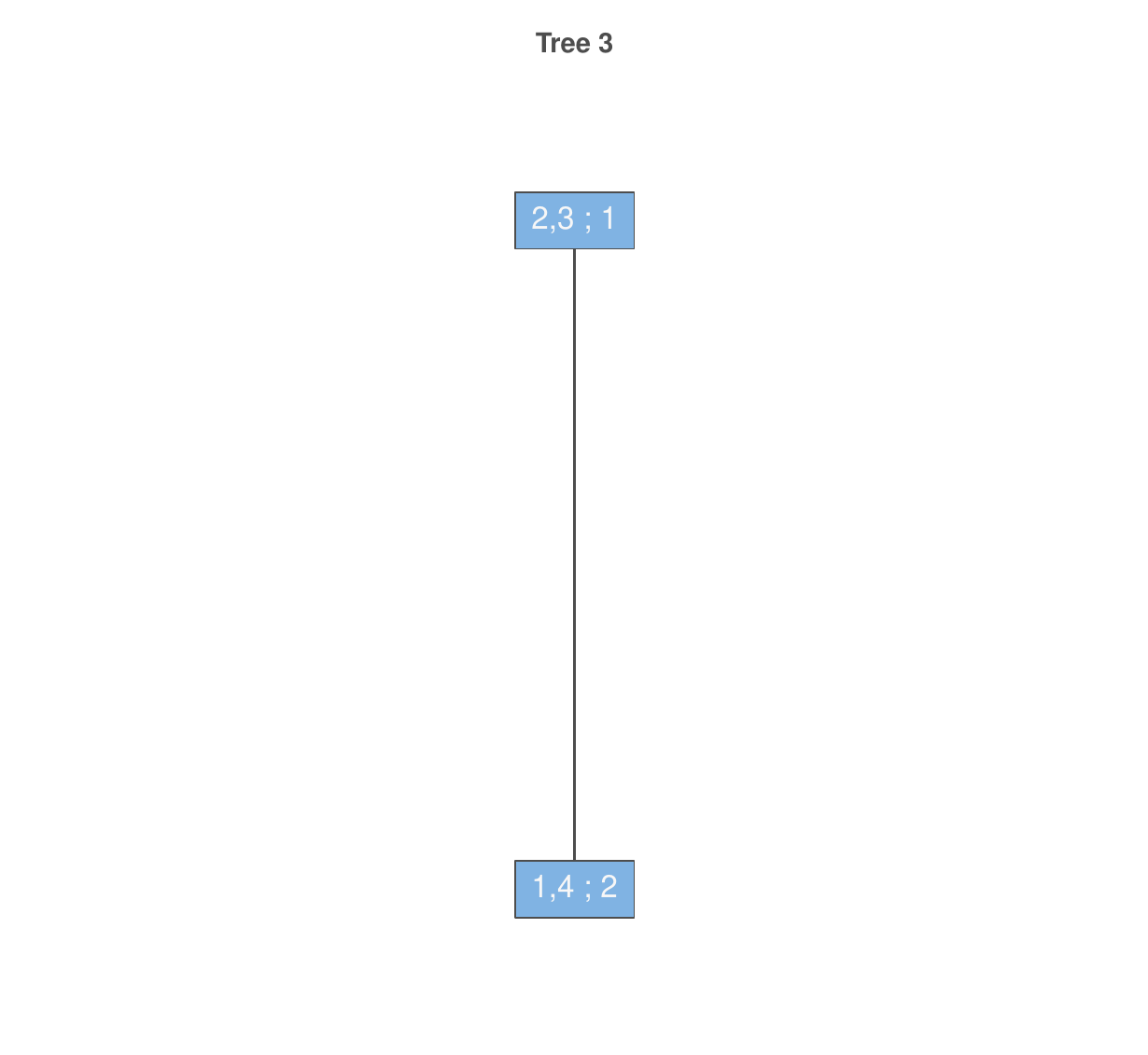}
        \caption{R-vine trees corresponding to the four-dimensional R-vine copula model.}
        \label{4drvt}
\end{figure}

The second example is a four-dimensional R-vine copula whose tree expression is displayed in Figure~\ref{4drvt}. Similarly, the relationship between $\bm{X}$ and $\bm{V}$ can be expressed as
\begin{align*}
      & V_1 = X_1, \\
      & V_2 = h_{2 \vert 1}(X_2 \vert X_1), \\
      & V_3 = h_{3 \vert 2 ;1}\left(h_{3\vert 1}(X_3 \vert X_1)\vert h_{2\vert 1}(X_2\vert X_1)\right), \\
      & V_4 = h_{4\vert 3;1,2}\left(h_{4\vert 1;2}\left(h_{4\vert 2}(X_4\vert X_2) \vert h_{1\vert 2}(X_1 \vert X_2)\right) \vert h_{3 \vert 2 ;1}\left(h_{3\vert 1}(X_3 \vert X_1)\vert h_{2\vert 1}(X_2\vert X_1)\right)\right).
\end{align*}
Then, the conditional inverse transformation is
\begin{equation*}
      \begin{aligned}
            & X_1 = V_1, \\
            & X_2 = h_{2 \vert 1}^{-1}(V_2 \vert X_1), \\
            & X_3 = h_{3\vert 1}^{-1}\left(h_{3\vert 2;1}^{-1}(V_3 \vert h_{2\vert 1}(X_2\vert X_1)) \vert X_1\right), \\
            & X_4 = h_{4\vert 2}^{-1}\left(h_{4\vert 1;2}^{-1}\left(h_{4\vert 3;1,2}^{-1}\left(V_4 \vert h_{3 \vert 2 ;1}\left(h_{3\vert 1}(X_3 \vert X_1)\vert h_{2\vert 1}(X_2\vert X_1)\right)\right) \vert h_{1\vert 2}(X_1\vert X_2)\right)\vert X_2\right),
      \end{aligned}
      \label{4dcit}
\end{equation*}
which completes the Step 1 of the importance sampling algorithm via conditional inverse method. Continuing with the rest of the steps in Subsection \ref{ISgeneral} will finish the importance sampling.

\begin{figure}[]
      \small
      \centering
        \includegraphics[scale=0.50]{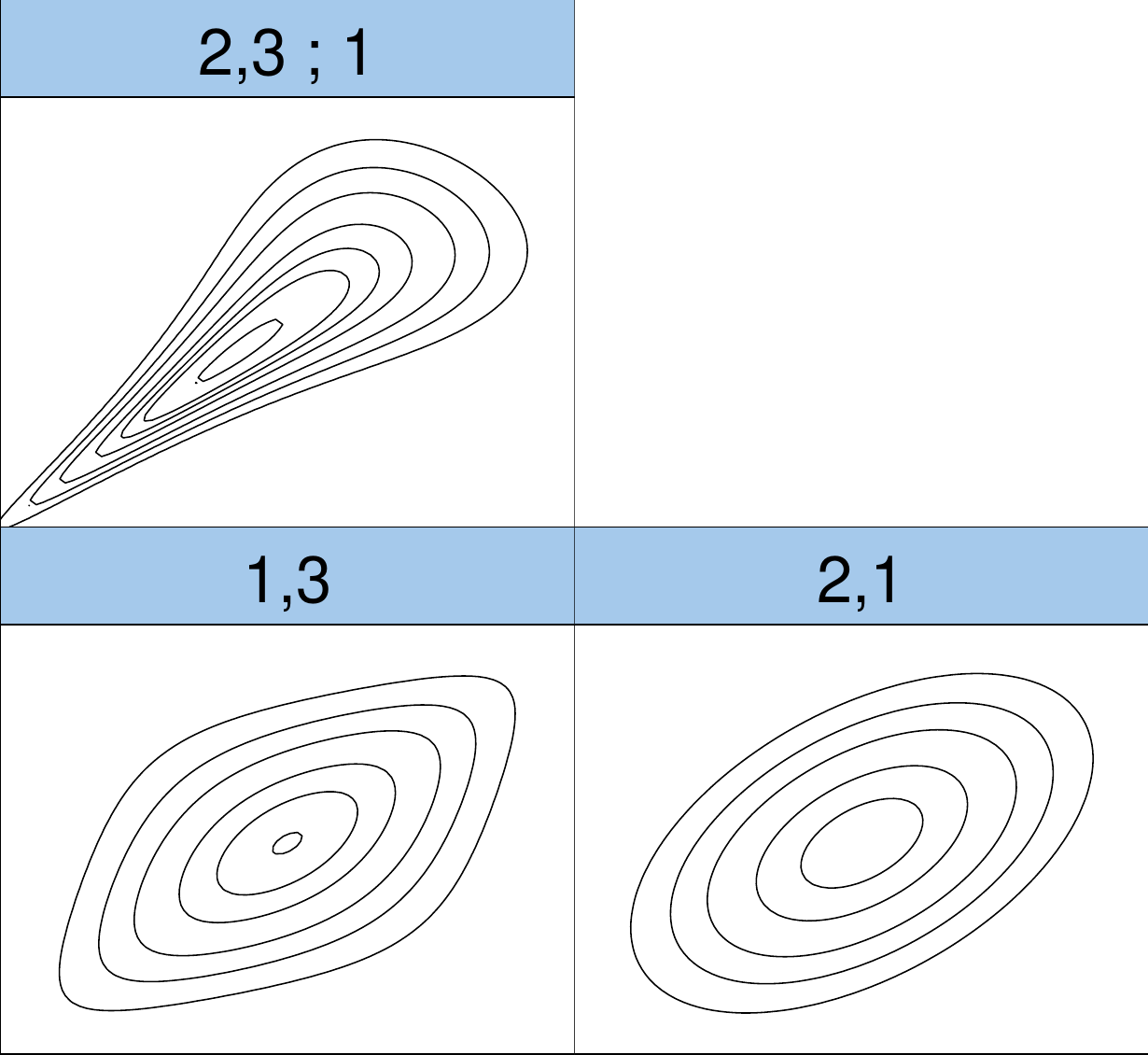}
        \caption{Contour plots corresponding to the three-dimensional R-vine copula model.}
        \label{3dcont}
\end{figure}

\begin{figure}[]
      \small
      \centering
        \includegraphics[scale=0.55]{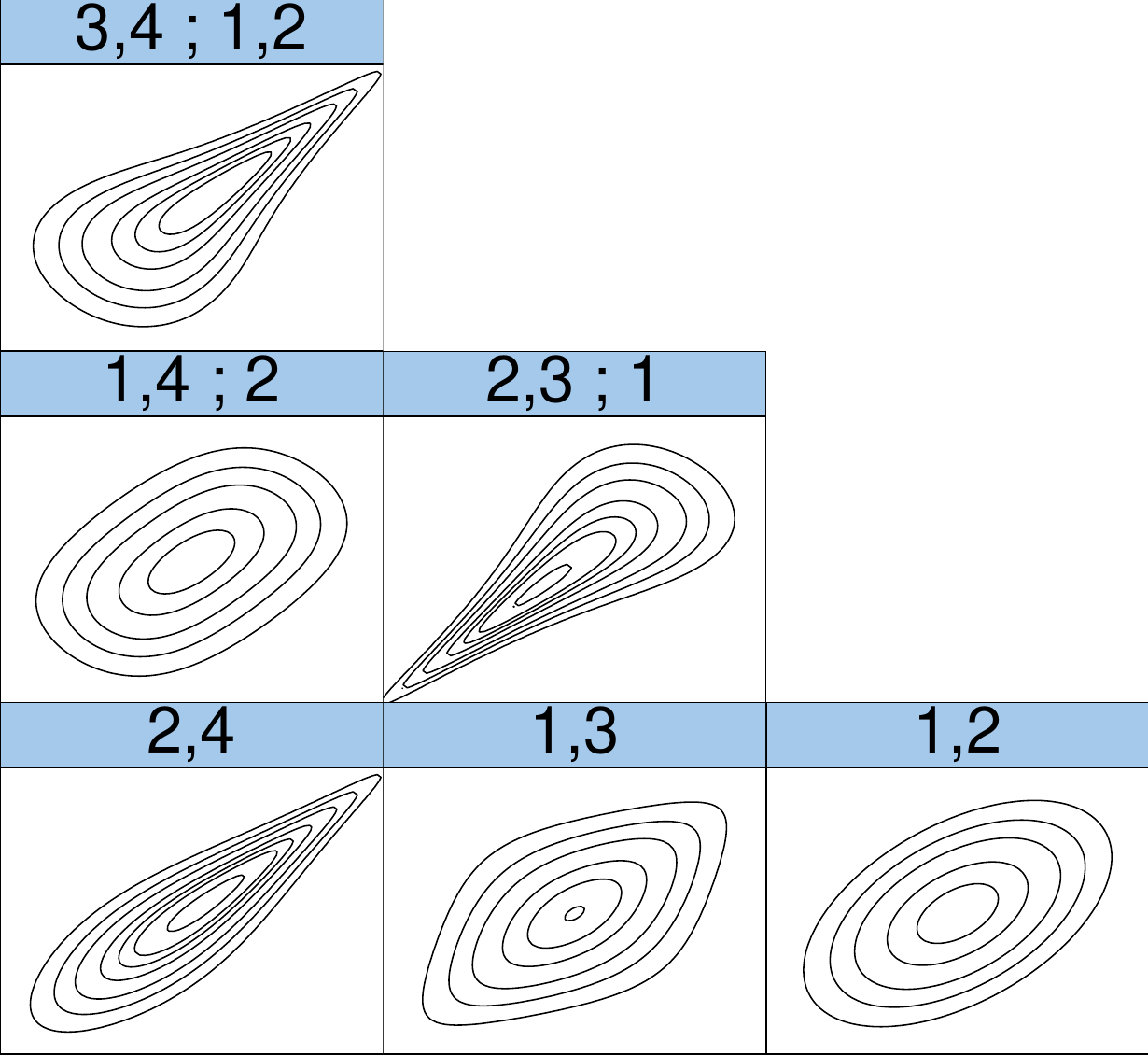}
        \caption{Contour plots corresponding to the four-dimensional R-vine copula model.}
        \label{4dcont}
\end{figure}

Tables \ref{3dset} and \ref{4dset} show our settings and Figures \ref{3dcont} and \ref{4dcont} display contour plots of corresponding bivariate copulas. The sample size $n$ and the number of replications $M$ are still set to $500$ and $5,000$, respectively. The simulation results are reported in Tables \ref{3dsim} and \ref{4dsim}. Note from Tables \ref{3dsim} and \ref{4dsim} that: (1) The conditional inverse method with the exponential twisting and the conditional inverse method with the hazard rate twisting both outperform the crude Monte Carlo method in terms of variance reduction, where IS$_{t1}$ outperforms IS$_{t3}$, and the relative efficiency of IS$_{t1}$ and IS$_{t3}$ with respect to the Naive estimator increases as the estimated probability decreases. (2) In terms of the WNRV metric which considers the variance reduction and computational time simultaneously, overall speaking, IS$_{t1}$ still outperforms the Naive estimate, and this advantage becomes more apparent as the estimated probability decreases, but IS$_{t3}$ fails to beat the Naive estimator due to extensive calculation as well as not so good variance reduction effect.

\begin{table}[]
      \caption{Model settings for the three-dimensional R-vine copula model.}
      \label{3dset}\centering
      \begin{tabular}{ccccc}
            \toprule
            bivariate copula & family & parameters \\
            \toprule
            $C_{1,2}$ & Gaussian & $\rho = 0.5$ \\
            \hline
            $C_{1,3}$ & $t$ & $df = 5$, $\rho = 0.5$ \\
            \hline
            $C_{2,3;1}$ & Clayton & $\delta = 3$ \\
            \bottomrule
      \end{tabular}
\end{table}

\begin{table}[]
      \caption{Model settings for the four-dimensional R-vine copula model.}
      \label{4dset}\centering
      \begin{tabular}{ccccc}
            \toprule
            bivariate copula & family & parameters \\
            \toprule
            $C_{1,2}$ & Gaussian & $\rho = 0.5$ \\
            \hline
            $C_{1,3}$ & $t$ & $df = 5$, $\rho = 0.5$ \\
            \hline
            $C_{2,4}$ & Gumbel & $\delta = 3$ \\
            \hline
            $C_{2,3;1}$ & Clayton & $\delta = 3$ \\
            \hline
            $C_{1,4;2}$ & Frank & $\delta = 3$ \\
            \hline
            $C_{3,4;1,2}$ & Joe & $\delta = 3$ \\
            \bottomrule
      \end{tabular}
\end{table}

\begin{table}[]\footnotesize
      \caption{Numerical results for three-dimensional R-vine copula whose margins are all $U(0,1)$.}
      \label{3dsim}\centering
      \begin{tabular}{ccccc}
            \toprule
            $p$ &	0.9    &	0.95	& 0.975  \\
            \toprule
            Naive estimator  &2.40E-02	&8.73E-03	&3.16E-03 \\
            sd(Naive)       &6.76E-03	&4.16E-03	&2.51E-03 \\
            WNRV(Naive)     &2.19E-04	&6.21E-04	&1.75E-03 \\
            \hline
            $\theta_{t1}$       	& (20.19, 3.77, 1.09) & (41.35, 4.88, 0.51) & (91.92, 7.75, 1.02)   \\
            IS$_{t1}$ estimator       & 2.42E-02          & 8.65E-03          & 3.17E-03  \\
            sd(IS$_{t1}$)            & 1.29E-03          & 4.95E-04	        & 1.77E-04  \\
            WNRV(IS$_{t1}$)          & 5.10E-04          & 4.62E-04	        & 5.34E-04  \\
            sd$\_$eff(Naive, IS$_{t1}$)         & 5.26          & 8.39 	       & 14.19      \\
            WNRV$\_$eff(Naive, IS$_{t1}$)        & 0.43          & 1.35 	       & 3.27      \\
            \hline
            $\theta_{t3}$               &0.35       	  &0.47              &0.55 \\
            IS$_{t3}$ estimator        &2.41E-02	      &8.63E-03          &3.15E-03 \\
            sd(IS$_{t3}$)             &3.89E-03	      &1.71E-03          &7.45E-04 \\
            WNRV(IS$_{t3}$)           &2.18E-03	      &3.35E-03          &4.66E-03 \\
            sd$\_$eff(Naive, IS$_{t3}$)           &1.74             &2.42 	         &3.36 \\
            WNRV$\_$eff(Naive, IS$_{t3}$)         &0.10             &0.19 	         &0.37 \\
            \hline
            sd$\_$eff(IS$_{t3}$, IS$_{t1}$)     & 3.03    & 3.46    & 4.22  \\
            WNRV$\_$eff(IS$_{t3}$, IS$_{t1}$)   & 4.26    & 7.26    & 8.72  \\
            \bottomrule
      \end{tabular}
\end{table}

\begin{table}[H]\footnotesize
      \caption{Numerical results for four-dimensional R-vine copula whose margins are all $U(0,1)$.}
      \label{4dsim}\centering
      \begin{tabular}{cccc}
            \toprule
            $p$  &	0.9    &	0.95  &     0.975 \\
            \toprule
            Naive estimator 	 &2.33E-02	&8.36E-03 & 3.04E-03\\
            sd(Naive)       &6.74E-03	&4.00E-03 & 2.49E-03\\
            WNRV(Naive)     &7.55E-04	&2.04E-03 & 6.19E-03\\
            \hline
            $\theta_{t1}$       	& (20.53, 4.24, 1.05, 0.13) & (46.03, 6.91, 1.41, 0.11) &  (91.44, 7.55, 1.46, 0.15) \\
            IS$_{t1}$ estimator      & 2.34E-02          & 8.40E-03  &  3.07E-03\\
            sd(IS$_{t1}$)             & 1.22E-03          & 4.63E-04  & 1.73E-04\\
            WNRV(IS$_{t1}$)           & 6.32E-04          & 7.95E-04  & 4.10E-03\\
            sd$\_$eff(Naive, IS$_{t1}$)          & 5.52          & 8.64  &   14.45 \\
            WNRV$\_$eff(Naive, IS$_{t1}$)        & 1.19          & 2.57  &   1.51 \\
            \hline
            $\theta_{t3}$               &0.22       	  &0.34 &  0.43 \\
            IS$_{t3}$ estimator        &2.34E-02	      &8.36E-03 & 3.08E-03\\
            sd(IS$_{t3}$)         	    &4.57E-03	      &2.07E-03 & 9.37E-04\\
            WNRV(IS$_{t3}$)       	    &7.80E-03	      &1.20E-02 & 1.82E-02\\
            sd$\_$eff(Naive, IS$_{t3}$)   	        &1.47             &1.94 &  2.66 \\
            WNRV$\_$eff(Naive, IS$_{t3}$)         &0.10             &0.17 &  0.34 \\
            \hline
            sd$\_$eff(IS$_{t3}$, IS$_{t1}$)      & 3.74    & 4.46  &  5.43  \\
            WNRV$\_$eff(IS$_{t3}$, IS$_{t1}$)   & 12.34    & 15.15  &  4.45  \\
            \bottomrule
      \end{tabular}
\end{table}

\section{Conclusions}\label{Conclusion}

In this paper, we propose an importance sampling algorithm for rare event simulations in copula models. The key features of our method are threefold. First, we apply the transform likelihood ratio method as an alternative tilting probability family. Second, we characterize the optimal tilting parameter via a conjugate probability measure. Third, the theoretical results of logarithmic efficiency and bounded relative error are proved for some commonly-used copula models under the case of simple rare events. Although the original event becomes more complicated after transformation, the transformed random variables are more tractable. Simulation studies confirm the theoretical results that our method always gains a substantial variance reduction in comparison to the crude Monte Carlo method. Simulation studies also indicate that, for Gaussian, $t$ and Clayton copulas, the importance sampling using the conditional inverse method with the exponential twisting always outperforms the crude Monte Carlo method in terms of the WNRV metric which considers the variance reduction and computational time simultaneously. Although the conditional inverse method may be difficult to be applied to general multivariate copulas, it is still valid for plenty of multivariate copulas such as R-vine copulas. It seems that variance reduction and computational simplicity are hard to achieve simultaneously. A more challenging project is to figure out other importance sampling algorithms that have strength in both computational speed and accuracy, and this will be the focus of our future study.

\section*{Appendix 1: Proof of Theorem \ref{thm_theta}}\label{appendix1}

Consider an exponential tilting technique such that
\begin{align*}
\frac{\d Q}{\d P}( {\bm V})=\frac{e^{\theta^{'} {\bm V}}}{E_{P}[e^{\theta^{'} {\bm V}}]}=e^{\theta^{'} {\bm V} -\psi(\theta)},
\end{align*}
where $\theta= (\theta_{1},\ldots,\theta_{d})^{'}$ and $\psi(\theta)=\ln E_{P}[e^{\theta^{'} {\bm V}}]$ is the cumulant-generating function of ${\bm V}$. Recall that
\begin{align*}
G(\theta) := E_{P}\left[1\{g({\bm V})\in A\}e^{-\theta^{'} {\bm V} +\psi(\theta)}\right]
\end{align*}
from equation~(\ref{2ndmoment}). For a standard minimization problem, let the first-order derivative of $G(\theta)$ be equal to $0$ and obtain a system of equations as follows:
\begin{align*}
\frac{\partial}{\partial \theta_{i}}E_{P}\left[1\{g({\bm V})\in A\}e^{-\theta^{'} {\bm V} +\psi(\theta)}\right]=0, \quad  \hbox{for}~ i=1, \ldots, d.
\end{align*}
Under condition (\ref{eq12}), through an effortless calculation, we have
\begin{align*}
\frac{ E_{P}\left[1\{g({\bm V})\in A\}e^{-\theta^{'} {\bm V}}V_{i}\right] }{ E_{P}\left[1\{g({\bm V})\in  A\}e^{-\theta^{'} {\bm V}}\right] } = \frac{\partial \psi(\theta)}{\partial \theta_{i}}, \quad  \hbox{for}~ i=1, \ldots, d.
\end{align*}
Then, if $-\theta\in \Theta$, the left side of the above equation can be rewritten using the $\bar{Q}$-measure as
\begin{align*}
& \frac{ E_{P}\left[1\{g({\bm V})\in A\}e^{-\theta^{'} {\bm V}}V_{i}\right] }{ E_{P}\left[1\{g({\bm V})\in A\}e^{-\theta^{'} {\bm V}}\right] } = \frac{ \int 1\{g({v})\in A\}v_{i}e^{-\theta^{'} {v}} \d P }{ \int 1\{g({v})\in A\} e^{-\theta^{'} {v}} \d P } \notag \\
=& \frac{ \int 1\{g( {v})\in A\} v_{i} \d\bar{Q} \times e^{\bar{\psi}(\theta)} }{ \int 1\{g({v})\in A\}  \d\bar{Q} \times e^{\bar{\psi}(\theta)}}
= \frac{ E_{\bar{Q}}[V_{i}1\{g({\bm V})\in A\}] }{ E_{\bar{Q}}[1\{g({\bm V})\in A\}] }  \notag \\
=& E_{\bar{Q}}[V_{i}|g({\bm V})\in A]=: E_{\bar{Q}_{\theta}}[V_{i}|g({\bm V})\in A].
\end{align*}
Therefore, if the solution exists, the optimal $\theta$ satisfies
$$ E_{\bar{Q}_{\theta}}[V_{i}|g({\bm V})\in A] = \frac{\partial \psi(\theta)}{\partial \theta_{i}}\quad  \hbox{for}~ i=1, \ldots, d$$
if $-\theta\in \Theta$.

To show the uniqueness of the solution provided its existence, we consider the second-order derivative of $G(\theta)$. Under assumption~(\ref{eq12}), we have
\begin{align*}
\frac{\partial^{2} G(\theta)}{\partial \theta_{i}\partial \theta_{j}} =  & E_{P}\left[ 1\{g({\bm V})\in  A\} \left(-V_{i}+\frac{\partial \psi(\theta)}{\partial \theta_{i}}\right)
\left(-V_{j}+\frac{\partial \psi(\theta)}{\partial \theta_{j}}\right)e^{-\theta^{'} {\bm V} +\psi(\theta)}\right] \notag \\
 &+ E_{P}\left[ 1\{g({\bm V})\in  A\}\frac{\partial^{2} \psi(\theta)}{\partial \theta_{i} \partial \theta_{j}}e^{-\theta^{'} {\bm V} +\psi(\theta)}\right].
\end{align*}
Let the first- and second-order partial derivatives of $\psi(\theta)$ be
\begin{align*}
\nabla\psi(\theta_{k}):=\frac{\displaystyle \partial \psi(\theta)}{\displaystyle \partial \theta_{k}},  \quad \nabla^{2}\psi(\theta_{k}):=\frac{ \partial^{2} \psi(\theta)}{ \partial \theta_{k}^{2}},
 \quad \hbox{and}\quad \nabla^{2}\psi(\theta_{k},\theta_{l}):=\frac{ \partial^{2} \psi(\theta)}{ \partial \theta_{k} \partial \theta_{l}}.
\end{align*}
Then, for $i\neq j$,
\begin{align*}
\frac{\partial^{2} G(\theta)}{\partial \theta_{i}\partial \theta_{j}}=E_{P}\left[ 1\{g( {\bm V})\in  A\}\left\{\left[-V_{i}+\nabla\psi(\theta_{i})\right]
\left[-V_{j}+\nabla\psi(\theta_{j})\right]+\nabla^{2}\psi(\theta_{i},\theta_{j})\right\} e^{-\theta^{'} {\bm V} +\psi(\theta)} \right],
\end{align*}
and for $i=j$,
\begin{align*}
\frac{\partial^{2} G(\theta)}{\partial \theta_{i}^2}=E_{P}\left[ 1\{g({\bm V})\in  A\}\left\{\left[-V_{i}+\nabla\psi(\theta_{i})\right]^{2}+\nabla^{2}\psi(\theta_{i})\right\}
e^{-\theta^{'} {\bm V} +\psi(\theta)} \right].
\end{align*}
Hence,
\begin{align*}
\frac{\partial^{2} G(\theta)}{\partial \theta \partial \theta^{'}}=E_{P}\left[1\{g({\bm V})\in  A\}\left\{( {\bm V}-\nabla\psi(\theta))
({\bm V}-\nabla\psi(\theta))^{'}+\nabla^{2}\psi(\theta)\right\}e^{-\theta^{'} {\bm V} +\psi(\theta)}\right],
\end{align*}
where $\nabla\psi(\theta)=(\nabla\psi(\theta_{1}),\ldots,\nabla\psi(\theta_{d}))^{'}$ and
$\nabla^{2}\psi(\theta)=(\nabla^{2}\psi(\theta_{i},\theta_{j}))_{1\le i,j\le d}$. Clearly, the matrix $\frac{\partial^{2} G(\theta)}{\partial \theta \partial \theta^{'}}$ is positive definite. The proof is completed. $\hfill \Box$

\section*{Appendix 2: Proofs of Propositions \ref{Pro1}--\ref{Pro4}, Corollary \ref{Cor1}, and Theorems \ref{SET}--\ref{LEC}}\label{appendix2}

\noindent
{\bf Proof of Proposition \ref{Pro1}}~~
This is based on Theorem~\ref{thm_theta} and a change of variables via the conditional inverse method, ${\bm V}=(V_1, \ldots, V_d)^{'}$ with $V_1, \ldots, V_d$ being independent uniform random variables over $(0,1)$ (note that, under this situation, $\Theta=\mathbb{R}^{d}$). Thus, $\d P=1$ for ${\bm V}= {\bm v}\in (0,1)^{d}$ and
\begin{align*}
\d Q = \frac{\d Q}{\d P}\d P = e^{\theta^{'} {\bm V}-\psi(\theta)} = \prod_{i=1}^{d}\frac{\theta_{i}e^{\theta_{i}V_{i}}}{e^{\theta_{i}}-1},
\end{align*}
where $\psi(\theta) = \ln \left(\prod_{i=1}^{d}\frac{\displaystyle e^{\theta_{i}}-1}{\displaystyle \theta_{i}}\right)$. Note that here the alternative measure is an independently conjugate truncated exponential on $(0,1)$ with rate parameter $\theta=(\theta_1, \ldots, \theta_d)^{'}$. The conjugate truncated exponential random variate $V_i$ means that $-V_i$ is the truncated exponential random variate.

The optimal tilting point $\theta_{o}$ is obtained by minimizing the second-order moment of the importance sampling estimator, and Theorem~1 demonstrates that $\theta_{o}$ is unique and satisfies a system of equations~(\ref{eq_theta}) provided its existence. Let $\d\bar{Q}_{\theta_{o}}=\left(\prod_{i=1}^{d}\frac{\theta_{oi}}{1-e^{-\theta_{oi}}}\right)e^{-\theta_{o}^{'}{\bm V}}\d P$. The denominator term of equation~(\ref{eq_theta}) is
\begin{align*}
E_{P}\left[1\{g({\bm V})\in A\}e^{-\theta_{o}^{'} {\bm V}}\right] = \int_{\{g( {v})\in A\}}e^{-\theta_{o}^{'} {v}}\d {v} = \left(\prod_{i=1}^{d}\frac{1-e^{-\theta_{oi}}}{\theta_{oi}}\right) E_{\bar{Q}_{\theta_{o}}}\left[1\{g({\bm V})\in A\}\right].
\end{align*}
Similarly, the numerator term is
\begin{align*}
E_{P}\left[1\{g({\bm V})\in A\}e^{-\theta_{o}^{'} {\bm V}}V_{i}\right] = \left(\prod_{i=1}^{d}\frac{1-e^{-\theta_{oi}}}{\theta_{oi}}\right) E_{\bar{Q}_{\theta_{o}}}\left[1\{g( {\bm V})\in A\}V_{i}\right]
\end{align*}
for $i=1, \ldots, d$. Therefore, the left side of equations~(\ref{eq_theta}) becomes $E_{\bar{Q}_{\theta_{o}}}[V_{i}|g({\bm V})\in A]$ for $i=1, \ldots, d$. Note that the $\bar{Q}_{\theta_{o}}$-measure is a product of independently truncated exponential on $(0,1)$ with rate parameters
$\theta_{oi}, i=1,\ldots,d$.

Next, we shall prove that the optimal tilting point $\theta_{o}$ indeed exists. We first show that $G(\theta)$ is a convex function. Note that the cumulant-generating function is convex, and $\psi(\theta)$ is the cumulant-generating function of ${\bm V}$. Therefore, for any given $\lambda \in(0,1)$, and $\theta_1, \theta_2 \in  {\Theta} \subset \mathbb{R}^{d}$, we have
\begin{align*}
& G(\lambda \theta_1 + (1 - \lambda)\theta_2) \notag \\
=& E_P\bigg[ 1\{g(\bm V)\in A\} e^{ - (\lambda \theta_1 + (1 - \lambda)\theta_2)^{'} \bm V + \psi(\lambda \theta_1 + (1 -
\lambda)\theta_2)} \bigg] \notag \\
\leq & E_P\bigg[ 1\{g(\bm V)\in A\} e^{ - (\lambda \theta_1 + (1 - \lambda)\theta_2)^{'} \bm V + \lambda \psi(\theta_1) + (1 -
\lambda)\psi(\theta_2)} \bigg]\notag \\
\leq & \lambda E_P\bigg[ 1\{g(\bm V)\in A\}  e^{ -  \theta_1^{'} \bm V + \psi(\theta_1)}\bigg]  + (1 - \lambda)
  E_P\bigg[ 1\{g(\bm V)\in A\}  e^{-\theta_2^{'} \bm V  + \psi(\theta_2)} \bigg] \notag \\
=& \lambda G(\theta_1) + (1 - \lambda )  G(\theta_2).
\end{align*}

To obtain the global minimum of $G(\theta)$, we will combine the result whereby if {\it $f: \Theta \mapsto (-\infty,\infty]$ is a convex function, then a local minimum of $f$ over $\Theta$ is also a global minimum}, with Theorem~VI.3.4. of Ellis \cite{Ellis1985}, which states that {\it $G(\theta)$ is differentiable at $\theta \in int(\Theta)$ if and only if the $d$ partial derivatives $\frac{\displaystyle \partial G(\theta)}{\displaystyle \partial \theta_{i}}$ for $i =1,\ldots, d$ exist at $\theta \in int(\Theta)$ and are finite.} Therefore, for the solution of $\frac{\displaystyle \partial G(\theta)}{\displaystyle \partial \theta} = 0$, we need only to show that $\frac{\displaystyle \partial G(\theta)}{\displaystyle \partial \theta_{i}} = 0$ has a solution, for $i =1, \ldots, d$. To this end, we seek to show that $\frac{\displaystyle \partial \psi(\theta)}{\displaystyle \partial \theta_{i}}$ is strictly increasing while $\frac{ E_{P}\left[1\{g({\bm V})\in A\}e^{-\theta^{'} {\bm V}}V_{i}\right] }{ E_{P}\left[1\{g( {\bm V})\in A\}e^{-\theta^{'} {\bm V}}\right] }$ is strictly decreasing with respect to every $\theta_i$ ($i=1,\ldots, d$), and that they have an intersection point for $i=1, \ldots, d$.

Recall again that the cumulant-generating function is convex. Thus, for all $\theta_{i}$ such that $\theta\in \Theta=\mathbb{R}^d$, the second-order partial derivative of $\psi(\theta)$ with respect to $\theta_{i}$ is positive, that is, $\frac{\displaystyle \partial^{2} \psi(\theta)}{\displaystyle \partial \theta_{i}^{2}}>0$, which implies that $\frac{\displaystyle \partial \psi(\theta)}{\displaystyle \partial \theta_{i}}$ is strictly increasing with respect to $\theta_{i}$ for $i=1, \ldots, d$.

Likewise, let the conditional measure of $P$ on the set $\{g({\bm V})\in A\}$ be $P_{A}$, defined as
\begin{align*}
\d{P}_{A}({\bm V}) = \frac{1\{g({\bm V})\in A\}e^{-\theta^{'} {\bm V}}\d P({\bm V})} {\int 1\{g({\bm V})\in A\}e^{-\theta^{'} {\bm V}}\d P({\bm V})}.
\end{align*}
Then for all $\theta_{i}$ such that $\theta\in \Theta$, we have
\begin{align*}
& \frac{\partial}{\partial \theta_{i}} \frac{ E_{P}\left[1\{g({\bm V})\in A\}e^{-\theta^{'} {\bm V}}V_{i}\right] }
{ E_{P}\left[1\{g({\bm V})\in  A\}e^{-\theta^{'} {\bm V}}\right] } \notag \\
=& -\frac{E_{P}\left[1\{g({\bm V})\in A\}e^{-\theta^{'} {\bm V}}V_{i}^{2}\right]}{E_{P}\left[1\{g({\bm V})\in  A\}e^{-\theta^{'} {\bm V}}\right]} +
\frac{E_{P}^{2}\left[1\{g({\bm V})\in A\}e^{-\theta^{'} {\bm V}}V_{i}\right]}{E_{P}^{2}\left[1\{g({\bm V})\in  A\}e^{-\theta^{'} {\bm V}}\right]} \notag\\
  =& -\mathrm{Var}_{P_{A}}(V_{i}) < 0
\end{align*}
since $V_i$ is non-degenerate. This implies that $\frac{ E_{P}\left[1\{g({\bm V})\in A\}e^{-\theta^{'} {\bm V}}V_{i}\right] }{ E_{P}\left[1\{g(\bm {\bm V})\in A\}e^{-\theta^{'} {\bm V}}\right]}$ is strictly decreasing with respect to $\theta_{i}$.

Denote $\theta_{i, {\rm max}}=\sup\{\theta_i: \Psi(\theta)<\infty\}$ for $i=1, \ldots, d$. Then, $\theta_{i, {\rm max}}=\infty$ for $i=1, \ldots, d$. Note that
$$E(V_i)=\frac{\displaystyle \partial \psi(\theta)}{\displaystyle \partial \theta_{i}}{\bigg|}_{\theta=0}\quad {\rm and}\quad E[V_i|g({\bm V}) \in A ]=\frac{ E_{P}\left[1\{g({\bm V})\in A\}e^{-\theta^{'} {\bm V}}V_{i}\right] }{ E_{P}\left[1\{g({\bm V})\in  A\}e^{-\theta^{'} {\bm V}}\right]}{\bigg|}_{\theta=0}.$$
Since $V_i\sim U(0,1)$ and $1\{g({\bm V})\in A\}e^{-\theta^{'} {\bm V}}V_i < 1\{g({\bm V})\in  A\}e^{-\theta^{'} {\bm V}}$ almost surely, we have
$$\lim_{\theta_i\rightarrow \theta_{i, {\rm max}}}\frac{ E_{P}\left[1\{g({\bm V})\in A\}e^{-\theta^{'} {\bm V}}V_i\right] }{ E_{P}\left[1\{g({\bm V})\in  A\}e^{-\theta^{'} {\bm V}}\right] }< 1.$$
In addition, recalling that $\psi(\theta) = \ln \left(\prod_{i=1}^{d}\frac{e^{\theta_{i}}-1}{\theta_{i}}\right)$, we have
$$\lim_{\theta_i\rightarrow \theta_{i, {\rm max}}}\frac{\partial \psi(\theta)}{\partial \theta_i}=\lim_{\theta_i\rightarrow \theta_{i, {\rm max}}}\left(\frac{e^{\theta_i}}{e^{\theta_i} -1}-\frac{1}{\theta_i}\right)=1.$$
As a result, the assumption $E[V_i|g({\bm V}) \in A ] > E(V_i)$ implies that $\frac{ E_{P}\left[1\{g({\bm V})\in A\}e^{-\theta^{'} {\bm V}}V_i\right] }{E_{P}\left[1\{g({\bm V})\in  A\}e^{-\theta^{'} {\bm V}}\right] }$ and $\frac{\partial \psi(\theta)}{\partial \theta_i}$ must have an intersection point which is larger than $0$ for $i=1, \ldots, d$. $\hfill \Box$ \\

\noindent
{\bf Proof of Proposition \ref{Pro2}}~~
Based on Theorem~\ref{thm_theta} and a change of variables in terms of the characteristic of Gaussian copulas, ${\bm V}$ is a multivariate normal random variable $\mathrm{MN}(0,\Sigma)$ (note that, under this situation, $\Theta=\mathbb{R}^{d}$). Thus,
\begin{align*}
\d Q & = \frac{\d Q}{\d P}\d P = e^{\theta^{'} {\bm V}-\psi(\theta)}\frac{1}{(2\pi)^{d/2}|\Sigma|^{1/2}}\exp\left\{ -\frac{1}{2} {\bm V}^{'}\Sigma^{-1} {\bm V}\right\} \notag \\
& =\frac{1}{(2\pi)^{d/2}|\Sigma|^{1/2}}\exp\left\{ -\frac{1}{2}({\bm V}-\Sigma\theta)^{'}\Sigma^{-1}({\bm V}-\Sigma\theta)\right\},
\end{align*}
where $\psi(\theta) = \frac{1}{2}\theta^{'}\Sigma\theta$, meaning that an alternative measure is $\mathrm{MN}(\Sigma\theta,\Sigma)$.

The optimal tilting point $\theta_{o}$ is obtained by minimizing the second-order moment of the importance sampling estimator, and Theorem~1 demonstrates that $\theta_{o}$ is unique and satisfies the system of equations~(\ref{eq_theta}) provided its existence. The denominator term of equation~(\ref{eq_theta}) is
\begin{align*}
&E_{P}\left[1\{g({\bm V})\in A\}e^{-\theta_{o}^{'} {\bm V}}\right]\notag \\
=& \int_{\{g({v})\in A\}}\frac{1}{(2\pi)^{d/2}|\Sigma|^{1/2}}\exp\left\{ -\frac{1}{2}( {v}+\Sigma\theta_{o})^{'}\Sigma^{-1}( {v}+\Sigma\theta_{o})\right\}\d {v}
\times \exp\left\{\frac{1}{2}\theta_{o}^{'}\Sigma\theta_{o}\right\} \notag \\
=& \exp\left\{\frac{1}{2}\theta_{o}^{'}\Sigma\theta_{o}\right\} E_{\bar{Q}_{\theta_{o}}}[1\{g({\bm V})\in A\}].
\end{align*}
Similarly, the numerator term of equation~(\ref{eq_theta}) is
\begin{align*}
E_{P}\left[1\{g({\bm V})\in A\}e^{-\theta_{o}^{'} {\bm V}}V_{i}\right] = \exp\left\{\frac{1}{2}\theta_{o}^{'}\Sigma\theta_{o}\right\} E_{\bar{Q}_{\theta_{o}}}[V_{i}1\{g({\bm V})\in A\}]
\end{align*}
for $i=1, \ldots, d$, where ${\bm V}$, under the $\bar{Q}_{\theta_{o}}$-measure, is $\mathrm{MN}(-\Sigma\theta_{o},\Sigma)$. Therefore, the left side of equation~(\ref{eq_theta}) becomes $E_{\bar{Q}_{\theta_{o}}}[V_{i}|g({\bm V})\in A]$ for $i=1,\ldots,d$.

Next, we shall prove that the optimal tilting point $\theta_{o}$ indeed exists. By checking the proof of Proposition~\ref{Pro1}, it suffices to show that $\frac{ E_{P}\left[1\{g({\bm V})\in A\}e^{-\theta^{'} {\bm V}}V_i\right] }{E_{P}\left[1\{g({\bm V})\in  A\}e^{-\theta^{'} {\bm V}}\right] }$ and $\frac{\partial \psi(\theta)}{\partial \theta_i}$ must have an intersection point for $i=1, \ldots, d$. Recall $\Theta=\mathbb{R}^d$ and $\psi(\theta) = \frac{1}{2}\theta^{'}\Sigma\theta=\frac{1}{2}\sum_{i=1}^d\sum_{j=1}^d \sigma_{ij}\theta_i\theta_j$ (by denoting $\Sigma=(\sigma_{ij})_{1\le i,j\le d}$), and denote $\theta_{i, {\rm max}}=\sup\{\theta_i: \Psi(\theta)<\infty\}$ for $i=1, \ldots, d$. Then, $\theta_{i, {\rm max}}=\infty$ for $i=1, \ldots, d$. Therefore,
$$\lim_{\theta_i\rightarrow \theta_{i, {\rm max}}}\frac{\partial \psi(\theta)}{\partial \theta_i}=\lim_{\theta_i\rightarrow \theta_{i, {\rm max}}}\left(\sum_{j=1, j\neq i}^d\sigma_{ij}\theta_j+\sigma_{ii}\theta_i\right)=\infty \quad \hbox{for}~ i=1, \ldots, d.$$
Moreover, it is not difficult to see that
\begin{align*}
\lim_{\theta_i\rightarrow \theta_{i, {\rm max}}}\frac{ E_{P}\left[1\{g({\bm V})\in A\}e^{-\theta^{'} {\bm V}}V_{i}\right] }{ E_{P}\left[1\{g({\bm V})\in  A\}e^{-\theta^{'} {\bm V}}\right]}<\infty \quad \hbox{for}~ i=1, \ldots, d.
\end{align*}
The above arguments together with the assumption $E[V_i|g({\bm V}) \in A ] > E(V_i)$ imply that $\frac{ E_{P}\left[1\{g({\bm V})\in A\}e^{-\theta^{'} {\bm V}}V_i\right]}{ E_{P}\left[1\{g({\bm V})\in  A\}e^{-\theta^{'} {\bm V}}\right] }$ and $\frac{\partial \psi(\theta)}{\partial \theta_i}$ must have an intersection point which is larger than $0$ for $i=1, \ldots, d$ by noting that
$$E(V_i)=\frac{\displaystyle \partial \psi(\theta)}{\displaystyle \partial \theta_{i}}{\bigg|}_{\theta=0}\quad {\rm and}\quad E[V_i|g({\bm V}) \in A ]=\frac{ E_{P}\left[1\{g({\bm V})\in A\}e^{-\theta^{'} {\bm V}}V_{i}\right] }{ E_{P}\left[1\{g({\bm V})\in  A\}e^{-\theta^{'} {\bm V}}\right]}{\bigg|}_{\theta=0}.$$
The proof is completed. $\hfill \Box$ \\

\noindent
{\bf Proof of Corollary \ref{Cor1}}~~
For $A=\{V_{1}>a_{1}^{*},\ldots, V_{d}>a_{d}^{*}\}$, where $a_{i}^{*}=\Phi^{-1}(F_{i}(a_{i}))$ for $i=1, \ldots ,d$, the left side of equation~(\ref{eq_theta_2}) becomes the first-order lower truncated tail moment function of the multivariate normal random variate ${\bm V}$. The moment can be computed with the integral:
\begin{align*}
E_{\bar{Q}_{\theta_{o}}}[V_{i}|V_{1}>a_{1}^{*},\ldots, V_{d}>a_{d}^{*}] &= \int_{a_{1}^{*}}^{\infty}\ldots \int_{a_{d}^{*}}^{\infty}
\frac{ v_{i}f_{{\bm V}}( {v}) }{ \bar{F}_{{\bm V}}( {a}^{*}) }\d v_{1}\ldots \d v_{d},
\end{align*}
where $f_{{\bm V}}(\cdot)$ is a joint density of ${\bm V}$ and $\bar{F}_{{\bm V}}(\cdot)$ is the survival function of ${\bm V}$ under the $\bar{Q}_{\theta_{o}}$-measure. In this case, $\bar{F}_{{\bm V}}({a}^{*})=\bar{\Phi}_{d}(a^{*}+\Sigma\theta_{o};\Sigma)$ is the survival function of a $d$-variate normal random variable with mean $0$ and covariance matrix $\Sigma$. Note that
\begin{align}
&\int_{a_{1}^{*}}^{\infty}\ldots \int_{a_{d}^{*}}^{\infty}v_{i}f_{ {\bm V}}( {v})\d v_{1}\ldots \d v_{d} \notag\\
=& \int_{ {a}^{*}}^{\infty}v_{i}(2\pi)^{-\frac{d}{2}}|\Sigma |^{-\frac{1}{2}}\exp \left\{-\frac{1}{2}( {v}+\Sigma\theta_{o})^{'}\Sigma^{-1}( {v}+\Sigma\theta_{o}) \right\}\d {v}\notag\\
=& \int_{ {a}^{*}+\Sigma\theta_{o}}^{\infty}(v_{i}-\rho_{i\cdot}\theta_{o})(2\pi)^{-\frac{d}{2}}|\Sigma |^{-\frac{1}{2}}\exp \left\{-\frac{1}{2} {v}^{'}\Sigma^{-1} {v} \right\}\d {v}\notag\\
=& \int_{ {a}^{*}+\Sigma\theta_{o}}^{\infty}v_{i}(2\pi)^{-\frac{d}{2}}|\Sigma |^{-\frac{1}{2}}\exp \left\{-\frac{1}{2} {v}^{'}\Sigma^{-1} {v} \right\}\d {v} - \rho_{i\cdot}\theta_{o}
\bar{\Phi}_{d}(a^{*}+\Sigma\theta_{o};\Sigma) \notag\\
=& \sum_{q=1}^{d}\rho_{iq} \phi(a_{q}^*+\rho_{q\cdot}\theta_{o})\bar{\Phi}_{d-1}(A_{qs};\Sigma_{q}) - \rho_{i\cdot}\theta_{o} \bar{\Phi}_{d}(a^{*}+\Sigma\theta_{o};\Sigma), \label{tallis}
\end{align}
where the first term of equation~(\ref{tallis}) results from Tallis \cite{T1961}, in which $\rho_{iq}$ is the $(i,q)$-th component of $\Sigma$, $\rho_{ q \cdot}$ is the $q$-th row vector of $\Sigma$, $A_{qs}=\frac{a_{s}^*-\rho_{sq}a_{q}^*}{\sqrt{1-\rho_{sq}^{2}}}$ for $s\neq q$ in $\bar{\Phi}_{d-1}$, and $\Sigma_{q}$ is the matrix of the first-order partial correlation coefficients of $V_{s}$ for $s\neq q$.

Furthermore, based on the assumption of $\Sigma$, the right side of equation~(\ref{eq_theta_2}) is $\rho_{i\cdot}\theta_{o}$. Then we complete this proof by a technical calculation, rewriting equation~(\ref{eq_theta_2}) as
\begin{align*}
\frac{\sum_{q=1}^{d}\rho_{iq}\phi(a_{q}^*+\rho_{q\cdot}\theta_{o})\bar{\Phi}_{d-1}(A_{qs};\Sigma_{q})}{\bar{\Phi}_{d}({a}^*+\Sigma\theta_{o};\Sigma)}-\rho_{i\cdot}\theta_{o} =\rho_{i\cdot}\theta_{o}.
\end{align*}
The proof is completed.  $\hfill \Box$\\


\noindent
{\bf Proof of Theorem \ref{SET}}~~
Denote
$$
\Psi(\theta, M) = E_P\left[\exp\left\{\theta^{'} \bm{X} + \bm{X}^{'} M \bm{X} \right\}\right].
$$
The second-order moment of $1\{\bm{X} > (p,p)^{'}\}\frac{\d P}{\d Q_{\theta, M}}({\bm X})$ under $Q_{\theta, M}$-measure is
$$
\begin{aligned}
      G(p, \theta, M) &:= E_{P}\left[1\{\bm{X} >  (p,p)^{'}\}\frac{\d P}{\d Q_{\theta, M}}({\bm X})\right] \\
      & = \int_{\{x >  (p,p)^{'}\}} \frac{\d P}{\d Q_{\theta,M}}(x)\d P(x) \\
      & = \frac{1}{2\pi} \Psi(\theta,M) \int_{\{x >  (p,p)^{'}\}} \exp\left\{-\frac{1}{2}x^{'} x - \theta^{'} x - x^{'} M x\right\} \d x \\
      & = \frac{1}{2\pi} \Psi(\theta,M) \int_{\{x >  (p,p)^{'}\}} \exp\left\{-(x-\mu)^{'} (M + \frac{1}{2}I_2) (x-\mu) + \mu^{'} (M + \frac{1}{2}I_2) \mu\right\} \d x \\
      & = \frac{1}{2\pi} \Psi(\theta,M) \exp\left\{\mu^{'} (M + \frac{1}{2}I_2) \mu\right\} \int_{\{x >  (p,p)^{'}\}} \exp\left\{-(x-\mu)^{'} (M + \frac{1}{2}I_2) (x-\mu)\right\} \d x \\
      & = \frac{1}{2\pi} \Psi(\theta,M) \exp\left\{\mu^{'} (M + \frac{1}{2}I_2) \mu\right\} \int_{\{x >  (p,p)^{'} - \mu\}} \exp\left\{-x^{'} (M + \frac{1}{2}I_2) x\right\} \d x, \\
\end{aligned}
$$
where $\mu = -\frac{1}{2}(M + \frac{1}{2}I_2)^{-1} \theta$. To insure that $\Psi(\theta,M) < \infty$, $M$ must satisfy that $M \preceq \frac{1}{2} I_2$, which has been assumed in Theorem \ref{SET}. Suppose that $(\theta_o, M_o)$ is the minimum point of $G(p, \theta, M)$ for fixed $p$, and $\theta_*$ is the minimum point of $G(p, \theta, 0)$ for fixed $p$. Then, it is clear that
$$
G(p, \theta_o, M_o) \le G(p, \theta_*, 0) \le G(p, (p,p)^{'}, 0).
$$
Therefore, for completing the proof it suffices to show that for any $\epsilon > 0$,
$$
\lim_{p \to \infty} \frac{1}{u^{2-\epsilon}} G(p,(p,p)^{'},0) = 0.
$$
Note that
\begin{align*}
      G(p,(p,p)^{'},0) & = \frac{1}{2\pi} \Psi((p,p)^{'},0) \exp\left\{\frac{1}{2}(p,p) (p,p)^{'}\right\} \int_{\{x > 2 (p,p)^{'}\}} \exp\left\{-\frac{1}{2}x^{'} x\right\} \d x \\
      & = C_1 \exp\left\{\frac{1}{2}(p,p) (p,p)^{'}\right\}\exp\left\{\frac{1}{2}(p,p) (p,p)^{'}\right\} \int_{\{x > 2 (p,p)^{'}\}} \exp\left\{-\frac{1}{2}x^{'} x\right\} \d x \\
      & = C_2 \exp\{2p^2\}[1-\Phi(2p)]^2 \\
      & \sim C_3 \exp\{2p^2\} \left[ \frac{1}{2p} \exp\left\{-\frac{1}{2}(2p)^2\right\}\right]^2\quad ({\rm as}~ p\rightarrow \infty) \\
      & = \frac{C_3}{4p^2} \exp\left\{-2p^2\right\},
\end{align*}
where $C_1, C_2$ and $C_3$ are three positive constants. Note that
$$u = P(\bm{X} >  (p,p)^{'}) = [1-\Phi(p)]^2 \sim C_4 p^{-2}\exp\{-p^2\}\quad ({\rm as}~ p\rightarrow \infty),$$
where $C_4$ is a positive constant. Then, the above arguments imply that
\begin{align*}
\lim_{p \to \infty} \frac{1}{u^{2-\epsilon}} G(p, (p,p)^{'}, 0) =& \lim_{p \to \infty} \frac{\frac{C_3}{4p^2} \exp\left\{-2p^2\right\}}{\left[\frac{C_4}{p^2}\exp\{-p^2\}\right]^{2-\epsilon}} \\
=& \lim_{p \to \infty} \frac{C_3}{4C_4^{2-\epsilon}}  p^{2-2\epsilon} \exp\{-\epsilon p^2\} \\
=& 0.
\end{align*}
This further yields
$$\lim_{p \to \infty}\frac{\mathrm{Var}_{Q_{\theta_o, M_o}}(\hat{u})}{u^{2-\epsilon}}=\frac{1}{n}\cdot\lim_{p \to \infty}\frac{G(p, \theta_o, M_o)-u^2}{u^{2-\epsilon}}=0<\infty$$
since $u\to 0$ as $p \to \infty$. The proof is completed.  $\hfill \Box$\\


\noindent
{\bf Proof of Proposition \ref{Pro3}}~~
Through equations~(\ref{event_t1}) and (\ref{event_t2}) we let $W_{i}=\sqrt{\frac{Y}{\nu}}Z_{i}-\frac{Y}{\nu}a_{i}^{*}$ for $i=1, \ldots, d$. Then, it is true that $\{X_{1}>a_{1},\ldots, X_{d}>a_{d}\} = \{\bm W>0\}$. Consider an exponential tilting on ${\bm W}=(W_{1},\ldots, W_{d})^{'}$, whose likelihood ratio is
\begin{align*}
\frac{\d Q}{\d P} = e^{\theta^{'} {\bm W}-\psi(\theta)},
\end{align*}
where $\theta=(\theta_{1},\ldots, \theta_{d})^{'}$ and $\psi(\theta)=\ln E_P[e^{\theta^{'} {\bm W}}]$ is the cumulant-generating function of $\bm W$. Note that
\begin{align*}
\d P = \frac{1}{(2\pi)^{d/2} |\Sigma|^{1/2}}\exp\left\{-\frac{1}{2} {\bm Z}^{'}\Sigma^{-1} {\bm Z}\right\}\frac{1}{\Gamma(\nu /2)2^{\nu /2}}Y^{\frac{\nu}{2}-1}\exp\left\{-\frac{Y}{2}\right\},
\end{align*}
where ${\bm Z}=(Z_{1},\ldots, Z_{d})^{'}$ and the moment-generating function of ${\bm W}$ is
\begin{align*}
\Psi(\theta)=E_P[e^{\theta^{'} {\bm W}}] = E_P \left\{ E_P [ e^{\theta^{'} {\bm W}}|Y ] \right\},
\end{align*}
in which the term $E_P[e^{\theta^{'} {\bm W}}|Y] = \exp\left\{(\frac{1}{2}\theta^{'}\Sigma\theta-\theta^{'}a^{*})\frac{Y}{\nu}\right\}$ with $a^{*}=(a_{1}^{*},\ldots, a_{d}^{*})^{'}$. Hence,
\begin{align*}
\Psi(\theta)=E_P[e^{\theta^{'} {\bm W}}] = \left[1-\frac{2}{\nu}(\frac{1}{2}\theta^{'}\Sigma\theta-\theta^{'}a^{*})\right]^{-\nu /2},\quad \theta\in \Theta,
\end{align*}
and
\begin{align*}
\psi(\theta)=\ln \Psi(\theta)=-\frac{\nu}{2} \ln\left[1-\frac{2}{\nu}(\frac{1}{2}\theta^{'}\Sigma\theta-\theta^{'}a^{*})\right],\quad \theta\in \Theta.
\end{align*}
Note that here $\Theta\neq \mathbb{R}^{d}$, because the moment-generating function of a $\chi^2$ random variable only exists on a subset of $\mathbb{R}$. Actually, here
$$\Theta=\left\{\theta:  1-\frac{2}{\nu}(\frac{1}{2}\theta^{'}\Sigma\theta-\theta^{'}a^{*})>0\right\},$$
which is compact. Then via a few techniques, the $Q$-measure is
\begin{align*}
\d Q=& \frac{1}{(2\pi)^{d/2} |\Sigma|^{1/2}}\exp\left\{-\frac{1}{2}({\bm Z}-\sqrt{\frac{Y}{\nu}}\Sigma\theta)^{'}\Sigma^{-1}( {\bm Z}-\sqrt{\frac{Y}{\nu}}\Sigma\theta)\right\} \notag \\
&  \times\frac{(1+\frac{2}{\nu}\theta^{'}a^{*}-\frac{1}{\nu}\theta^{'}\Sigma\theta)^{\nu /2}}{\Gamma(\nu /2)2^{\nu /2}}Y^{\frac{\nu}{2}-1}\exp\left\{-\frac{Y}{2}(1+\frac{2}{\nu}\theta^{'}a^{*}-\frac{1}{\nu}\theta^{'}\Sigma\theta)\right\},
\end{align*}
that is, conditional on $Y=y$, $\bm Z=(Z_{1},\ldots, Z_{d})^{'}$ is $\mathrm{MN}(\sqrt{\frac{y}{\nu}}\Sigma\theta,\Sigma)$, and $Y$ is a Gamma random variable with the shape parameter $\nu/2$ and the inverse scale parameter $(1+\frac{2}{\nu}\theta^{'}a^{*}-\frac{1}{\nu}\theta^{'}\Sigma\theta)/2$.

According to Theorem~\ref{thm_theta}, we must specify the optimal tilting point $\theta_{o}$ by minimizing $E_{P}\left[1\{{\bm W}> {0}\}\frac{\d P}{\d Q}({\bm W})\right]$, which is equal to solving the equations $\frac{\partial }{\partial \theta_{oi}}E_{P}\left[1\{{\bm W}> {0}\}\frac{\d P}{\d Q}({\bm W})\right]=0$ for $i=1, \ldots, d$, that is,
\begin{eqnarray}\label{equation_t}
\frac{E_{P}\left[1\{ {\bm W}> {0}\}e^{-\theta_{o}^{'} {\bm W}}W_{i}\right]}{E_{P}\left[1\{ {\bm W}> {0}\}e^{-\theta_{o}^{'} {\bm W}}\right]}=\frac{\partial \psi(\theta_{o})}{\partial \theta_{oi}} \quad \hbox{for}~ i=1, \ldots, d.
\end{eqnarray}

If $-\theta_o\in \Theta$, we next show that $\theta_o$ also satisfies (\ref{eq14}). We proceed to calculate the two expectations in equation
(\ref{equation_t}). First,
\begin{align*}
& E_{P}\left[1\{\bm W> {0}\}e^{-\theta_{o}^{'} {\bm W}}\right] \notag\\
=& \int_{\mathbb{R}^{d}}\int_{0}^{\infty} 1\left\{ \sqrt{\frac{y}{\nu}}z-\frac{y}{\nu}a^{*} > {0}\right\} \frac{1}{ (2\pi)^{d/2} |\Sigma|^{1/2}}\exp\left\{-\frac{1}{2}( {z}+\sqrt{\frac{y}{\nu}}\Sigma\theta_{o})^{'}\Sigma^{-1}( {z}+\sqrt{\frac{y}{\nu}}\Sigma\theta_{o})\right\}\notag \\
& \times \frac{(1-\frac{2}{\nu}\theta_{o}^{'}a^{*}-\frac{1}{\nu}\theta_{o}^{'}\Sigma\theta_{o})^{\nu /2}}{\Gamma(\nu /2)2^{\nu /2}}y^{\frac{\nu}{2}-1}\exp\left\{-\frac{y}{2}(1-\frac{2}{\nu}\theta_{o}^{'}a^{*}-\frac{1}{\nu}\theta_{o}^{'}\Sigma\theta_{o})\right\}\d y \d {z} \notag \\
&  \times\left[ 1- \frac{2}{\nu}(\frac{1}{2}\theta_{o}^{'}\Sigma\theta_{o}+\theta_{o}^{'}a^{*})\right]^{-\nu /2} \notag \\
=& E_{\bar{Q}_{\theta_{o}}}[1\{{\bm W}> {0}\}]\times \left[ 1- \frac{2}{\nu}(\frac{1}{2}\theta_{o}^{'}\Sigma\theta_{o}+\theta_{o}^{'}a^{*})\right]^{-\nu /2},
\end{align*}
where the $\bar{Q}_{\theta_{o}}$-measure is that, conditional $Y=y$, ${\bm Z}\sim \mathrm{MN}(-\sqrt{\frac{y}{\nu}}\Sigma\theta_{o},\Sigma)$, and $Y\sim \gammadist(\nu/2, 1-\frac{2}{\nu}\theta_{o}^{'}a^{*}-\frac{1}{\nu}\theta_{o}^{'}\Sigma\theta_{o}/2)$. Similarly,
\begin{align*}
E_{P}\left[1\{{\bm W}> {0}\}W_{i}e^{-\theta_{o}^{'} {\bm W}}\right]=E_{\bar{Q}_{\theta_{o}}}[W_{i}1\{{\bm W}> {0}\}]\times \left[ 1- \frac{2}{\nu}(\frac{1}{2}\theta_{o}^{'}\Sigma\theta_{o}+\theta_{o}^{'}a^{*})\right]^{-\nu /2}.
\end{align*}
Hence, the optimal tilting point $\theta_{o}$ is the solution to the following equations
\begin{align*}
E_{\bar{Q}_{\theta_{o}}}[W_{i}| {\bm W}> {0}]=\frac{\partial \psi(\theta_{o})}{\partial \theta_{oi}}, \quad  \hbox{for}~ i=1, \ldots, d.
\end{align*}

Next, we shall prove that $\theta_{o}$ indeed exists. Checking the proof of Proposition \ref{Pro1} suffices to show that
$\frac{ E_{P}\left[1\{g({\bm W})\in A\}e^{-\theta^{'} {\bm W}}W_i\right] }{ E_{P}\left[1\{g({\bm W})\in  A\}e^{-\theta^{'} {\bm W}}\right] }$ and
$\frac{\partial \psi(\theta)}{\partial \theta_i}$ must have an intersection point for $i=1, \ldots, d$. Denote $\theta_{i, {\rm max}}=\sup\{\theta_i: \Psi(\theta)<\infty\}$ for $i=1, \ldots, d$. Then, $0<\theta_{i, {\rm max}}<\infty$ for $i=1, \ldots, d$ since $\Theta$ is compact. Moreover, $\Psi(\theta)\rightarrow \infty$ as $\theta_i\rightarrow \theta_{i, {\rm max}}$. Recalling that $\psi(\theta)=\ln \Psi(\theta)=-\frac{\nu}{2} \ln\left[1-\frac{2}{\nu}(\frac{1}{2}\theta^{'}\Sigma\theta-\theta^{'}a^{*})\right]$, we have $1-\frac{2}{\nu}(\frac{1}{2}\theta^{'}\Sigma\theta-\theta^{'}a^{*})\rightarrow 0$ as $\theta_i\rightarrow \theta_{i, {\rm max}}$. Therefore,
\begin{eqnarray*}
\lim_{\theta_i\rightarrow \theta_{i, {\rm max}}}\frac{\partial \psi(\theta)}{\partial \theta_i}=\lim_{\theta_i\rightarrow \theta_{i, {\rm max}}}\frac{\sum_{j=1,j\neq i}^d \sigma_{ij}\theta_j+\sigma_{ii}\theta_i-a_i^{*}}{1-\frac{2}{\nu}(\frac{1}{2}\theta^{'}\Sigma\theta-\theta^{'}a^{*})}= \infty \quad \hbox{for}~ i=1, \ldots, d.
\end{eqnarray*}
Moreover,
\begin{align*}
\lim_{\theta_i\rightarrow \theta_{i, {\rm max}}}\frac{ E_{P}\left[1\{g({\bm W})\in A\}e^{-\theta^{'} {\bm W}}W_i\right] }{ E_{P}\left[1\{g({\bm W})\in  A\}e^{-\theta^{'} {\bm W}}\right] }<\infty \quad \hbox{for}~ i=1, \ldots, d
\end{align*}
by some algebraic manipulations. The above arguments imply that $\frac{ E_{P}\left[1\{g({\bm W})\in A\}e^{-\theta^{'} {\bm W}}W_i\right] }{ E_{P}\left[1\{g({\bm W})\in  A\}e^{-\theta^{'} {\bm
W}}\right] }$ and $\frac{\partial \psi(\theta)}{\partial \theta_i}$ must have an intersection point for $i=1, \ldots, d$ by the assumption $E[W_i|g({\bm W}) \in A ] > E(W_i)$ and noting that
$$E(W_i)=\frac{\displaystyle \partial \psi(\theta)}{\displaystyle \partial \theta_{i}}{\bigg|}_{\theta=0}\quad {\rm and}\quad E[W_i|g({\bm W}) \in A ]=\frac{ E_{P}\left[1\{g({\bm W})\in A\}e^{-\theta^{'} {\bm W}}W_{i}\right] }{ E_{P}\left[1\{g({\bm W})\in  A\}e^{-\theta^{'} {\bm W}}\right]}{\bigg|}_{\theta=0}.$$
The proof is completed.  $\hfill \Box$ \\


\noindent
{\bf Proof of Theorem \ref{LEt}}~~
Recall that $\Psi(\theta) = E_P[e^{\theta^{'} \bm{W}}]$. According to Proposition \ref{Pro3}, we have
$$
\Psi(\theta) = \left[1-\frac{1}{\nu}\theta^{'}\theta + \frac{2}{\nu}\theta^{'} (p,p)^{'}\right]^{-\nu /2},\quad \theta\in \Theta,
$$
where $\Theta$ is the domain of $\theta$ that satisfies $\Psi(\theta) < \infty$. The second-order moment of $1\{\bm{W} >  0\}\frac{\d P}{\d Q_{\theta}}({\bm W})$ under $Q_{\theta}$-measure is
\begin{align*}
       G(p, \theta) := & E_{P}\left[1\{\bm{W} >  0\}\frac{\d P}{\d Q_{\theta}}({\bm W})\right] \\
      = & \frac{1}{2\pi} 2^{-\frac{\nu}{2}} \frac{1}{\Gamma(\frac{\nu}{2})} \Psi(\theta) \int_{\{\sqrt{\frac{y}{\nu}}z - \frac{y}{\nu}(p,p)^{'} > 0\}} y^{\frac{\nu}{2} - 1} \\
      & \times \exp\left\{-\theta^{'} (\sqrt{\frac{y}{\nu}}z - \frac{y}{\nu}(p,p)^{'}) - \frac{1}{2}z^{'} z - \frac{1}{2}y\right\} \d z \d y \\
      = & C_1 \Psi(\theta) \int_{\{z > \sqrt{\frac{y}{\nu}}(p,p)^{'}\}} y^{\frac{\nu}{2} - 1} \exp\left\{-\theta^{'} (\sqrt{\frac{y}{\nu}}z - \frac{y}{\nu}(p,p)^{'}) - \frac{1}{2}z^{'} z - \frac{1}{2}y\right\} \d z \d y \\
      = & C_1 \Psi(\theta) \int_{0}^{\infty}  y^{\frac{\nu}{2} - 1} \exp\left\{-\frac{1}{2}y + \theta^{'}(p,p)^{'}\frac{y}{\nu} + \frac{1}{2}\theta^{'} \theta \frac{y}{\nu}\right\} \\
      & \times \int_{\{z > \sqrt{\frac{y}{\nu}}(p,p)^{'}\}} \exp\left\{- \frac{1}{2}(z + \sqrt{\frac{y}{\nu}}\theta)^{'} (z + \sqrt{\frac{y}{\nu}}\theta) \right\} \d z \d y \\
      = & C_1 \Psi(\theta) \int_{0}^{\infty}  y^{\frac{\nu}{2} - 1} \exp\left\{-\frac{1}{2}y + \theta^{'}(p,p)^{'}\frac{y}{\nu} + \frac{1}{2}\theta^{'} \theta \frac{y}{\nu}\right\} \\
      & \times \int_{\{z > \sqrt{\frac{y}{\nu}}((p,p)^{'} + \theta)\}} \exp\left\{- \frac{1}{2}z^{'} z \right\} \d z \d y,
\end{align*}
where $C_1$ is a positive constant. Then we will use the following inequalities to bound Gaussian tails,
\begin{align}\label{GT}
      \frac{C_0}{t} \exp\left\{-\frac{t^2}{2}\right\} &\le (\frac{1}{t} - \frac{1}{t^3}) \exp\left\{-\frac{t^2}{2}\right\} \notag \\
      &\le \int_{t}^{\infty} \exp\left\{-\frac{s^2}{2}\right\} \d s \notag \\
      &\le \frac{1}{t} \exp\left\{-\frac{t^2}{2}\right\}, \quad \forall t > 1/\sqrt{1-C_0}~ {\rm with}~ 0<C_0<1.
\end{align}
Applying the third inequality of (\ref{GT}), we have that for all $p\in (0,\infty)$ and any $\theta=(\theta_1, \theta_2)^{'}\in \Theta$,
\begin{align}
       G(p, \theta) \le & \frac{C_1 \nu \Psi(\theta)}{(p+\theta_1)(p+\theta_2)} \int_{0}^{\infty}  y^{\frac{\nu}{2} - 2} \notag \\
      & \times \exp\left\{-\frac{1}{2}y + \theta^{'}(p,p)^{'}\frac{y}{\nu} + \frac{1}{2}\theta^{'} \theta \frac{y}{\nu} -\frac{y}{2\nu}(p+\theta_1)^2 - \frac{y}{2\nu}(p+\theta_2)^2\right\}\d y \notag \\
      = & \frac{C_1 \nu \Psi(\theta)}{(p+\theta_1)(p+\theta_2)} \int_{0}^{\infty}  y^{\frac{\nu}{2} - 2} \exp\left\{-\frac{1}{2}y -\frac{y}{\nu}p^{2} \right\}\d y \notag \\
      = & \frac{C_1 \nu \Psi(\theta)}{(p+\theta_1)(p+\theta_2)} \left(\frac{1}{2} + \frac{p^2}{\nu}\right)^{-\frac{\nu}{2}+1} \Gamma\left(\frac{\nu}{2}-1\right) \notag \\
      = & \frac{C_2[\nu - \theta^{'}\theta + 2\theta^{'}(p,p)^{'}]^{-\nu/2}(\nu + 2p^{2})^{-\nu/2+1}}{(p+\theta_1)(p+\theta_2)}, \label{num}
\end{align}
where $C_2$ is a positive constant. Next, we use the first inequality of (\ref{GT}) to derive a lower bound for $u = P(\bm{W} > 0)$. For all $p\in (0,\infty)$, we have
\begin{align}
      u = & P(\bm{W} > 0) \notag \\
      = & \frac{1}{2\pi} 2^{-\frac{\nu}{2}} \frac{1}{\Gamma(\frac{\nu}{2})}  \int_{\{\sqrt{\frac{y}{\nu}}z - \frac{y}{\nu}(p,p)^{'} > 0\}} y^{\frac{\nu}{2} - 1} \exp\left\{- \frac{1}{2}z^{'} z - \frac{1}{2}y\right\} \d z \d y \notag \\
      = & \frac{1}{2\pi} 2^{-\frac{\nu}{2}} \frac{1}{\Gamma(\frac{\nu}{2})}  \int_{0}^{\infty} y^{\frac{\nu}{2} - 1} \exp\left\{- \frac{1}{2}y\right\} \int_{\{z > \sqrt{\frac{y}{\nu}}(p,p)^{'}\}} \exp\left\{-\frac{1}{2}z^{'} z\right\}\d z \d y \notag \\
      \ge & \frac{1}{2\pi} 2^{-\frac{\nu}{2}} \frac{1}{\Gamma(\frac{\nu}{2})}  \int_{\frac{\nu}{(1-C_0)p^2}}^{\infty} y^{\frac{\nu}{2} - 1} \exp\left\{- \frac{1}{2}y\right\} \int_{\{z > \sqrt{\frac{y}{\nu}}(p,p)^{'}\}} \exp\left\{-\frac{1}{2}z^{'} z\right\}\d z \d y \notag \\
     \ge & \frac{C_0^2}{2\pi} 2^{-\frac{\nu}{2}} \frac{1}{\Gamma(\frac{\nu}{2})}  \int_{\frac{\nu}{(1-C_0)p^2}}^{\infty} y^{\frac{\nu}{2} - 1} \exp\left\{- \frac{1}{2}y\right\}  \left(\sqrt{\frac{\nu}{y}}p^{-1}\right)^2\exp\left\{-\frac{y}{\nu}p^2\right\} \d y \notag \\
      = & \frac{C_0^2\nu}{2\pi} 2^{-\frac{\nu}{2}} \frac{1}{\Gamma(\frac{\nu}{2})}\int_{\frac{\nu}{(1-C_0)p^2}}^{\infty} p^{-2}y^{\frac{\nu}{2}-2} \exp\left\{ -\left(\frac{1}{2} + \frac{p^2}{\nu}\right)y \right\} \d y  \notag \\
      = & C_3 p^{-2}\int_{\frac{\nu}{(1-C_0)p^2}}^{\infty} y^{\frac{\nu}{2}-2} \exp\left\{ -\left(\frac{1}{2} + \frac{p^2}{\nu}\right)y \right\} \d y, \notag
\end{align}
where $C_3$ is a positive constant. Denote
$$
H(p) = \int_{\frac{\nu}{(1-C_0)p^2}}^{\infty} y^{\frac{\nu}{2}-2} \exp\left\{ -\left(\frac{1}{2} + \frac{p^2}{\nu}\right)y \right\} \d y.
$$
Then,
\begin{align}\label{den1}
u \ge C_3 p^{-2} H(p),
\end{align}
and we have
\begin{align}
      \frac{\partial H(p)}{\partial p} = & \int_{\frac{\nu}{(1-C_0)p^2}}^{\infty} y^{\frac{\nu}{2}-2}\left(-\frac{2p}{\nu}y\right) \exp\left\{ -\left(\frac{1}{2} + \frac{p^2}{\nu}\right)y \right\} \d y  \notag \\
      & - \left[-\frac{2\nu}{(1-C_0)p^{3}}\right]\left[\frac{\nu}{(1-C_0)p^2}\right]^{\frac{\nu}{2}-2}\exp\left\{-\left(\frac{1}{2} + \frac{p^2}{\nu}\right)\frac{\nu}{(1-C_0)p^2}\right\} \notag \\
      = & -\frac{2p}{\nu}\int_{\frac{\nu}{(1-C_0)p^2}}^{\infty} y^{\frac{\nu}{2}-1} \exp\left\{ -\left(\frac{1}{2} + \frac{p^2}{\nu}\right)y \right\} \d y \notag \\
      & + 2\left(\frac{\nu}{1-C_0}\right)^{\frac{\nu}{2}-1}p^{-\nu+1} \exp\left\{-\left(\frac{1}{2} + \frac{p^2}{\nu}\right)\frac{\nu}{(1-C_0)p^2}\right\} \notag \\
      \sim & -\frac{2p}{\nu}\left(\frac{1}{2} + \frac{p^2}{\nu}\right)^{-\frac{\nu}{2}} \int_{\frac{1}{1-C_0}}^\infty z^{\frac{\nu}{2}-1}\exp\left\{-z\right\}\d z\notag \\
      & + 2\left(\frac{\nu}{1-C_0}\right)^{\frac{\nu}{2}-1}p^{-\nu+1} \exp\left\{-\left(\frac{1}{2} + \frac{p^2}{\nu}\right)\frac{\nu}{(1-C_0)p^2}\right\} \quad ({\rm as}~ p \to \infty) \notag \\
      \sim & - 2\nu^{\frac{\nu}{2}-1} p^{-\nu + 1} \int_{\frac{1}{1-C_0}}^\infty z^{\frac{\nu}{2}-1}\exp\left\{-z\right\}\d z \notag \\
      & + 2\left(\frac{\nu}{1-C_0}\right)^{\frac{\nu}{2}-1}p^{-\nu+1} \exp\left\{-\frac{1}{1-C_0}\right\} \quad ({\rm as}~ p \to \infty) \notag \\
      = & C_4 p^{-\nu+1}, \label{den2}
\end{align}
where
$$C_4=2\left(\frac{\nu}{1-C_0}\right)^{\frac{\nu}{2}-1}\exp\left\{-\frac{1}{1-C_0}\right\}- 2\nu^{\frac{\nu}{2}-1} \int_{\frac{1}{1-C_0}}^\infty z^{\frac{\nu}{2}-1}\exp\left\{-z\right\}\d z<0$$
for all $0 < C_0 < 1$. Let $\theta_* = (\theta_{*1}, \theta_{*2})^{'} = (p, p)^{'}$. It is easy to check that $\theta_*\in \Theta$ for all large $p$. By inequality (\ref{num}), we have
\begin{equation}
G(p, \theta_*)\le \frac{C_2[\nu - \theta_*^{'}\theta_* + 2\theta_*^{'}(p,p)^{'}]^{-\nu/2}(\nu + 2p^{2})^{-\nu/2+1}}{(p+\theta_{*1})(p+\theta_{*2})} \sim C_2 2^{-\nu-1} p^{-2\nu}\quad ({\rm as}~ p \to \infty). \label{num1}
\end{equation}
Then, combining (\ref{num1}) with (\ref{den1}) and (\ref{den2}) and using L'Hospital's rule lead to
\begin{align*}
\lim_{p \to \infty} \frac{1}{u} \sqrt{G(p, \theta_o)} & \le \lim_{p \to \infty} \frac{1}{u} \sqrt{G(p, \theta_*)}  \\
& \le \lim_{p \to \infty}\frac{\sqrt{C_2 2^{-\nu-1} p^{-2\nu}}}{C_3 p^{-2}H(p)} \\
& = \lim_{p \to \infty} \frac{\sqrt{C_2 2^{-\nu-1}} p^{-\nu+2}}{C_3 H(p)} \\
& = \lim_{p \to \infty} \frac{\sqrt{C_2 2^{-\nu-1}}(-\nu+2) p^{-\nu+1}}{C_3 \frac{\partial H(p)}{\partial p}} \\
& = \lim_{p \to \infty} \frac{\sqrt{C_2 2^{-\nu-1}}(-\nu+2) p^{-\nu+1}}{C_3 C_4 p^{-\nu+1}} \\
& = \frac{\sqrt{C_2 2^{-\nu-1}}(-\nu+2)}{C_3 C_4} < \infty.
\end{align*}
This together with the fact $\mathrm{Var}_{Q_{\theta_o}}(\hat{u})=\frac{1}{n}[G(p, \theta_o)-u^2]$ imply
$$\lim_{p \to \infty}\frac{\mathrm{Var}_{Q_{\theta_o}}(\hat{u})}{u^2}<\infty.$$
The proof is completed.  $\hfill \Box$\\


\noindent
{\bf Proof of Proposition \ref{Pro5}}~~ By
$$
\psi(\theta) = -\frac{\nu}{2}\ln\left[1-\frac{1}{\nu}\theta^{'}\theta + \frac{2}{\nu}\theta^{'} (p,p)^{'}\right],
$$
it is easy to see that $\tilde{\theta}_o=\underset{\theta\in \tilde{\Theta}}{\arg\min} \psi(\theta)=(p,p)^{'}$, which is just the point $\theta_*$ in the proof of Theorem \ref{LEt}. Then, Proposition \ref{Pro5} directly follows by the proof of Theorem \ref{LEt}. $\hfill \Box$ \\

\noindent
{\bf Proof of Proposition \ref{Pro4}}~~ Note that, under the $P$-measure, $\tilde{\bm V}=(W,V_1,\ldots,V_d)^{'}$ is a $(d+1)$-variate independent
random variable, with $W\sim \gammadist(\frac{1}{\delta},1)$ and $V_{i}\sim U(0,1)$ for $i=1,\ldots,d$. Thus,
$$\Psi(\theta)=E_P[e^{\theta^{'}\tilde{\bm V}}]=(1-\theta_{W})^{-\frac{1}{\delta}}\prod_{i=1}^d \frac{e^{\theta_{i}}-1}{\theta_{i}},$$
and
$$\Theta=\{\theta: \Psi(\theta)<\infty\}=(-\infty,1)\times \mathbb{R}^d.$$
Moreover,
\begin{align*}
\d Q =& \frac{\d Q}{\d P}\d P = e^{\theta^{'} \tilde{\bm V}-\psi({\theta})}\d P \nonumber\\
=& e^{{\theta_{W}}{W}+{\theta_{1}}{V_1}+\ldots+{\theta_{d}}{V_d}}\times (1-\theta_{W})^{\frac{1}{\delta}}\prod_{i=1}^d \frac{\theta_{i}}{e^{\theta_{i}}-1}\times \frac{1}{\Gamma(\frac{1}{\delta})}W^{\frac{1}{\delta}-1}e^{-W}\nonumber\\
=& \frac{(1-\theta_{W})^{\frac{1}{\delta}}}{\Gamma(\frac{1}{\delta})} W^{\frac{1}{\delta}-1}e^{-(1-\theta_{W})W}\cdot \prod_{i=1}^d\frac{\theta_{i}e^{\theta_{i}V_i}}{e^{\theta_{i}}-1},
\end{align*}
meaning that an alternative measure for $W$ is $\gammadist(\frac{1}{\delta}, 1-\theta_{W})$, and an alternative measure for $V_i$ is the conjugate truncated exponential on $(0,1)$ with rate parameters $\theta_{i}$ for $i=1, \ldots, d$. By Theorem \ref{thm_theta}, the optimal tilting point $\theta_{o}=(\theta_{oW}, \theta_{o1}, \ldots, \theta_{od})^{'}$, provided its existence, is unique and satisfies
\begin{eqnarray*}
\begin{cases}
\displaystyle \frac{ E_{P}\left[1\{g(\tilde{\bm V})\in A\}e^{-\theta_o^{'} \tilde{\bm V}}W\right] }{ E_{P}\left[1\{g( \tilde{\bm V})\in  A\}e^{-\theta_o^{'} \tilde{\bm V}}\right] } = \frac{\partial \psi(\theta_o)}{\partial \theta_{oW}},\\
\displaystyle \frac{ E_{P}\left[1\{g(\tilde{\bm V})\in A\}e^{-\theta_o^{'} \tilde{\bm V}}V_{i}\right] }{ E_{P}\left[1\{g( \tilde{\bm V})\in  A\}e^{-\theta_o^{'} \tilde{\bm V}}\right] } = \frac{\partial \psi(\theta_o)}{\partial \theta_{oi}}, \quad \hbox{for}~ i=1, \ldots, d,
\end{cases}
\end{eqnarray*}
where
$$\psi(\theta_{o}) = -\frac{1}{\delta}\ln (1-\theta_{oW})+\sum_{i=1}^{d} \ln  \left(\frac{e^{\theta_{oi}}-1}{\theta_{oi}}\right).$$
By Theorem \ref{thm_theta} again, if $-\theta\in \Theta$, then $\theta_{o}$ also satisfies
\begin{eqnarray*}
\begin{cases}
\displaystyle E_{\bar{Q}_{\theta_{o}}}[W|g(\tilde{\bm V})\in A] = \frac{\partial \psi(\theta_{o})}{\partial \theta_{oW}},\\
\displaystyle E_{\bar{Q}_{\theta_{o}}}[V_{i}|g(\tilde{\bm V})\in A] = \frac{\partial \psi(\theta_{o})}{\partial \theta_{oi}}, \quad \hbox{for}~ i=1, \ldots, d,
\end{cases}
\end{eqnarray*}
where, under the $\bar{Q}_{\theta_{o}}$-measure, $W$ is $\gammadist(\frac{1}{\delta}, 1+\theta_{oW})$ and $V_i$ is the truncated exponential on $(0,1)$ with rate parameters $\theta_{oi}$ for $i=1, \ldots, d$.

Next, we shall prove that $\theta_{o}$ indeed exists. Checking the proof of Proposition \ref{Pro1} suffices to show that $\frac{ E_{P}\left[1\{g(\tilde{\bm V})\in A\}e^{-\theta^{'} \tilde{\bm V}}W\right] }{ E_{P}\left[1\{g( \tilde{\bm V})\in  A\}e^{-\theta^{'} \tilde{\bm V}}\right] }$ and $\frac{\partial \psi(\theta)}{\partial \theta_W}$ must have an intersection point, as well as $\frac{ E_{P}\left[1\{g(\tilde{\bm V})\in A\}e^{-\theta^{'} \tilde{\bm V}}V_i\right] }{ E_{P}\left[1\{g(\tilde{\bm V})\in  A\}e^{-\theta^{'} \tilde{\bm V}}\right] }$ and $\frac{\partial \psi(\theta)}{\partial \theta_i}$ also must have an intersection point for $i=1, \ldots, d$. Denote $\theta_{W, {\rm max}}=\sup\{\theta_W: \Psi(\theta)<\infty\}$ and $\theta_{i, {\rm max}}=\sup\{\theta_i: \Psi(\theta)<\infty\}$ for $i=1, \ldots, d$. Then, $\theta_{W, {\rm max}}=1$ and $\theta_{i, {\rm max}}=\infty$ for $i=1, \ldots, d$. Observe that
\begin{eqnarray*}
\lim_{\theta_W\rightarrow \theta_{W, {\rm max}}} \frac{\partial \psi(\theta)}{\partial \theta_W}=\lim_{\theta_W\rightarrow \theta_{W, {\rm max}}} \frac{1}{\delta(1-\theta_W)}=\infty,
\end{eqnarray*}
\begin{eqnarray*}
\lim_{\theta_i\rightarrow \theta_{i, {\rm max}}}\frac{\partial \psi(\theta)}{\partial \theta_i}=\lim_{\theta_i\rightarrow \theta_{i, {\rm max}}}\left(\frac{e^{\theta_i}}{e^{\theta_i} -1}-\frac{1}{\theta_i}\right)=1 \quad \hbox{for}~ i=1, \ldots, d,
\end{eqnarray*}
\begin{align*}
\lim_{\theta_W\rightarrow \theta_{W, {\rm max}}}\frac{ E_{P}\left[1\{g(\tilde{\bm V})\in A\}e^{-\theta^{'} \tilde{\bm V}}W\right] }{ E_{P}\left[1\{g( \tilde{\bm V})\in A\}e^{-\theta^{'} \tilde{\bm V}}\right] }<\infty,
\end{align*}
\begin{align*}
\lim_{\theta_i\rightarrow \theta_{i, {\rm max}}} \frac{ E_{P}\left[1\{g(\tilde{\bm V})\in A\}e^{-\theta^{'} \tilde{\bm V}}V_i\right] }{ E_{P}\left[1\{g(\tilde{\bm V})\in  A\}e^{-\theta^{'} \tilde{\bm V}}\right] }< 1 \quad \hbox{for}~ i=1, \ldots, d.
\end{align*}
Moreover, note that
$$E(W)=\frac{\displaystyle \partial \psi(\theta)}{\displaystyle \partial \theta_{W}}{\bigg|}_{\theta=0}, \quad E[W|g( \tilde{\bm V}) \in A ]=\frac{ E_{P}\left[1\{g(\tilde{\bm V})\in A\}e^{-\theta^{'} \tilde{\bm V}}W\right] }{ E_{P}\left[1\{g( \tilde{\bm V})\in A\}e^{-\theta^{'} \tilde{\bm V}}\right] } {\bigg|}_{\theta=0},$$
and
$$E(V_i)=\frac{\displaystyle \partial \psi(\theta)}{\displaystyle \partial \theta_{i}}{\bigg|}_{\theta=0}, \quad E[V_i|g( \tilde{\bm V}) \in A ]=\frac{ E_{P}\left[1\{g(\tilde{\bm V})\in A\}e^{-\theta^{'} \tilde{\bm V}}V_i\right] }{ E_{P}\left[1\{g(\tilde{\bm V})\in  A\}e^{-\theta^{'} \tilde{\bm V}}\right] } {\bigg|}_{\theta=0}\quad \hbox{for}~ i=1, \ldots, d.$$
Then, we claim that $\frac{ E_{P}\left[1\{g(\tilde{\bm V})\in A\}e^{-\theta^{'} \tilde{\bm V}}W\right] }{ E_{P}\left[1\{g( \tilde{\bm V})\in A\}e^{-\theta^{'} \tilde{\bm V}}\right] }$ and $\frac{\partial \psi(\theta)}{\partial \theta_W}$ must have an intersection point by the assumption $E[W|g( \tilde{\bm V}) \in A ] > E(W)$, as well as $\frac{E_{P}\left[1\{g(\tilde{\bm V})\in A\}e^{-\theta^{'} \tilde{\bm V}}V_i\right] }{ E_{P}\left[1\{g(\tilde{\bm V})\in  A\}e^{-\theta^{'} \tilde{\bm V}}\right] }$ and $\frac{\partial \psi(\theta)}{\partial \theta_i}$ also must have an intersection point by the assumption $E[V_i|g( \tilde{\bm V}) \in A ] > E(V_i)$ for $i=1, \ldots, d$. The proof is completed.  $\hfill \Box$ \\


\noindent
{\bf Proof of Theorem \ref{LEC}}~~
Recall that $\Psi(\theta) = E_P[e^{\theta^{'} \tilde{\bm{V}}}]$. According to Proposition \ref{Pro4}, we have
$$
\Psi(\theta) = (1-\theta_{W})^{-\frac{1}{\delta}} \frac{(e^{\theta_{1}}-1)(e^{\theta_2}-1)}{\theta_{1}\theta_2}, \quad \theta \in (-\infty, 1) \times \mathbb{R}^2.
$$
Notably, when $\theta_1 = \theta_2 = 0$, $\Psi((\theta_W, 0, 0)^{'}) = (1-\theta_W)^{-1/\delta}$. Let $p^* = \zeta(F(p)) \ge 0$, then the event $\{g(\tilde{\bm{V}}) > (p, p)^{'}\}$ is equivalent to $\{V_1 > e^{-Wp^*}, V_2 > e^{-Wp^*}\}$ and $p^* \to 0$ as $p \to \infty$.
The second-order moment of $1\{V_1 > e^{-Wp^*}, V_2 > e^{-Wp^*}\}\frac{\d P}{\d Q_{\theta}}(\tilde{\bm V})$ under $Q_{\theta}$-measure is
\begin{align*}
      G(p^*, \theta):=& E_{P}\left[1\{V_1 > e^{-Wp^*}, V_2 > e^{-Wp^*}\}\frac{\d P}{\d Q_{\theta}}(\tilde{\bm V})\right] \\
      = & \Psi(\theta) E_P \left[\frac{e^{-\theta_W W}}{\theta_1 \theta_2}\left(e^{-\theta_1 e^{-Wp^*}} - e^{-\theta_1}\right)\left(e^{-\theta_2 e^{-Wp^*}} - e^{-\theta_2}\right)\right].
\end{align*}
If $\theta_1 > 0$ and $\theta_2 > 0$, applying the inequality $e^x \ge 1+x$ with $x \in \mathbb{R}$ and noting that $Wp^*\ge 0$ almost surely, we have
\begin{align*}
      G(p^*, \theta) &\le \Psi(\theta) E_P \left[\frac{e^{-\theta_W W}}{\theta_1 \theta_2}\left(e^{-\theta_1 (1-Wp^*)} - e^{-\theta_1}\right)\left(e^{-\theta_2 (1-Wp^*)} - e^{-\theta_2}\right)\right] \\
      &= \Psi(\theta) \frac{e^{-\theta_1-\theta_2}}{\theta_1\theta_2}E_P \left[e^{-\theta_W W}\left(e^{Wp^*(\theta_1+\theta_2)} - e^{Wp^*\theta_1} - e^{Wp^*\theta_2} + 1\right)\right] \\
      &=: \bar{G}(p^*, \theta).
\end{align*}
Define $\theta_* = (0, \frac{1}{4}(p^*)^{-1}, \frac{1}{4}(p^*)^{-1})^{'}$. Then, by the definition of $\theta_o$, we have that
$$
G(p^*, \theta_o) \le G(p^*, \theta_*) \le \bar{G}(p^*, \theta_*).
$$
Therefore, for completing the proof it suffices to show that
\begin{align}\label{eq20}
\lim_{p^* \to 0} \frac{\bar{G}(p^*, \theta_*)}{u^2} < \infty
\end{align}
since $\mathrm{Var}_{Q_{\theta_o}}(\hat{u})=\frac{1}{n}[G(p^*, \theta_o)-u^2]$. Through some algebraic manipulations, we obtain that
\begin{align}\label{eq21}
      \bar{G}(p^*, \theta_*) = & (4p^*)^4\left[\exp\left\{-\frac{1}{2}(p^*)^{-1}\right\} - 2\exp\left\{-\frac{1}{4}(p^*)^{-1}\right\} + 1\right] E_P \left[e^{W/2} - 2e^{W/4} + 1\right] \nonumber\\
      = & C_1 (p^*)^4\left[\exp\left\{-\frac{1}{2}(p^*)^{-1}\right\} - 2\exp\left\{-\frac{1}{4}(p^*)^{-1}\right\} + 1\right] \nonumber\\
      \sim & C_1 (p^*)^4\quad ({\rm as}~ p^*\rightarrow 0),
\end{align}
where $C_1$ is a positive constant, and
\begin{align}\label{eq22}
      u = & u(p^*) = E_{P}\left[1\{V_1 > e^{-Wp^*}, V_2 > e^{-Wp^*}\}\right] \nonumber\\
      = & E_P \left[\left(1 - e^{-Wp^*}\right)^2\right] \nonumber\\
      = & (1 + 2p^*)^{-\frac{1}{\delta}} - 2(1 + p^*)^{-\frac{1}{\delta}} + 1 \nonumber\\
      \sim & C_2 (p^*)^2\quad ({\rm as}~ p^*\rightarrow 0),
\end{align}
where $C_2$ is also a positive constant. It is clear that (\ref{eq20}) follows by (\ref{eq21}) and (\ref{eq22}). The proof is completed.  $\hfill \Box$\\


\medskip

\bibliographystyle{plain}

\end{document}